\documentclass{jfm}
\usepackage[english]{babel}
\usepackage[utf8x]{inputenc}
\usepackage[T1]{fontenc}
\usepackage[a4paper,top=3cm,bottom=2cm,left=3cm,right=3cm,marginparwidth=1.75cm]{geometry}
\usepackage{amsmath}
\usepackage{graphicx}
\graphicspath{{figures/}}
\usepackage[colorinlistoftodos]{todonotes}
\usepackage[colorlinks=true, allcolors=blue]{hyperref}
\usepackage{xcolor}
\usepackage{pgfplots}

\usepackage{siunitx}
\usepackage{algorithm,amsmath,algpseudocode,algcompatible,setspace}

\usepackage{csquotes}

\graphicspath{{tikz/},{}}

\definecolor{RYB1}{RGB}{207, 37, 37}
\definecolor{RYB2}{RGB}{37, 91, 207}
\definecolor{RYB3}{RGB}{37, 207, 91}
\definecolor{RYB4}{RGB}{163,26,145}
\definecolor{RYB5}{RGB}{255, 167, 38}
\definecolor{RYB6}{RGB}{128, 177, 211}
\definecolor{RYB}{RGB}{179, 222, 105}
\definecolor{RYBblack}{RGB}{0, 0, 0}

\pgfplotscreateplotcyclelist{newcolors}{
{RYB1,every mark/.append style={fill=RYB1,mark size={1}},mark=o},
{RYB3,every mark/.append style={fill=RYB3,mark size={2}},mark=triangle},
{RYB2,every mark/.append style={fill=RYB2,mark size={1}},mark=square},
{RYB4,every mark/.append style={fill=RYB4,mark size={2}},mark=diamond},
{RYB5,every mark/.append style={fill=RYB5,mark size={2}},mark=pentagon},
{RYB6,every mark/.append style={fill=RYB6,mark size={2}},mark=10-pointed star},
{RYB6,every mark/.append style={fill=RYB6,mark size={2}},mark=+},
}

\pgfplotsset{
    standard/.style={
    compat=1.8,
        scale only axis,
        tick label style={font=\footnotesize},
        legend style={font=\scriptsize},
        width=0.5\textwidth,
        enlarge x limits=0.05,
        enlarge y limits=0.05,
    cycle list name=newcolors}
}

\title{Non-linear input/output analysis:\\ application to boundary layer transition}

\author{Georgios Rigas\aff{1}, Denis Sipp\aff{2}, Tim Colonius\aff{1}}

\affiliation{
\aff{1} Division of Engineering and Applied Science, California Institute of Technology, Pasadena, USA
\aff{2} DAAA, ONERA, Universit\'{e} Paris Saclay, 8 rue des Vertugadins, 92190 Meudon, France}

\begin{document}

\maketitle

\begin{abstract}

We extend linear input/output (resolvent) analysis to take into account nonlinear triadic interactions by considering a finite number of harmonics in the frequency domain using the harmonic balance method. Forcing mechanisms that maximize the drag are calculated using a gradient-based ascent algorithm. By including nonlinearity in the analysis, the proposed frequency-domain framework identifies the worst-case disturbances for laminar-turbulent transition.  We demonstrate the framework on a flat-plate boundary layer by considering three-dimensional spanwise-periodic perturbations triggered by a few optimal forcing modes of finite amplitude. Two types of volumetric forcing are considered, one corresponding to a single frequency/spanwise-wavenumber pair, and a multi-harmonic where a harmonic frequency and wavenumber are also added.  Depending on the forcing strategy, we recover a range of transition scenarios associated with K-type and H-type mechanisms, including oblique and Tollmien-Schlichting waves, streaks and their breakdown. We show that nonlinearity plays a critical role in optimizing growth by combining and redistributing energy between the linear mechanisms and the higher perturbation harmonics. With a very limited range of frequencies and wavenumbers, the calculations appear to reach the early stages of the turbulent regime through the generation and breakdown of hairpin and quasi-streamwise staggered vortices. 


\end{abstract}


\section{Introduction}

Methods for prediction of instability and transition have evolved considerably during the past several decades.  Advances, driven by increases in computer speed and memory, include the availability of high-fidelity DNS and LES solutions for canonical wall-bounded flows \citep{sayadi2013direct}, the recognition of transient growth (non-modal instability) as a key mechanism and mathematical formulations for {\it optimal} disturbances in linear and nonlinear frameworks \citep{SchmidHenningson2001,schmid2007nonmodal,kerswell2018nonlinear}, and generalization of parallel-flow analysis to global approaches to flows that are inhomogeneous in two or more directions \citep{theofilis}.

Most of the stability studies concern linearised evolution of perturbations.  For stable base flows, the physical mechanisms associated to linear growth mechanisms (modal and non-modal) and receptivity can be clarified by finding initial conditions in the time domain, or volumetric forcings, in the frequency domain, that maximize, for example, the kinetic energy of perturbations \citep{SchmidHenningson2001}.  The frequency-space problem is also called linear resolvent analysis or input/output analysis in the literature.  In these analyses, adjoint methods are used to maximize a specific cost function. \citet{trefethen1993hydrodynamic,jovanovic2005componentwise} showed that the computation of the optimal forcings and responses of the resolvent operator extracts the pseudo-resonances of a flowfield, that is the frequencies and spatial distributions of forcings that optimally trigger linear responses in a system. In a setup where the streamwise direction is also discretized (in addition to the cross-stream direction), accurate methods to extract the optimal features from the global resolvent have first been carried out with time-stepper approaches by \cite{blackburn2008convective,aakervik2008global,monokrousos2010global} and more recently with sparse direct LU methods by  \cite{sipp2010dynamics,brandt2011effect,rigas2017one,SchmidtJFM2017}, among others.

Determining the growth of finite-amplitude perturbations is, of course, more challenging. In practice, the direct solution of the 3D Navier-Stokes equations in the time domain is most commonly employed. For example, \cite{rist1995direct} and \cite{bake2002turbulence} reproduced experimental results evidencing different forms of transition in the flat-plate boundary layer.  More recently nonlinear transitional mechanisms have been studied by employing gradient-based techniques to find the smallest amplitude optimal initial conditions that trigger transition to turbulence \citep{cherubini2010rapid,cherubini2011minimal,pringle2012minimal,monokrousos2011nonequilibrium,kerswell2018nonlinear}. The optimal perturbation is calculated over a finite time interval and the one with the lowest energy is known as the minimal seed in the time domain.  The results still depend on the specific metric (cost function) used to measure the growth; common choices include perturbation kinetic energy \citep{pringle2012minimal,cherubini2011minimal}, integral skin friction coefficient \citep{jahanbakhshi2018nonlinearly}, dissipation \citep{monokrousos2011nonequilibrium}, and mean shear \citep{karp2017secondary}. 


The search for the minimal seed, while theoretically interesting, has no direct experimental counterpart.  By analog with the linear approaches, it is experimentally more natural to model transition from laminar to turbulent flow as a stationary process where disturbances are continually supplied to the system from the environment, i.e. to consider the receptivity problem.  For linear growth, this results in the aforementioned resolvent (or input/output) analysis that provides, in the frequency domain, a transfer function between inputs, for example environmental noise characterized by spatially localized spectral co-variance tensors, and outputs, for example the structure of the resulting amplified flow structures, and the net gain between them.
 
In order to deal with finite-amplitude perturbations in the frequency domain, the stability and numerical tools have to be extended to account for nonlinearity. Previous attempts in this regard have been limited to the nonlinear parabolized stability equations \citep[NPSE,][]{bertolotti1992linear,chang1994oblique}.  While such calculations showed good agreement with DNS for the very early stages of transition, they require specific inlet conditions to be specified and these are typically based on modal solutions to the local (parallel) spatial stability problem.  Furthermore, numerical instabilities and  robustness issues, associated with the minimum step restriction, have limited the applicability of both PSE and NPSE \citep{towne2019critical}, and cast doubt on whether PSE can be used to identify optimal inlet conditions or volumetric forcing. The aforementioned work on non-modal mechanisms relies on cooperative amplification of modes with disparate wavelengths, which raises further questions about the appropriateness of PSE ansatz.  

A natural generalization in order to calculate finite-amplitude perturbations in the frequency domain is to seek solutions to the full Navier-Stokes equations under the form of an expansion consisting of a mean-flow solution, a fundamental mode and $p$ harmonics of the fundamental, but without the parabolizing approximations inherent to PSE.  Such an approach, known in literature as the harmonic balance method \citep[HBM,][]{khalil2002nonlinear,fabrepractical} is a general method to find periodic or quasi-periodic solutions.  HBM has been used previously in fluid mechanics primarily in the context of turbomachinery \citep{hall2002computation,gopinath2007three, sicot2012time}, where one seeks a mean flow and harmonics associated with the externally imposed blade passing frequency. When used with $p=0$, HBM also recovers the \emph{self-consistent model} introduced by \citet{MLugo2014} and \citet{mantivc2016self} for the cylinder wake and backstep flow, respectively.

In this paper, using HBM we explore the optimal nonlinear amplification problem in the frequency domain, and we use the method to identify and analyze transition scenarios for the flat plate boundary layer.  We begin in \S\ref{sec:bl} by briefly reviewing the literature on boundary layer transition.  In \S\ref{sec:theory}, we propose a solution strategy for the following optimization problem. Given an amplitude $ A $,  a time-period and spanwise-wavelength associated respectively to the fundamental frequency $ \omega $ and fundamental wavenumber $ \beta $, we look for a spatial distribution of a time-periodic (of period $ 2\pi/\omega $) and spanwise-periodic (of period $2\pi/\beta$) volumetric forcing of amplitude $ A $ that triggers a solution maximising the mean skin friction coefficient (integrated over the wall).  In \S\ref{sec:Rist} we validate the HBM solver by reproducing a K-type transition scenario previously studied using DNS \citep{rist1995direct}, while in \S\ref{sec:linear_resolvent}, we validate the optimization procedure by reproducing previously reported linear optimal solutions.  Finally, in section~\ref{sec:nonlinear_resolvent} we calculate nonlinear optimal reponses and forcings that maximize the skin friction coefficient. By varying $A$, $\omega$, $\beta$ and the forcing component combinations, we identify a range of optimal transition scenarios.  We summarize our results in \S\ref{sec:conclusions}, and discuss prospects for transition prediction using HBM.


\section{Boundary layer transition: a brief review}
\label{sec:bl}

Early studies on zero-pressure gradient boundary layer transition have been mainly focused on the modal amplification of Tollmien-Schlichting (TS) waves. The primary TS-waves develop three-dimensional secondary instabilities, and subsequently break down to turbulence. The analysis of transition mechanisms resulting from the secondary instability of TS-waves has identified two main routes:
\begin{enumerate}
    \item The fundamental K-type transition, which involves a 2D TS-wave $(\omega,0)$ and two oblique waves of the same frequency $ (\omega,\pm \beta)$. Such a resonance has first been evidenced by \cite{klebanoff1962three}.
    \item The subharmonic H-type transition, triggered by a 2D TS-wave $(\omega,0)$ and two subharmonic oblique waves $ (\omega/2,\pm \beta)$. It has been experimentally observed by  \citet{kachanov1977nonlinear,kachanov1984resonant}.
\end{enumerate}

 In both cases, the oblique waves are strongly amplified, leading to $\Lambda$-shaped patterns composed of strong longitudinal vortices \citep{rist1995direct,berlin1999numerical,sayadi2013direct}. In the case of $H-$type transition, the $\Lambda-$patterns are staggered while they are aligned in the case of $K-$type transition \cite{herbert1988secondary, kachanov1994physical}. Many of the above transition characteristics can be  explained by linear modal stability analysis. \cite{herbert1988secondary} examined the secondary stability characteristics of the periodic flow (Blasius flow with superimposed TS waves) using linear Floquet analysis in a local framework. The analysis showed that the growth of three-dimensional subharmonic frequency waves (seen for H-type) is favoured over fundamental waves (K-type). 
 

More recent work shows that disturbances can undergo significant transient growth that leads to faster transition to turbulence, even at subcritical Reynolds numbers, and potentially bypassing transition through TS waves. A linear resolvent analysis for the Blasius boundary layer has been performed by \cite{monokrousos2010global} to identify optimal forcing in the frequency domain. Peaks of the optimal gain in the frequency/spanwise wavenumber space were linked to modal and non-modal instabilities. The analysis showed that maximum energy amplification is due to steady three-dimensional disturbances. The optimal forcing consists of streamwise vortices (rolls) and the response of streamwise elongated vortices, known as streaks. The amplification is a purely non-modal mechanism through the linear lift-up mechanism \citep{landahl1980note,butler1992three}. The non-modal analysis also shows that oblique TS waves are more amplified than the 2D ones, though these are linearly suboptimal to the aforementioned lift up mechanism.

Due to early observations that streaks can be significantly amplified and provide an alternative bypass route to turbulence, various studies have focused on the secondary instability of boundary layers distorted by streaks. \cite{andersson2001breakdown} performed an inviscid, secondary instability analysis of the optimally amplified boundary-layer streaks in a linear framework.  Depending on the symmetries of the perturbed flow, varicose or sinuous oscillations of the low-speed streaks are possible, with the latter being the most unstable one. Once the streaks reach certain amplitude and become unstable, breakdown to a turbulent flow is observed \citep{brandt2002transition}. The sinuous mode has been linked to the spanwise shear which leads to the formation of streamwise vortices around the low-speed streaks. On the other hand, the varicose mode has been associated with wall-normal shear and the formation of symmetric hairpin vortices \citep{asai2002instability}. 

An alternative bypass scenario for transition relies on oblique waves \citep{schmid1992new}. In this scenario, streamwise-aligned vortices are generated by non-linear interaction between a pair of oblique waves with equal angle but opposite sign in the flow direction \citep[]{schmid1992new,reddy1998stability,berlin1999numerical}. These vortices, in turn, induce streamwise streaks through the lift-up mechanism. The subsequent stages of transition to turbulence are similar to the ones described above for the streak breakdown. The initial stages of the nonlinear interaction of the oblique waves have been described also using NLPSE. \cite{chang1994oblique}  showed that the oblique waves are a dominant mechanism at low  supersonic speeds. Similarly to the incompressible regime, the nonlinear interaction of a pair of oblique waves results in the evolution of a streamwise vortex. This stage was described by a wave–vortex triad consisting of the oblique waves and a streamwise vortex whereby the oblique waves grow linearly while nonlinear effects  result in the rapid growth of the vortex mode.

\section{Nonlinear input/output analysis: theory and algorithms} \label{sec:theory}

\begin{figure}
\centering
\includegraphics[width=0.8\textwidth]{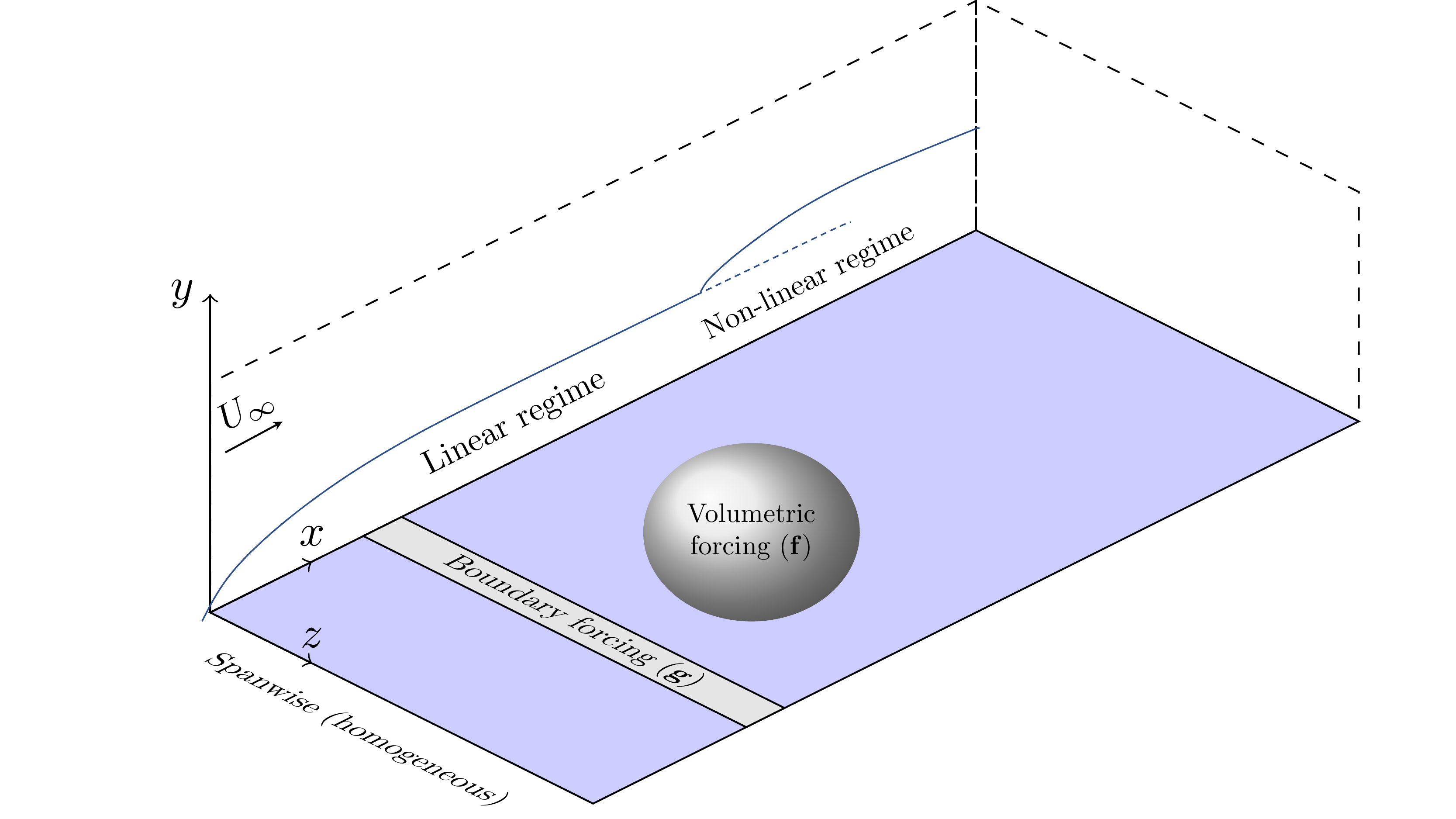}
\caption{Schematic of the zero-pressure gradient flat-plate set-up. Transition of the laminar boundary layer is triggered here either by boundary forcing (here wall blowing and suction) or volumetric momentum forcing. }
\label{fig:cartoon}
\end{figure}

In order to extend the linear input/output (resolvent) analysis to finite-amplitude perturbations, we need to proceed in two steps:
\begin{enumerate}
    \item Devise a method to find, for a given time- and spanwise-periodic finite amplitude forcing, a  time- and spanwise-periodic solution with the same periods that is solution to the forced nonlinear Navier-Stokes equations. For this,  we will follow the HBM framework. The theory and numerical algorithms are presented in \S \ref{sec:HBM}.
    \item Devise a method to search, over a fixed set of forcing and response frequencies, for an optimal forcing with a finite overall amplitude, $ A $, that maximizes a given cost-functional. Similarly to the optimization strategies followed in the time-domain \citep{kerswell2018nonlinear}, we use gradient-based strategies to find local maxima and optimal solutions in a few iterations (\S \ref{sec:optim}). 
\end{enumerate}
\subsection{Nonlinear input/output relation in frequency space with Harmonic Balance Method} \label{sec:HBM}

The flow under consideration is the zero-pressure gradient boundary layer flow, shown schematically in figure~\ref{fig:cartoon}. The spanwise direction $z$ is treated as homogeneous and, without loss of generality, we will assume that the forcing and response are $z$-periodic, in addition to being $t$-periodic.

We consider the forced three-dimensional incompressible Navier-Stokes equations:
\begin{subeqnarray} \label{eq:goveq}
    \partial_t \mathbf{u}+\mathbf{u}\cdot\nabla\mathbf{u}&=&-\nabla p+\nu\Delta\mathbf{u}+\mathbf{f}(\mathbf{x},t) \\
    \nabla \cdot \mathbf{u}&=& 0 \\
    \mathbf{u}&=&\mathbf{g}(\mathbf{x},t) \mbox{ on } \partial \Omega_f,
\end{subeqnarray}
where $ \mathbf{f} $ is a volumetric time-dependent momentum forcing and $ \mathbf{g} $ a time-dependent forcing on some boundary $ \partial \Omega_f $.  
We apply no-slip boundary conditions along the plate and zero stress conditions at the outlet. At the inlet and at the upper boundary, we impose the Blasius profile.

The governing equations are discretised in the $x$ and $y$ spatial directions, using the finite-element method, while $z$ and $t$ are treated as continuous homogeneous directions. The discretization is carried out with the FreeFem++ software \citep{freefem}, with first-order $[P_{1b},P_{1b},P_{1b},P_{1}] $ (Mini) elements \citep{arnold1984stable} for a $\mathbf{w}=[u,v,w,p]$ element.  In the discrete state space, the forcing and state variables are then vectors depending  only on $z$ and $t$, while the explicit dependence on $x$ and $y$ defines the degrees of freedom of the vectors. If we consider the compound state vector $ \mathbf{w}=[\mathbf{u},p] $, where $ \mathbf{u}=[u, v, w]$ refers to the $x$, $y$ and $z$ velocity components,
the semi-discretized governing equations \eqref{eq:goveq} may be recast in the following form:
\begin{subeqnarray} \label{eq:gov1}
    \mathbf{M} \partial_t \mathbf{w} + \mathbf{L}\mathbf{w} + \frac{1}{2} \mathbf{N}(\mathbf{w},\mathbf{w})&=& \mathbf{M}\mathbf{P} \mathbf{f}(z,t)  \\
    \mathbf{w} &=&  \mathbf{P} \mathbf{{g}}(z,t) \mbox{ on } \partial \Omega_f,
\end{subeqnarray}
where $ \mathbf{P}$ is the prolongation matrix mapping a $ [u,v,w] $ velocity vector into a $ [u,v,w,0] $ velocity-pressure vector. The matrices $ \mathbf{M}$,  $ \mathbf{L} $ and the bilinear operator $ \mathbf{N}$ are defined as:
\begin{eqnarray*}
    \mathbf{M}=\left( \begin{array}{cc} \mathbf{M}' & \mathbf{0} \\ \mathbf{0} & 0 \end{array} \right),
    \quad
    \mathbf{L}=\left( \begin{array}{cc} -\nu \Delta () & \nabla() \\ \nabla \cdot ()  & 0 \end{array} \right),
    \quad
    \mathbf{N}(\mathbf{w}_1,\mathbf{w}_2)= \left( \begin{array}{c} \mathbf{u}_1 \cdot \nabla \mathbf{u}_2+\mathbf{u}_2 \cdot \nabla \mathbf{u}_1 \\ 0 \end{array}\right),
\end{eqnarray*}
where $ \mathbf{M} $ and $ \mathbf{M}' $ are the mass matrices associated to the spatial discretization, $ \mathbf{L}$ the Stokes operator and $ \mathbf{N}$ the symmetrized nonlinear convection operator.

The volume $ \mathbf{f}(z,t) $ and boundary $ \mathbf{g}(z,t) $ forcings are assumed to be $z$-periodic of wavelength
$ \lambda = 2\pi/ \beta $ and $t$-periodic of period $T=2\pi/\omega$. We assume that the state vector $\mathbf{w}(z,t)$ behaves the same way.  
When considering boundary layers in early-stage transition, \emph{i.e} with weak external forcing amplitude, it is reasonable to assume that the response of the system follows the time-periodicity and spatial symmetries of the external forcing. For high forcing amplitude, quasi-periodic limit-cycles may appear, an investigation which is beyond the scope of the present paper.

A Fourier expansion is introduced for the periodic forcing and state variables, which is truncated at  $M+1$ harmonics in $ z $ and $ N+1 $ harmonics in $ t $.
Hence
\begin{equation} \label{eq:expansion}
\mathbf{w}(z,t) =\sum_{\substack{-M \leq m \leq M\\ -N \leq n \leq N}} e^{i(m\beta z+n\omega t)} \mathbf{\hat{w}}_{mn},
\end{equation}
with similar expansions (not shown here) for $ \mathbf{f}(z,t) $ and $\mathbf{g}(z,t) $. 
Term $ \mathbf{\hat{w}}_{mn}$ (resp. $ \mathbf{\hat{f}}_{mn}$, $ \mathbf{\hat{g}}_{mn}$
)  represents the harmonic associated to $ e^{im\beta z+in\omega t} $ for $\mathbf{\hat{w}}$ (resp. $\mathbf{\hat{f}}$ and $\mathbf{\hat{g}}$).
For these variables to be real, the following symmetry holds:
$$\mathbf{\hat{w}}_{-m,-n}=\overline{\mathbf{\hat{w}}_{mn}}$$ for all $(m,n)$, which
induces that $\mathbf{\hat{w}}_{00}$ is real. The overbar $\overline{(\cdot)}$ denotes the complex conjugate.

The discrete Fourier-transform of \eqref{eq:gov1} yields the Harmonic-Balanced Navier-Stokes (HBNS), described by the following system of coupled equations
\begin{subequations} \label{eq:nleqs}
\begin{align} 
    \left[ in\omega \mathbf{M}  + \mathbf{L}_m  + \gamma_{00}^{mn} \mathbf{N}_0^m (\mathbf{\hat{w}}_{00}, \cdot) \right]\mathbf{\hat{w}}_{mn}
    & +\sum_{{\cal S}(m,n)} \gamma_{m_1n_1}^{m_2n_2}\mathbf{N}_{m_1}^{m_2} (\mathbf{\hat{w}}_{m_1n_1},\mathbf{\hat{w}}_{m_2n_2})  = \mathbf{M}\mathbf{P}\mathbf{\hat{f}}_{mn}, \label{eq:nleqs1} \\
    \mathbf{\hat{w}}_{mn} & =\mathbf{P}\mathbf{\hat{g}}_{mn}, \mbox{ on }\partial \Omega_f \label{eq:nleqs2},
\end{align}  
\end{subequations}
for all $(m,n)$ such that $-M\leq m \leq M $ and $ -N \leq n \leq N$, and the sum is over the set of indices
\begin{align}
    {\cal S}(m,n) = \left. \begin{cases} 
    m= m_1 + m_2 & -M \le m_1 \le m_2 \le M \\
    n = n_1+n_2 & -N \le n_1 \le n_2 \le N \end{cases} \ \right| \  (m_1,n_1) \ne (0,0), (m_2,n_2) \ne (0,0)
\end{align}
The coefficients $ \gamma_{m_1n_1}^{m_2n_2}=0.5 $ if $(m_1=m_2,n_1=n_2)$ and $1$ in the other cases. The linear matrix $\mathbf{L}_{m}$ and bilinear operator $\mathbf{N}_{m_1}^{m_2}$ are deduced from $\mathbf{L}$ and $\mathbf{N} $ by replacing $\partial_z$ derivatives  by $im\beta z$. We define the solution and forcing vectors, $\mathbf{\hat{w}}$, $ \mathbf{\hat{f}} $
and $ \mathbf{\hat{g}}$ whose elements correspond to the $(2M+1)\times(2N+1)$ complex unknowns.
Then, \eqref{eq:nleqs}  may be rewritten in compact form:
\begin{subeqnarray} \label{eq:nlfinal}
    \mathbf{R}(\mathbf{\hat{w}})&=&\mathbf{M}\mathbf{P}\mathbf{\hat{f}} \\
    \mathbf{\hat{w}}&=&\mathbf{P}\mathbf{\hat{g}}, \quad \mbox{ on }\partial \Omega_f,
\end{subeqnarray}
where we reuse the symbols $ \mathbf{M} $ and $ \mathbf{P} $ to now refer to block matrices composed from the individual equations.  
For given forcing terms $\mathbf{\hat{f}} $ and $\mathbf{\hat{g}} $, equations~(\ref{eq:nlfinal}) are $ (2M+1)\times(2N+1)$ complex nonlinear equations for the unknowns $\mathbf{\hat{w}}$. Due to the fact that the equation governing the $ (m,n) $ harmonic of $ \mathbf{\hat{w}} $ corresponds to the complex conjugate of the equation governing the $(-m,-n)$ harmonic, the solution will be symmetric, $\mathbf{\hat{w}}_{-m,-n}=\overline{\mathbf{\hat{w}}_{mn}}$, whenever the forcing is.

\subsubsection{Special cases} \label{sec:sc}

In order to get some insight into the structure of the governing equations, we consider two particular cases where the boundary forcing term, $\mathbf{\hat{g}}$, is set to zero for simplicity.

In the case where $M=N=1$, equations \eqref{eq:nleqs} reduce to:
\begin{subequations} \label{eq:nleqssc2}
\begin{align}
    \left[  \mathbf{L}_0  + \frac{1}{2}\mathbf{N}_0^0 (\mathbf{\hat{w}}_{00}, \cdot) \right]\mathbf{\hat{w}}_{00}
    &+  \mathbf{N}_{-1}^{1} (\overline{\mathbf{\hat{w}}_{10}},\mathbf{\hat{w}}_{10})+\mathbf{N}_{0}^{0} (\overline{\mathbf{\hat{w}}_{01}},\mathbf{\hat{w}}_{01})+ \mathbf{N}_{-1}^{1} (\overline{\mathbf{\hat{w}}_{11}},\mathbf{\hat{w}}_{11}) \nonumber \\
    & =\mathbf{M}\mathbf{P}\mathbf{\hat{f}}_{00}, \label{eq:nleqssc2_a}\\
    \left[  \mathbf{L}_1  + \mathbf{N}_0^1 (\mathbf{\hat{w}}_{00}, \cdot) \right]\mathbf{\hat{w}}_{10}
    &
    =\mathbf{M}\mathbf{P}\mathbf{\hat{f}}_{10}, \\
     \left[ i\omega \mathbf{M}  + \mathbf{L}_0  + \mathbf{N}_0^0 (\mathbf{\hat{w}}_{00}, \cdot) \right]\mathbf{\hat{w}}_{01}
    &=\mathbf{M}\mathbf{P}\mathbf{\hat{f}}_{01}, \\
     \left[ i\omega \mathbf{M}  + \mathbf{L}_1  + \mathbf{N}_0^1 (\mathbf{\hat{w}}_{00}, \cdot) \right]\mathbf{\hat{w}}_{11}
    & =\mathbf{M}\mathbf{P}\mathbf{\hat{f}}_{11}.
\end{align}    
\end{subequations}
For a boundary layer, the terms $\mathbf{\hat{w}}_{10}e^{i\beta z}$, $\mathbf{\hat{w}}_{01}e^{i\omega t}$ and $\mathbf{\hat{w}}_{11}e^{i\beta z+i\omega t}$ may represent, respectively, a streak, a 2D Tollmien-Schlichting wave and an oblique wave.  In this case, these components are {\it linearly} triggered by the forcing terms $\mathbf{\hat{f}}_{10}$, $\mathbf{\hat{f}}_{01}$ and $\mathbf{\hat{f}}_{11}$, whereupon they deform the mean flow through the nonlinear interactions in \eqref{eq:nleqssc2_a} (in addition to any mean flow forcing, $\mathbf{\hat{f}}_{00}$).  
The linear operators $in\omega \mathbf{M}  + \mathbf{L}_m  + \mathbf{N}_0^m (\mathbf{\hat{w}}_{00}, \cdot)$ are strictly damped and thus invertible.  Connections with the Restricted Nonlinear Model (RNL) introduced for the study of transition in streamwise invariant configurations  \citep{waleffe1997RNL,biau2008optimalRNL,farrell2012dynamicsRNL} become apparent. The equations governing the steady harmonics $\mathbf{\hat{w}}_{00}$ and $\mathbf{\hat{w}}_{10}$ (which comprise the streaks and the rolls), are related to the equation governing the streamwise averaged component of the flow in the RNL equation, while those 
governing $\mathbf{\hat{w}}_{01}$ and $\mathbf{\hat{w}}_{11}$ are related to the streamwise fluctuating part (one harmonic in $\omega$ being equivalent to one streamwise wavenumber). In the present approach the spanwise direction is treated as homogeneous while the streamwise direction is solved for, while for RNL model, the opposite is true. But for both models, nonlinear interactions only appear in the mean flow equation due to the low-order truncation.

In the case $M=0,N=2$, nonlinear interactions also appear at the fluctuation level:
\begin{subequations} \label{eq:nleqssc}
\begin{align}
    \left[  \mathbf{L}_0  + \frac{1}{2}\mathbf{N}_0^0 (\mathbf{\hat{w}}_{00}, \cdot) \right]\mathbf{\hat{w}}_{00}
    &+ \mathbf{N}_{0}^{0} (\overline{\mathbf{\hat{w}}_{01}},\mathbf{\hat{w}}_{01})+ \mathbf{N}_{0}^{0} (\overline{\mathbf{\hat{w}}_{02}},\mathbf{\hat{w}}_{02})&=&\mathbf{M}\mathbf{P}\mathbf{\hat{f}}_{00},\\
     \left[ i\omega \mathbf{M}  + \mathbf{L}_0  + \mathbf{N}_0^0 (\mathbf{\hat{w}}_{00}, \cdot) \right]\mathbf{\hat{w}}_{01}
    &+ \mathbf{N}_{0}^{0} (\overline{\mathbf{\hat{w}}_{01}},\mathbf{\hat{w}}_{02})&=&\mathbf{M}\mathbf{P}\mathbf{\hat{f}}_{01}, \label{eq:nlinterterm}\\
    \left[ 2i\omega \mathbf{M}  + \mathbf{L}_0  + \mathbf{N}_0^0 (\mathbf{\hat{w}}_{00}, \cdot) \right]\mathbf{\hat{w}}_{02}
    &+ \frac{1}{2}\mathbf{N}_{0}^{0} ({\mathbf{\hat{w}}_{01}},\mathbf{\hat{w}}_{01})
    &=&\mathbf{M}\mathbf{P}\mathbf{\hat{f}}_{02}.
\end{align}    
\end{subequations}
They correspond to the extension at second order of the self-consistent model \citep{mantivc2016self} for backward-facing step flow.  We recognize the dynamics of the three harmonics $\mathbf{\hat{w}}_{00}$, $e^{i\omega t}\mathbf{\hat{w}}_{01}$ and $e^{2i\omega t}\mathbf{\hat{w}}_{02}$, the nonlinear interactions ($\mathbf{N}_{0}^{0} (\overline{\mathbf{\hat{w}}_{01}},\mathbf{\hat{w}}_{01})+ \mathbf{N}_{0}^{0} (\overline{\mathbf{\hat{w}}_{02}},\mathbf{\hat{w}}_{02})$) and forcing term ($\mathbf{\hat{f}}_{00}$) generating the mean-flow deformation, the nonlinear interactions ($\mathbf{N}_{0}^{0} (\overline{\mathbf{\hat{w}}_{01}},\mathbf{\hat{w}}_{02})$ and $1/2\mathbf{N}_{0}^{0} ({\mathbf{\hat{w}}_{01}},\mathbf{\hat{w}}_{01})$) and forcing terms ($\mathbf{\hat{f}}_{01}$ and $\mathbf{\hat{f}}_{02}$) affecting the first and second harmonics ($e^{i\omega t}\mathbf{\hat{w}}_{01}$ and $e^{2i\omega t}\mathbf{\hat{w}}_{02}$).
If higher order truncations are considered, the complexity is increased by additional nonlinear interaction terms that affect both the mean-flow and the fluctuating harmonics (see for example the term $\mathbf{N}_{0}^{0} (\overline{\mathbf{\hat{w}}_{01}},\mathbf{\hat{w}}_{02})$ in eq. \ref{eq:nlinterterm}).

\subsubsection{Algorithms and numerical methods} \label{sec:HBMalg}

In order to solve the coupled nonlinear equations \eqref{eq:nlfinal} and calculate the response $\mathbf{\hat{w}}$, we use an iterative Newton algorithm. An initial guess $\mathbf{\hat{w}}_i$ may be improved according to $\mathbf{\hat{w}}_{i+1}=\mathbf{\hat{w}}_i-\mathbf{\delta\hat{w}}_i$ with:
\begin{subeqnarray}\label{eq:newton}
     \mathbf{A}\mathbf{\delta\hat{w}}_i&=&\mathbf{R}(\mathbf{\hat{w}}_i)-\mathbf{M}\mathbf{P}\mathbf{\hat{f}} \\
    \delta\mathbf{\hat{w}}_i&=& \mathbf{\hat{w}}_i-\mathbf{P}\mathbf{\hat{g}} \mbox{ on }\partial \Omega_f,
\end{subeqnarray}
where $ \mathbf{A}=\partial \mathbf{R}/\partial \mathbf{\hat{w}} $
is the Jacobian of operator $ \mathbf{R}$, given by
\begin{equation}
\left(
\begin{array}{cccc}
      \mathbf{L}_0 + \mathbf{N}_0^0    (\mathbf{\hat{w}}_{00},\cdot) & \mathbf{N}_0^0    (\mathbf{\hat{w}}_{0,-1},\cdot) & \mathbf{N}_0^0    (\mathbf{\hat{w}}_{01},\cdot) & \cdots \\
      \mathbf{N}_0^0    (\mathbf{\hat{w}}_{01},\cdot) & i \omega \mathbf{M} + \mathbf{L}_0 + \mathbf{N}_0^0    (\mathbf{\hat{w}}_{00},\cdot) & \mathbf{N}_0^0    (\mathbf{\hat{w}}_{02},\cdot) & \cdots \\
    \mathbf{N}_0^0    (\mathbf{\hat{w}}_{0,-1},\cdot) &\mathbf{N}_0^0    (\mathbf{\hat{w}}_{0,-2},\cdot) & 
     -i \omega \mathbf{M} + \mathbf{L}_{0} + \mathbf{N}_0^0    (\mathbf{\hat{w}}_{00},\cdot) & \cdots \\ \vdots & \vdots & \vdots &\ddots
\end{array}\right),
\end{equation}
where the off-diagonal blocks stem from non-linear interactions between harmonics, while the diagonal blocks correspond to Navier-Stokes equations linearized around the current mean-flow $ \hat{\mathbf{w}}_{00} $. This matrix is also known in the literature as the finite-dimensional block Hill-matrix \citep{lazarus2010harmonic}.

The linear problem \eqref{eq:newton} involves a large number of unknowns, equal to the number of harmonics $ (2N+1)(2M+1)$ times the number of degrees of freedom in a velocity-pressure vector on a two-dimensional computational mesh. If the number of retained harmonics is large, solution of the linear system  becomes the pacing item, primarily due to associated computer memory limitations rather operation counts, when a direct LU method is used.  Iterative solvers for HBM problems partially bypass these limitations \citep{hall2002computation,gopinath2007three, sicot2012time}.  In order to decrease the computational cost, we follow  \cite{moulin2019augmented} and use a preconditioned 
Generalized Minimal Residual (GMRES) algorithm that only requires matrix-vector products. We use a block-Jacobi preconditioner, where the blocks correspond to the harmonics: $ \mathbf{\hat{w}}_{00}$, $ (\mathbf{\hat{w}}_{01},\mathbf{\hat{w}}_{0,-1})$, etc. The block-Jacobi preconditioner is very efficient when the diagonal blocks of matrix $\mathbf{A}$ are dominant, that is when the nonlinear interactions between harmonics remain reasonably weak. This occurs when the amplitude $ A $ of the forcing remains small.
The code is parallel with each processor handling a block. In the block-Jacobi preconditioner, the linear system associated to the diagonal block of a given harmonic, for example
\begin{equation} \label{eq:blockdiag}
\left(
\begin{array}{cc}
    i n \omega \mathbf{M} + \mathbf{L}_m + \mathbf{N}_0^m    (\mathbf{\hat{w}}_{00},\cdot) & {\mathbf{A}'} \\
    \overline{\mathbf{A}'} & 
     -i n \omega \mathbf{M} + \mathbf{L}_{-m} + \mathbf{N}_0^{-m}    (\mathbf{\hat{w}}_{00},\cdot)
    \end{array}
\right),
\end{equation}
is solved by the processor handling the harmonic $(\hat{\mathbf{w}}_{mn},\hat{\mathbf{w}}_{-m,-n}) $ with a sparse direct LU method \citep{MUMPS:1}. For an efficient distributed implementation, we use the PETSc software \citep{petsc-web-page} with the scalable linear equation solver component (KSP). Since a single processor solves for a system involving matrix \eqref{eq:blockdiag}, the size of the mesh needs to remain reasonable. Should larger meshes be required, domain decomposition could be used to distribute each harmonic over several processors.

To obtain a good initial guess, we solve the linear problem, which uncouples the equations and may be solved with a direct LU. For larger $A$, we continue in steps from smaller $A$.  Likewise, we may increment $M$ and $N$ as the iteration proceeds.

\newcommand{\elim}[1]{
Note that the implementation can also be carried out with just real variables.
In such a case, instead of the structure \eqref{eq:structcmplx}, vector $ \mathbf{\hat{w}} $ contains
the $ \mathbf{\hat{w}}_{00} $
harmonic and the real and imaginary parts of the other harmonics:
\begin{equation}
    \mathbf{\hat{w}}=\left( \begin{array}{c} \mathbf{\hat{w}}_{00} \\
    \mathbf{\hat{w}}_{01r} \\
    \mathbf{\hat{w}}_{01i} \\
    \mathbf{\hat{w}}_{02r} \\
    \mathbf{\hat{w}}_{02i} \\
    \vdots \\
    \mathbf{\hat{w}}_{MNr} \\
    \mathbf{\hat{w}}_{MNi}
    \end{array} \right).
\end{equation}
The size of the new real unknown $ \mathbf{\hat{w}} $
is therefore the same as in the complex case.
The nonlinear $ \mathbf{R} $ operator and the $ \mathbf{A} $ matrix may be rewritten in real formats.
Doing so, the size of the $ \mathbf{A} $ matrix is left unchanged and it therefore seems advantageous to switch to a real formalism (a complex scalar multiplication is equivalent to four real scalar multiplications).
This is only partly true since the diagonal block of the real $ \mathbf{A} $ matrix is now:
\begin{equation} \label{eq:blockdiagreal}
\left(
\begin{array}{cc}
     \mathbf{L}_m^r + \mathbf{N}_0^{m,r}    (\mathbf{\hat{w}}_{00},\cdot) + {\mathbf{A}'^{r}} & - n \omega \mathbf{M}-\mathbf{L}_m^i-\mathbf{N}_0^{m,i}    (\mathbf{\hat{w}}_{00},\cdot)  + {\mathbf{A}'^{i}} \\
     n \omega \mathbf{M}+\mathbf{L}_m^i+\mathbf{N}_0^{m,i} + {\mathbf{A}'^{i}} &
     \mathbf{L}_m^r + \mathbf{N}_0^{m,r}    (\mathbf{\hat{w}}_{00},\cdot) -{\mathbf{A}'^{r}}
    \end{array}
\right).
\end{equation}
As can be seen, the number of non-zero elements of the matrix has roughly doubled, so that a real implementation may only reduce the cost by a small amount.
}

\subsubsection{Reflectional symmetry in  $z$}\label{sec:HBMsym}
For a reflectionally symmetric solution with respect to $z=0$, we restrict the forcing so that
\begin{subequations}\label{eq:symmetryz}
\begin{align}
    f_x(-z,t) = f_x(z,t) \ &\implies \ \hat{f}_{x}(-m,n) = {\hat{f}_{x}(m,n)}, \\
    f_y(-z,t) = f_y(z,t) \ &\implies \ \hat{f}_{y}(-m,n) = {\hat{f}_{y}(m,n)},  \\
    f_z(-z,t) = -f_z(z,t) \ &\implies \  \hat{f}_{z}(-m,n) = -{\hat{f}_{z}(m,n)}.
\end{align}
\end{subequations}
Imposing symmetry on $ \mathbf{f} $ and $\mathbf{g}$ requires that the spanwise velocity component must be set to zero at the inlet boundary.  Imposing the same symmetries on the solution reduces the number of unknowns by about a factor of 2.  These symmetric solutions, it must be stressed, may be unstable to asymmetrical disturbances.


\subsection{Optimal forcings (nonlinear resolvent)} \label{sec:optim}

For brevity, we only consider optimal volumetric forcings $ \mathbf{\hat{f}} $.  Inlet- and wall-forcings $ \mathbf{\hat{g}} $ can be handled in an analogous manner. We pose a procedure to find the forcing $ \mathbf{\hat{f}} $ that maximizes a positive, real-valued cost-functional $ J(\mathbf{\hat{w}}) $, under the constraint that $\mathbf{\hat{w}}$ is a solution to the HBNS nonlinear problem forced by $ \mathbf{\hat{f}} $ with finite amplitude $ A $.    To solve the constrained optimization, we consider the Lagrangian functional
\begin{equation}
    \mathcal{L}(\mathbf{\hat{w}},[\mathbf{\tilde{w}},\lambda],\mathbf{\hat{f}})=J(\mathbf{\hat{w}})-\mathbf{\tilde{w}}^*\left( \mathbf{R}(\mathbf{\hat{w}})-\mathbf{M}\mathbf{P}\hat{\mathbf{f}}\right)-\lambda\left(\hat{\mathbf{f}}^*\mathbf{Q}\hat{\mathbf{f}} -A^2 \right),
\end{equation}
where $ \mathbf{\tilde{w}} $ and $ \lambda $ are Lagrange multipliers enforcing the constraints. The $\lambda$-constraint is that the forcing $\mathbf{\hat{f}}$ must exhibit a prescribed amplitude $ A $:
\begin{equation} \label{eq:surface}
\mathbf{\hat{f}}^*\mathbf{Q}\mathbf{\hat{f}}=A^2,
\end{equation}
where $\mathbf{Q}$ is a positive-definite Hermitian matrix defining a norm on the forcing space $\mathbf{\hat{f}}$.  Proceeding in the usual way by zeroing the variations of ${\cal L}$ with respect to $\mathbf{\tilde{w}}$ and $\lambda$ yields the constraints, whereas variations w.r.t. $\mathbf{\tilde{w}}$ gives an equation for the adjoint state,
\begin{equation}
 \mathbf{A}^*\mathbf{\tilde{w}}=\frac{dJ}{d\hat{\mathbf{w}}},
\end{equation}
and variations w.r.t. $\mathbf{\hat{f}}$ lead to a relation
\begin{eqnarray}
     \underbrace{\mathbf{Q}^{-1}\mathbf{P}^*\mathbf{M}\mathbf{\tilde{w}}}_{\mathbf{\tilde{w}'}} -2\lambda \mathbf{\hat{f}} & = 0,
\end{eqnarray}
that shows that  $\mathbf{\hat{f}}$ needs to be parallel to $\mathbf{\tilde{w}'}$. A convergence criteria (to a local maximum) is that the angle $ \theta $ between these two vectors vanishes
        \begin{equation}
    \cos(\theta) = \frac{\mathbf{\hat{f}}^*\mathbf{Q} \mathbf{\tilde{w}'}}{A \gamma} = 1,
    \end{equation}
    where $\gamma=\sqrt{\mathbf{\tilde{w}'^*}\mathbf{Q} \mathbf{\tilde{w}'}}$.
    
Following \citet{kerswell2018nonlinear}, the algorithm for the update of $\mathbf{\hat{f}}$ is based on steepest ascent:
$$\mathbf{\hat{f}}_{\mbox{new}}=\mathbf{\hat{f}}+A \epsilon (\mathbf{\tilde{w}'} - 2\lambda \mathbf{\hat{f}}),$$
where the Lagrange parameter $\lambda$ is chosen such that it constraints the forcing energy $\mathbf{\hat{f}}_{\mbox{new}}^*\mathbf{Q}\mathbf{\hat{f}}_{\mbox{new}}=A^2$, and $ \epsilon $ governs the amplitude change between 
 $\mathbf{\hat{f}}$ and $\mathbf{\hat{f}}_{\mbox{new}}$.
 The parameter $ \epsilon $ may be chosen as  $\epsilon=c/\gamma$ where $0 < c \leq 1 $ to allow a solution for $ \lambda$.
 
 The explicit steps of the iterative procedure are detailed in algorithm \ref{nlalgorith}. 
 The parameter $c$ can be fixed to 1 if the guess $ \hat{\mathbf{f}}$ is close to the optimum. 
 If not, large derivatives of the cost functional (i.e transition) can lead to large drifts of $ \hat{\mathbf{f}}$, which may destabilize the Newton algorithm. In such a case, lower values of $c$ need to be imposed. In the present study, a good compromise was found with $c = 0.5$, for which
 most of the cases converged, without penalizing too much the number of iterations for the Newton method to converge. In a few cases, we had to decrease the value of $c$ down to $ c=0.2$. The stopping criterion was chosen so that the alignment $\theta$ is less than $\theta_c=1^\circ$.

\begin{algorithm}[t!]
  \caption{Nonlinear Optimization using HBNS}\label{nlalgorith}
  \begin{algorithmic}[1]
\setstretch{1.0}
    \State 
    Initialize. Set stopping criterion $\theta_c$. Let $\mathbf{\hat{f}}_n$ be an approximation of a maximum of $
    J(\mathbf{\hat{w}})$ such that 
    $$
    \mathbf{\hat{f}}_n^* \mathbf{Q}\mathbf{\hat{f}}_n=A^2. 
    $$

    \State \label{step:HBM}
    Solve the nonlinear HBNS system \eqref{eq:nlfinal} to determine the state  $\mathbf{\hat{w}}_n$, using the iterative Newton method and the iterative preconditioned GMRES algorithm (\S \ref{sec:HBMalg})
    $$
    \mathbf{R}(\mathbf{\hat{w}}_n)=\mathbf{M}\mathbf{P}\hat{\mathbf{f}}_n.
    $$

    \State
    Solve the linear system for the adjoint state $\mathbf{\tilde{w}}_n$, using the same iterative preconditioned GMRES algorithm (\S \ref{sec:HBMalg})
    $$
    \mathbf{A}^*\mathbf{\tilde{w}}_n=\left.\frac{dJ}{d\mathbf{\hat{w}}}\right|_{\mathbf{\hat{w}}_n}.
    $$

    \State
    Set
    $\mathbf{\tilde{w}'}_n=\mathbf{Q}^{-1}\mathbf{P}^*\mathbf{M}\mathbf{\tilde{w}}_n $, compute the norm $ \gamma_n=\sqrt{\mathbf{\tilde{w}}_n^{'*}\mathbf{Q} \mathbf{\tilde{w}}'_n}$ and evaluate alignment angle 
    $$
    \cos(\theta_n)=\mathbf{\hat{f}}_n^*\mathbf{Q} \mathbf{\tilde{w}}'_n/(A \gamma_n).
    $$
        
    \IF{$|\cos(\theta_n)|>\cos(\theta_c)$}
              \State 
              Break. 
              Return $(\mathbf{\hat{f}_n},\mathbf{\hat{w}_n}) $, which is a reasonable approximation of an extremum.
    \ELSE 
              \State
              Update $\mathbf{\hat{f}}$:
            \begin{eqnarray} \lambda_n=\frac{1+\epsilon_n \gamma_n \cos\theta_n -\sqrt{1-\epsilon_n^2\gamma_n^2\sin^2\theta_n}}{2A\epsilon_n} \label{eq:enforceampl}, \quad
            \epsilon_n=\frac{c }{\gamma_n}. \label{eq:cond}
            \end{eqnarray}
              $$
                \mathbf{\hat{f}}_{n+1} = \mathbf{\hat{f}}_{n}+ A\epsilon_n  (\mathbf{\tilde{w}}'_n -2\lambda_n \mathbf{\hat{f}}_n),
              $$
        
            \State
            Go to \ref{step:HBM}.
    \ENDIF
  \end{algorithmic}
\end{algorithm}

\section{HBNS: validation for controlled transition}\label{sec:Rist}
%
In this section, we validate the HBNS implementation described above against the DNS of controlled K-type transition by \citet{rist1995direct}.
We consider the free-stream velocity $ U_\infty $ and $ \nu/U_\infty $ as reference velocity and length scales  throughout the manuscript. For this specific choice we have $x \mapsto Re_x$. 
The computational domain for the zero-pressure flat-plate configuration is rectangular with the plate located at $y=0$, the upper boundary at $y=\num{1.2 e5} $, the inlet at $ x_i=\num{0.30e5}$ and the outlet at $ x_o=\num{2.52e5} $.

 The volumetric forcing $\mathbf{\hat f}$ is set to zero and perturbations are triggered through $\mathbf{\hat g}$ which is chosen to represent wall-normal forcing by local time-dependent blowing and suction within a narrow strip at the wall. 
Thus, in accordance with \citet{rist1995direct}, we impose $u=w=0$ and
\begin{eqnarray}
v(x,z,t)&=& \num{5e-3} \sin(\omega t) v_a(x) + \num{1.3e-4}  \cos(\beta z) v_s(x),
\end{eqnarray}
which represents a superposition of a 2D planar TS wave $(0,\omega)$ of frequency $ \omega=\num{11e-5} $ and a steady oblique wave $(\beta,0)$ of wavenumber $\beta=\num{42.3e-5}$.  The specific profiles of the wall-normal velocity of the unsteady and steady waves, which are localized between $x_1$ and $x_2$ on the wall boundary, are given by:
\begin{eqnarray}
 v_a(x)&=&\left\{ \begin{array}{ccc} 0 & , &  x \leq x_1 \\
 15.1875 \xi^5-35.4375\xi^4+20.25\xi^3 & , & 
 x_1 < x \leq x_m \\
 -v_a(2x_m-x) & , & 
 x_m < x \leq x_2 \\
 0 & , & x_2 < x
 \end{array}\right. \\
  v_s(x)&=&\left\{ \begin{array}{ccc} 0 & , &  x \leq x_1 \\
 -3 \xi^4+4\xi^3 & , & 
 x_1 < x \leq x_m \\
 v_s(2x_m-x) & , & 
 x_m < x \leq x_2  \\
 0 & , & x_2 < x
 \end{array}\right.
\end{eqnarray}
Here: $x_1=\num{1.3438e5}$, $ x_2=\num{1.5532e5}$, $ x_m=(x_1+x_2)/2$ and $\xi=\frac{x-x_1}{x_m-x_1}$. 

Due to the symmetry of the wall forcing, spanwise reflectional symmetry was assumed enforcing equations \eqref{eq:symmetryz}. The mean flow harmonic $\hat{\mathbf{w}}_{00}$ was initialized with the base-flow solution and the other harmonics were set to zero except the $(0,\omega)$ and $(\beta,0)$ harmonics, which were initialized with the linearized responses. For $M=N=2$ (9 harmonics in total) the solution of the HBNS system converged after 9 Newton iterations (residuals of the order of $ 10^{-10} $).
The $M=N=3$ (16 harmonics) solution was obtained using as initial guess the $M=N=2$ solution and converged after 4 iterations, whereas the $M=N=4$ (25 harmonics) solution was obtained from the $M=N=3$ one in 4 iterations.

\begin{figure}
\centering
\hspace{0.5cm}
\includegraphics[width=0.7\textwidth]{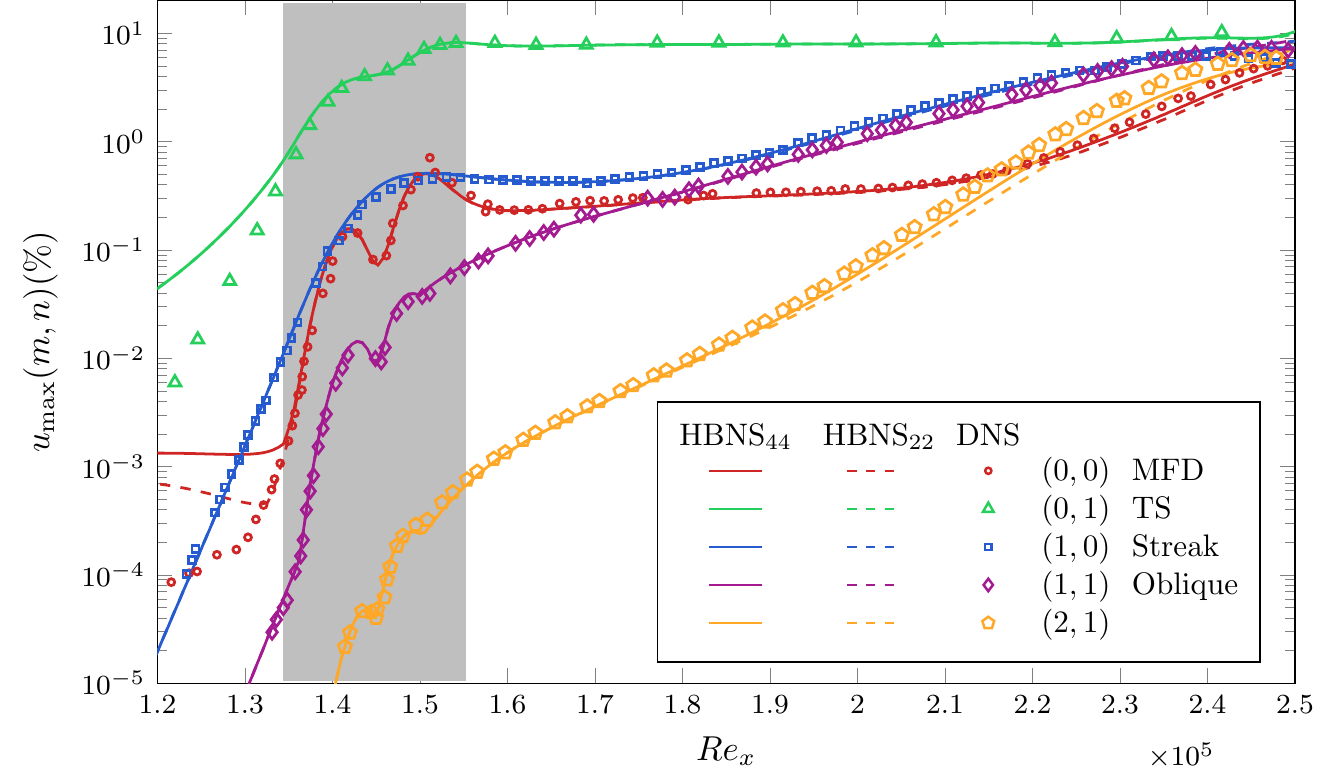}
\caption{K-type controlled transition. Comparison between DNS~\citep{rist1995direct} and Harmonic-Balanced Navier-Stokes retaining  $M=N=2$ (HBNS$_{22}$) and $M=N=4$ (HBNS$_{44}$) harmonics in spanwise/frequency. The grey region denotes the streamwise extent of the wall blowing and suction region  that triggers K-type transition. Note that to ease representation, we have plotted one fifth of the amplitude of harmonics $(0,\omega)$ and $(2\beta,\omega)$.}
\label{fig:comp}
\end{figure}

\begin{figure}
\centering
\includegraphics[width=0.8\textwidth,trim={0cm 28cm 0cm 15cm},clip]{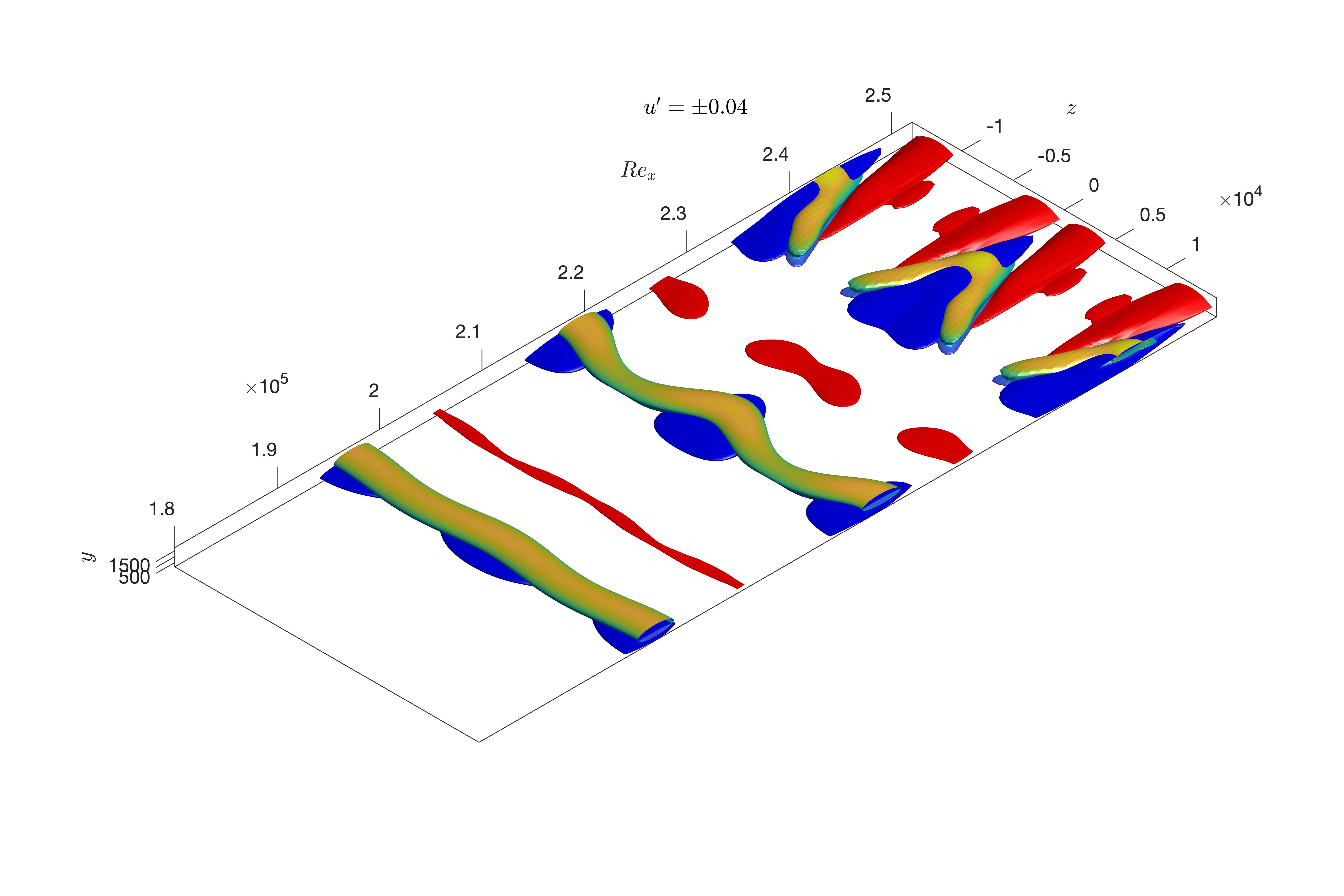}
\caption{K-type controlled transition with $\mathit{HBNS_{44}}$. Isosurfaces of pertubation velocity $u'=\pm 0.04$ (red: high speed, blue: low speed) and of the second invariant of the velocity gradient tensor, $Q$, coloured by the vertical distance from the wall ($Q = \num{2e-9}$).}
\label{fig:Qu_Rist}
\end{figure}

In figure~\ref{fig:comp}, we compare the amplitude of the first few harmonics from the HBNS against the DNS results obtained by \citet{rist1995direct}. A sensitivity analysis of the domain length and of the finite element discretization is given in appendix~\ref{app:sensitivityKtype}. For plotting the $(0,0)$ harmonic component, we have subtracted the base-flow solution, which leads to the mean flow deformation (MFD). The definition of the amplitudes of the different harmonics are described in appendix \ref{app:HarmonicAmplitudes}. The wall-normal forcing excites initially planar TS waves $(0,\omega)$ and streamwise vortices/streaks $(\beta,0)$ at a given frequency and spanwise wavelength. Oblique waves $(\beta,\pm\omega)$ and higher harmonics are generated through nonlinear interactions. Similarly, the self-interaction of the modes when they reach sufficiently high amplitudes, generates $(0,0)$ components that cause departure of the mean-flow harmonic from the base-flow solution. 
Even with $M=N=2$, good agreement is obtained for the fundamental $(0,\omega)$ and $(\beta,0)$ harmonics and for the oblique wave $(\beta,\omega)$. As the perturbations grow in the streamwise direction, the $M=N=4$ results are in slightly better with the DNS for the higher harmonic $(2\beta,\omega)$.

In figure~\ref{fig:Qu_Rist}, isosurfaces of streamwise velocity show low-speed velocity streaks (blue) developing in the streamwise direction. Isosurfaces of the $Q$-criterion, colored based on the normal distance from the wall, show $\Lambda$-vortices sitting on low-speed streaks.  They are elongated and move away from the wall as they propagate downstream, in accordance with \citet{rist1995direct}.

\begin{figure}
\vspace{0.5cm}
\centering
\includegraphics[width=0.5\textwidth]{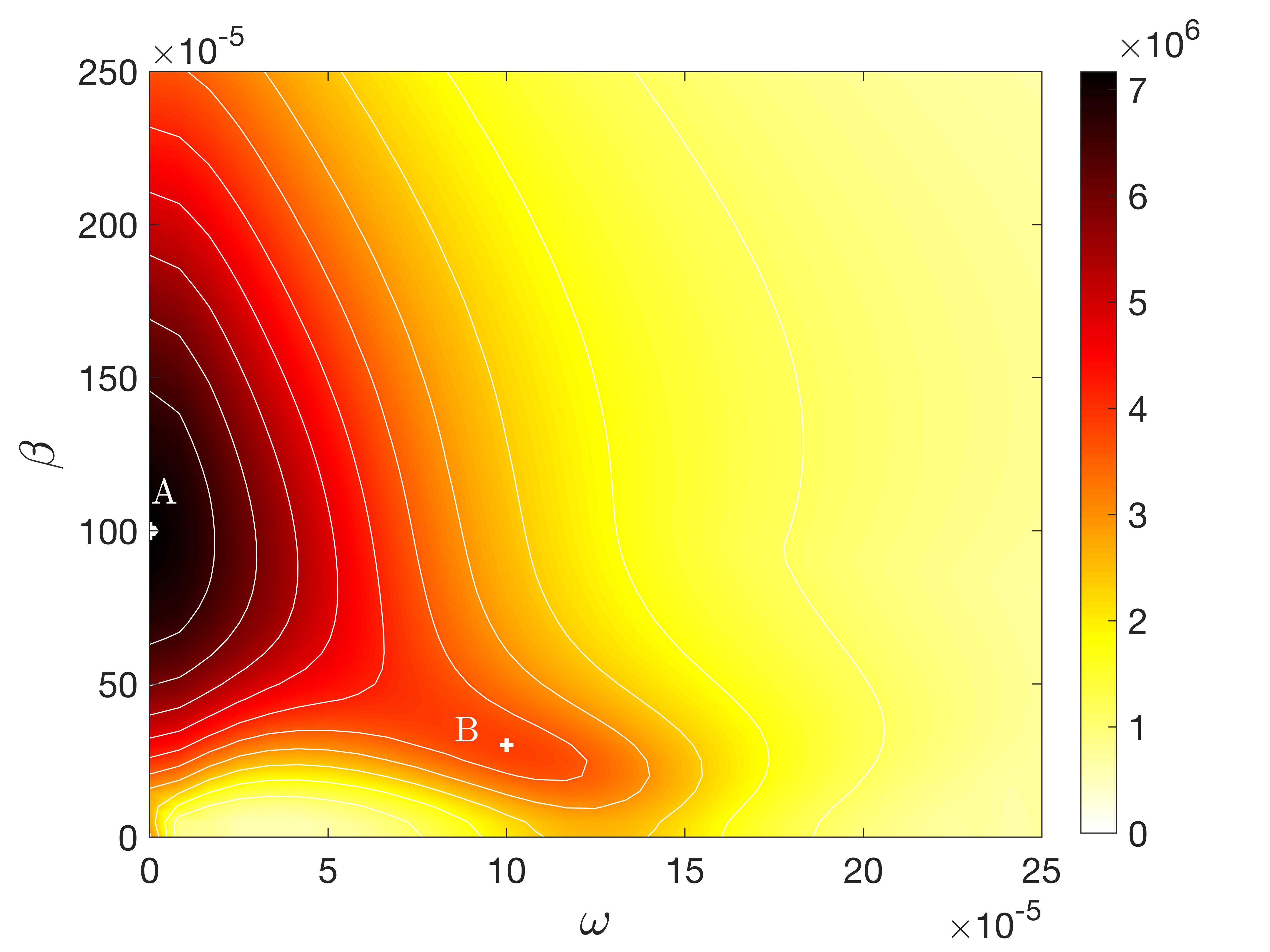}
\vspace{0.5cm}
\\{\scriptsize $|f'_v|_{max}=\pm0.5$ \hspace{1.5cm} {\bf A: streak}  \hspace{1.5cm} $|u'|_{max}=\pm0.8$ }\\
\vspace{0cm}
\includegraphics[width=0.85\textwidth,trim={15cm 11.4cm 15cm 8cm},clip]{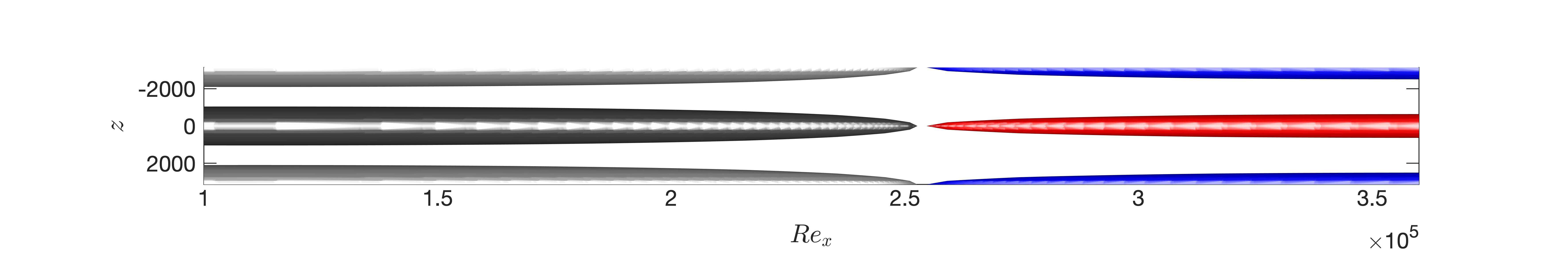}
\vspace{0.5cm}
\includegraphics[width=0.85\textwidth,trim={15cm 15cm 15cm 10cm},clip]{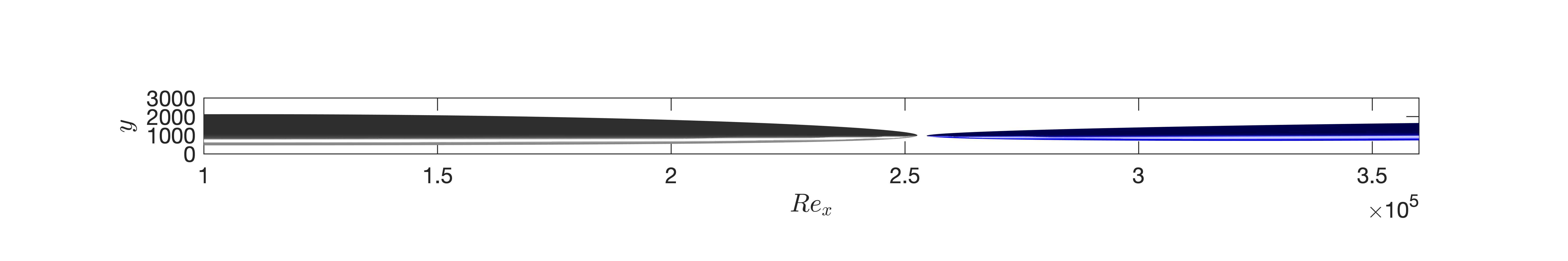}
\vspace{0.5cm}
\\{\scriptsize $|f'_u|_{max}=\pm0.8$ \hspace{1cm} {\bf B: oblique wave}  \hspace{1cm} $|u'|_{max}=\pm0.5$ }\\
\includegraphics[width=0.85\textwidth,trim={15cm 29cm 15cm 25cm},clip]{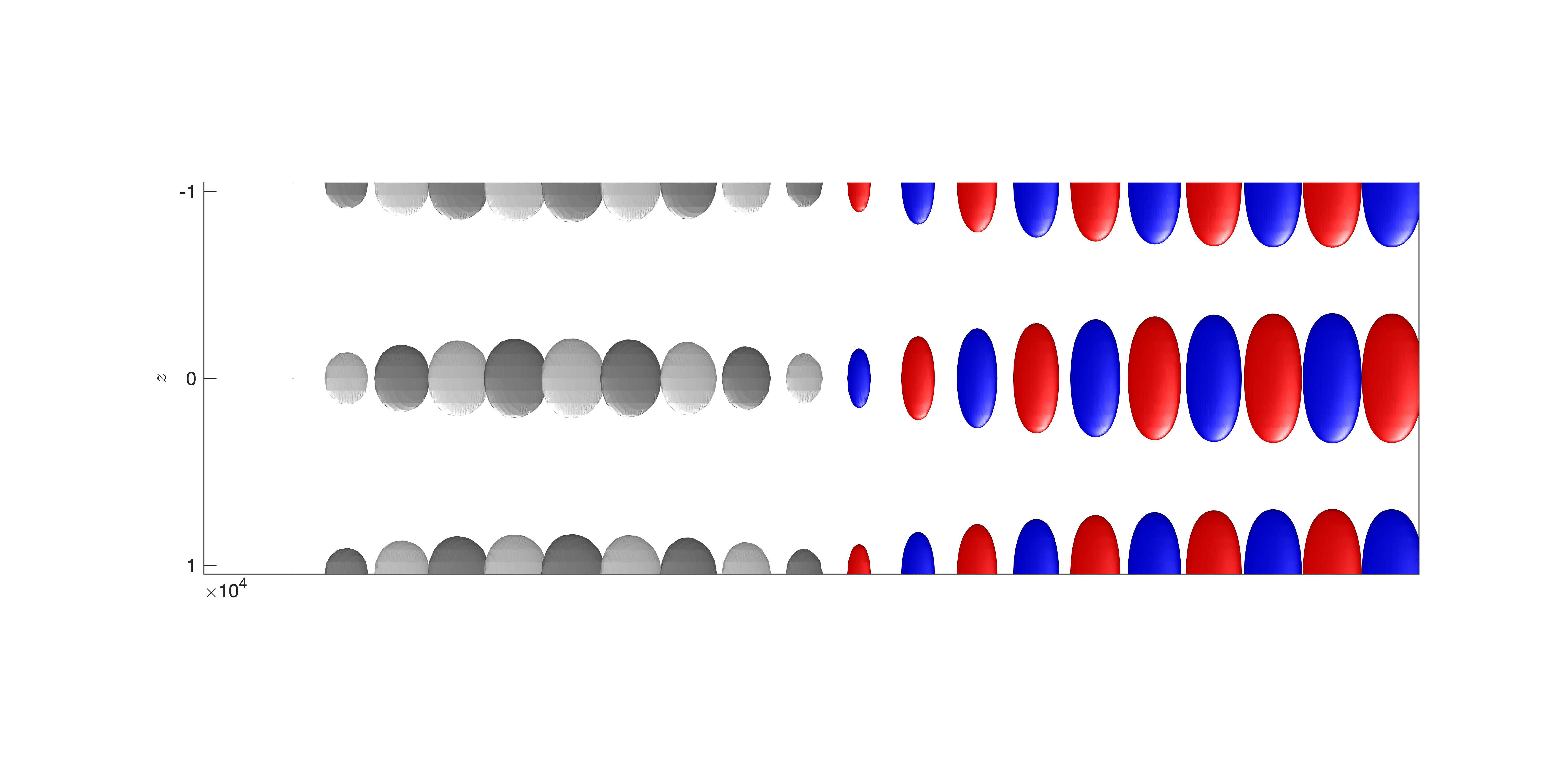}
\includegraphics[width=0.85\textwidth,trim={15cm 5cm 15cm 10cm},clip]{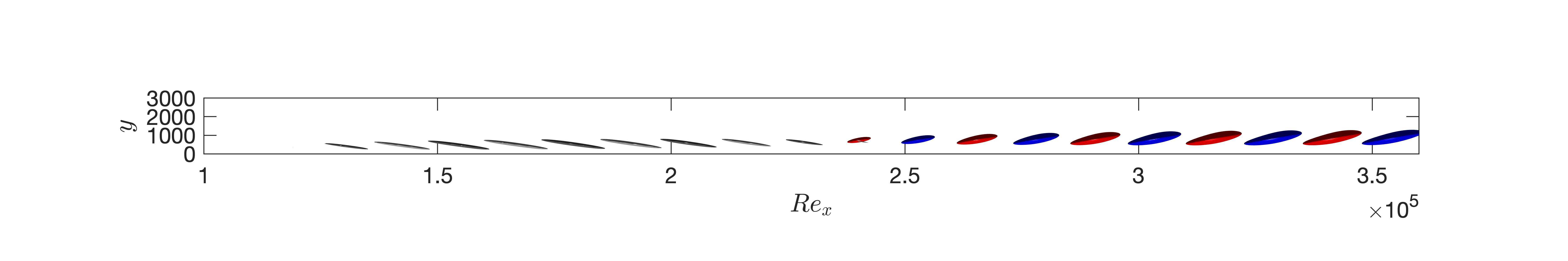}
\caption{Linear input/output (resolvent) analysis. Optimal gain (top). The two local maxima correspond to the amplification of streaks (A at $(\beta,\omega) =(100,0)\times10^{-5}$) and oblique waves (B at $(\beta,\omega) =(30,10)\times10^{-5}$). Optimal forcing (left; light gray:positive, dark grey: negative) and optimal response (right; red: positive, blue negative) for streaks and obliques waves. Top and side views are shown. The x-axis has been scaled by a factor of 4.}
\label{fig:LinearGain}
\end{figure}

\section{Linear input/output (resolvent) analysis}\label{sec:linear_resolvent}

Before performing nonlinear optimization, we briefly recall here results obtained by \citet{monokrousos2010global,brandt2011effect} concerning linear optimal forcing in the frequency domain that aim at maximizing energetic gains (resolvent analysis). Such results are important to understand and analyse the forthcoming nonlinear optimizations.
For this, we consider a generic volumetric forcing and no-slip boundary conditions on the wall.  The cost function for the linear optimization is the input/output kinetic energy gain of the fluctuations over the whole domain:
\begin{equation}
    J^{\mbox{lin}} \equiv \lambda=
    \frac{\mathbf{\hat{w}}^*\mathbf{Q}'\mathbf{\hat{w}}}{\mathbf{\hat{f}}^*\mathbf{Q}\mathbf{\hat{f}}},
\end{equation}
where
\begin{eqnarray} \label{eq:Q}
{\mathbf{f}}^*\mathbf{Q}{\mathbf{f}}=\iint (|{f}_x|^2+|{f}_y|^2|+|{f}_z|^2)\; d\Omega, \\
\mathbf{Q}'=\left( \begin{array}{cc}
\mathbf{Q} & 0 \\ 0 & 0
\end{array}
\right). \label{eq:Qp}
\end{eqnarray}
Such a linear optimization problem is efficiently solved by iterative methods \citep{sipp2013characterization}.
The mesh extends here from $ x_i=0.30 \times 10^5$ to $ x_o=3.60 \times 10^5 $. It comprises 116806 triangles, yielding 586178 degrees of freedom. The same mesh will be used in the next section dealing with nonlinear optimization (\S \ref{sec:nonlinear_resolvent}).

The linear optimal amplitude gain ($\sigma=\sqrt{\lambda}$) is shown in figure~\ref{fig:LinearGain}, as a function of frequency $\omega$ and spanwise wavenumber $\beta$. Two local maxima are observed, in agreement with \cite{monokrousos2010global}. The forcing and response mode shapes of the two linear optimal mechanisms are shown in the same figure.

The first local maximum at $(\beta,\omega) =(100,0)\times10^{-5}$, point A, is associated with the nonmodal lift-up mechanism. The optimal forcing corresponds to steady streamwise rolls ($v$, $w$ components; for the optimal forcing the $v$ component is shown), and the optimal response to streamwise streaks located further downstream ($u$ component).

The second local maximum at $(\beta,\omega) =(30,10)\times10^{-5}$, point B, corresponds to the amplification of oblique TS waves. The planar TS waves are not the most amplified ones due to the cooperative non-modal amplification through the Orr and lift-up mechanisms. It is clearly noticed that the optimal forcing is tilted upstream, against the mean shear so that the response takes advantage of the algebraic amplification through the Orr mechanism.

\section{Nonlinear input/output analysis}\label{sec:nonlinear_resolvent}
To uncover the optimal nonlinear mechanisms that promote transition, the nonlinear interactions of the modes and their impact on the mean flow is now incorporated in the analysis through the optimization approach developed in \S\ref{sec:optim}.

We choose as cost function the (squared) shear-stress of the mean-flow deviation, integrated over the wall. With the notation introduced above, this is:
$$ J(\hat{\mathbf{w}})=J(\hat{\mathbf{w}}_{00})=(\hat{\mathbf{w}}_{00}-\mathbf{w}_b)^*\mathbf{C}^*\mathbf{C}(\hat{\mathbf{w}}_{00}-\mathbf{w}_b),$$
with $ \mathbf{C}\mathbf{w} = \int_{y=0} \frac{\partial u}{\partial y} dx$ and $ \mathbf{w}_b$ is the base-flow. For this choice of cost function, we have:
$$
\frac{dJ}{d\hat{\mathbf{w}}_{00}}=2\mathbf{C}^*\mathbf{C}(\hat{\mathbf{w}}_{00}-\mathbf{w}_b)
$$
and  $0$ for the other harmonics. This cost function can be directly linked to the drag change exerted on the plate,
\begin{equation}
\Delta C_D = \frac{\nu J^{0.5}}{\frac{1}{2} U_\infty^2 L_p}, \label{eq:deltacd}
\end{equation}
where $ L_p=x_o$ is the plate length.
In other words, by maximizing the specific cost function $J$, we maximize the drag on the plate. 

\begin{figure}
\centering
\includegraphics[width=1\textwidth]{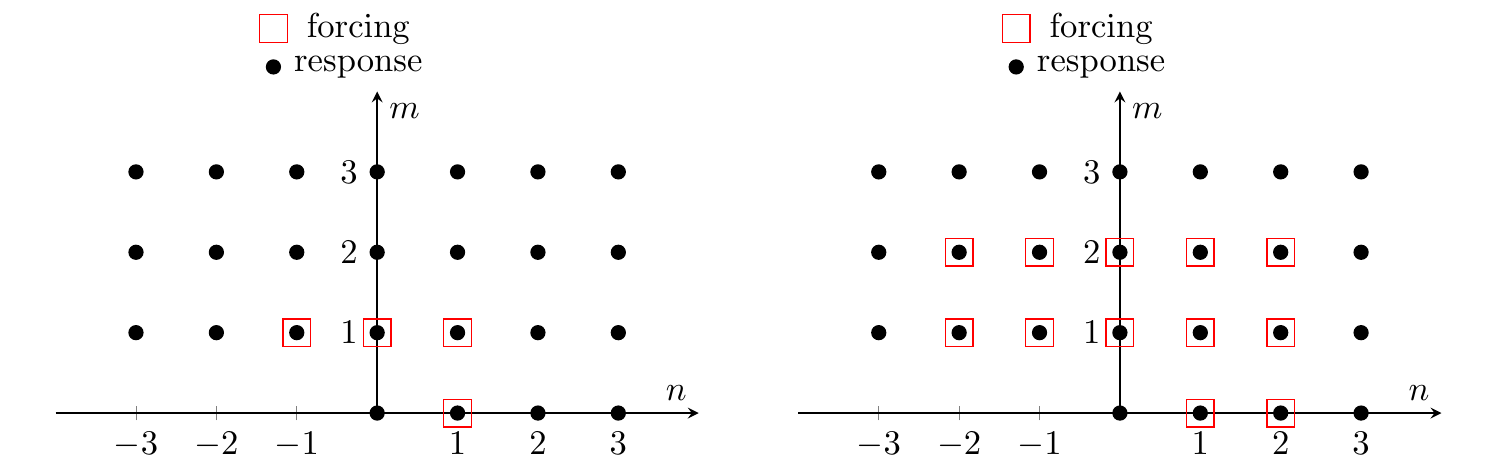}
\caption{Fundamental (left) and superharmonic (right) cases. The nonlinear optimization is restricted to forcing components $1\beta,\pm1\omega$, or $1\beta,2\beta,\pm1\omega,\pm2\omega$, and their oblique combinations, respectively.  Other forcing and response harmonics in the $(m,n)$ plane may be deduced from the real-value constraint, e.g. $ \mathbf{\hat{w}}_{-m-n}=\overline{\mathbf{\hat{w}}_{mn}}$. In case of reflectional symmetry in $z$, the $-n \omega$ components are linked to the $+n\omega$ ones.}
\label{fig:cases}
\end{figure}

The entries of matrix $\mathbf{P}$ allow selection of the forced equations and of a subset of forced harmonics.
As in the linear case, we will restrict the forcing to the momentum equations and exclude mass sources.
In order to preserve the mean-flow harmonic $\mathbf{\hat{w}}_{00}$ from direct modifications induced by steady forcing terms,
we set $\mathbf{\hat{f}}_{00}=0$ and exclude this mode from the optimization process.
Two types of forcing are then considered, which we refer to as \emph{fundamental} and \emph{superharmonic} cases, as depicted in figure~\ref{fig:cases}. For the first case, forcing is restricted to components $(m,n)=(\beta,\pm\omega)$, $(\beta,0)$, $(0,\pm \omega)$; we call this \emph{fundamental}, since forcing is allowed only at the primary forcing frequency and spanwise wavenumber. Each of these forcing components, can potentially lead to the amplification of a pair of unsteady oblique waves, steady streamwise streaks or vortices, and planar TS waves, respectively. For the superharmonic forcing case, we allow also the second forcing harmonics to be optimized, $ |m|\leq 2$ and $ |n|\leq 2$, except $m=n=0$. This allows forcings of fundamental harmonic and superharmonic components. For example, forcing $\hat{\mathbf{f}}_{02}$ is at twice the frequency of forcing  $\hat{\mathbf{f}}_{11}$. If the perturbation satisfies reflectional symmetry in $z$, all forcing and response harmonics $n<0$ are directly linked to those satisfying $ n > 0 $.  We note that in both cases, we solve separate optimization cases over a wide range of the fundamental forcing frequency, $\omega$ and $\beta$.



\subsection{Identification of optimal transition mechanisms: $z$-symmetric case with $ A=\num{7.07e-5}$ and $M=N=2$} \label{sec:toy}

\begin{figure}
\includegraphics[width=0.52\textwidth]{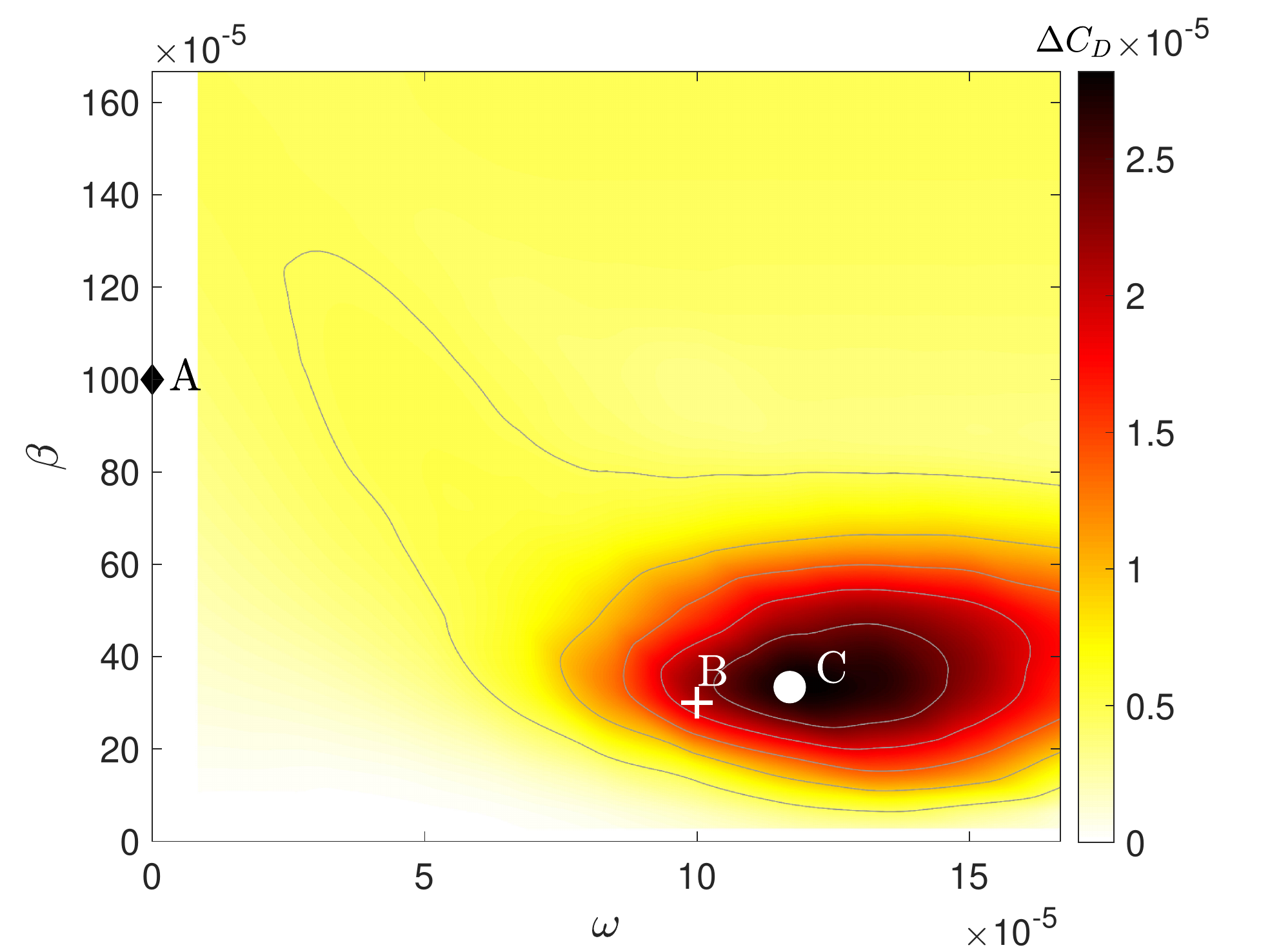}
\includegraphics[width=0.52\textwidth]{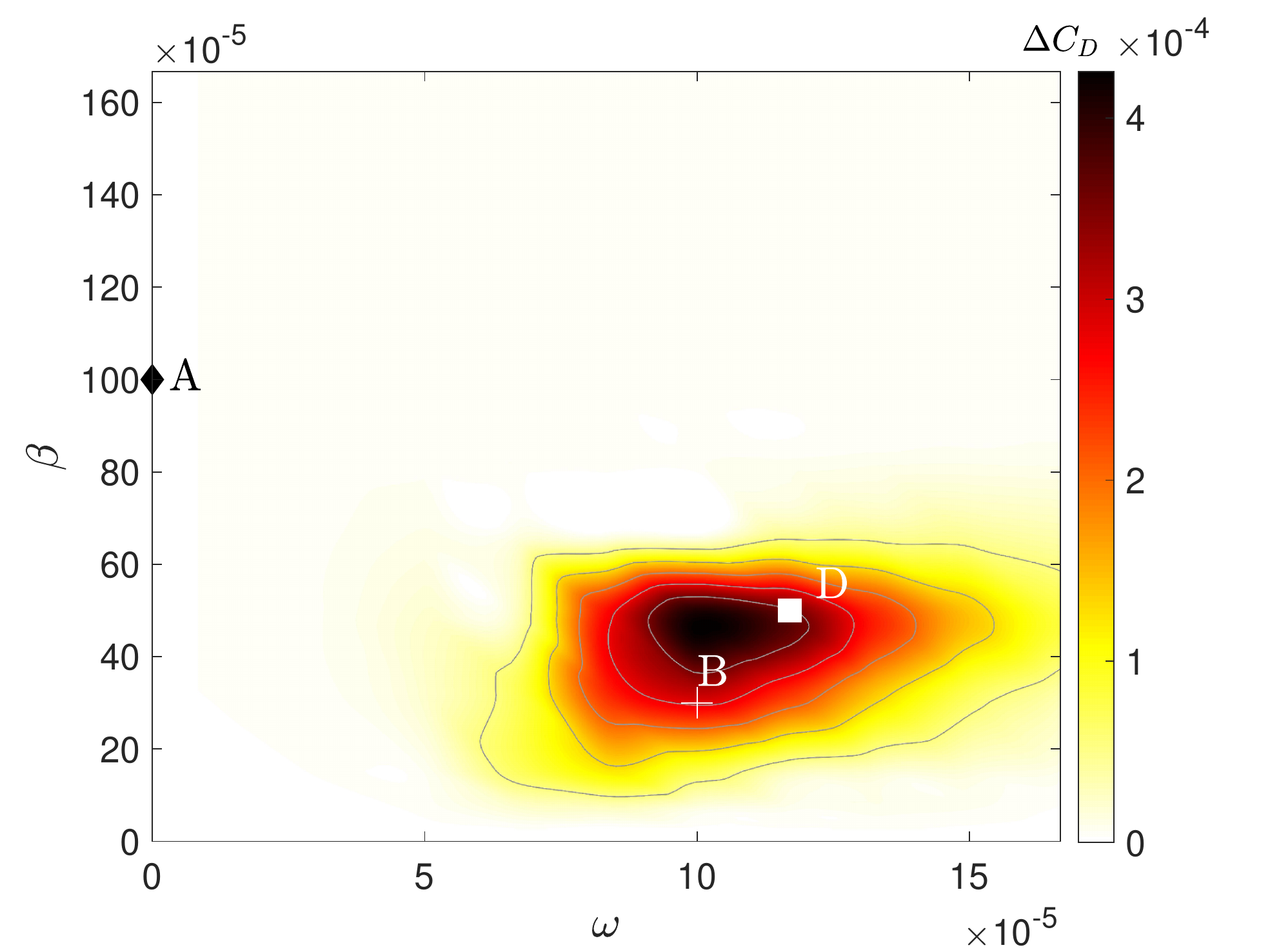}
\caption{Optimal drag change from nonlinear input/output analysis with fundamental (left) and superharmonic (right) forcing, $M=N=2$ and $A=\num{7.07e-5}$.  Fundamental maximum at point C: $\Delta C_{D,max}=\num{2.8e-5}$ at $(\beta,\omega) =(33.4,11.7)\times 10^{-5}$. Superharmonic maximum at point D: $\Delta C_{D,max}=\num{41.6e-5}$ at $(\beta,\omega) =(50,11.7)\times10^{-5}$. Both are close to the optimal linear amplification of the oblique waves (point B).}
\label{fig:J}
\end{figure}

We first consider a case with imposed spanwise symmetry on the forcing and response and $M=N=2$. For small enough amplitude $A$, we expect that the forcing and perturbation should exhibit the $z$-reflectional symmetry of the configuration. The small number of resulting modes allows for more expensive parametric studies over $\omega$ and $ \beta$, but, strictly speaking, the results are converged for sufficiently small $A$ so that higher-order harmonics may be neglected.  We select here $A=\num{7.07e-5}$ and in a subsequent sections we examine convergence as $A$ is increased and with retaining more modes, and verify {\it a posteriori} that the present results are reasonably well converged.

The cost function (expressed as mean drag perturbation via \eqref{eq:deltacd}) is shown in figure \ref{fig:J} for both the fundamental- and superharmonic-type forcings.  For the fundamental case, maximum drag increase 
is observed at $(\beta,\omega) =(33.4,11.7)\times10^{-5}$, whereas for the superharmonic case 
the maximum occurs at the same frequency but a slightly higher wavenumber, $(\beta,\omega) =(50,11.7)\times10^{-5}$. For the superharmonic case, the drag increase is approximately 14 times higher compared to the fundamental forcing. In both cases, the overall optimal frequency/wavenumber pairs are close to the point marked $B$ on the linear amplifcation plot (figure~\ref{fig:LinearGain}), which represents the local maximum in linear amplification of oblique waves.  While those waves are linearly less amplified than streaks (point A), they are nonlinearly superior. As will be shown in detail below, the nonlinear fundamental mechanism C and superharmonic mechanism D initially harness oblique wave amplification, and eventually lead, through nonlinearity, to redistribution of energy near A and a strong response related to the lift-up mechanisms producing streaks.

\subsubsection{Symmetric fundamental forcing}

\begin{figure}
\hspace{-0.8cm}
\includegraphics[width=1.1\textwidth,trim={0cm 0cm 0cm 0},clip]{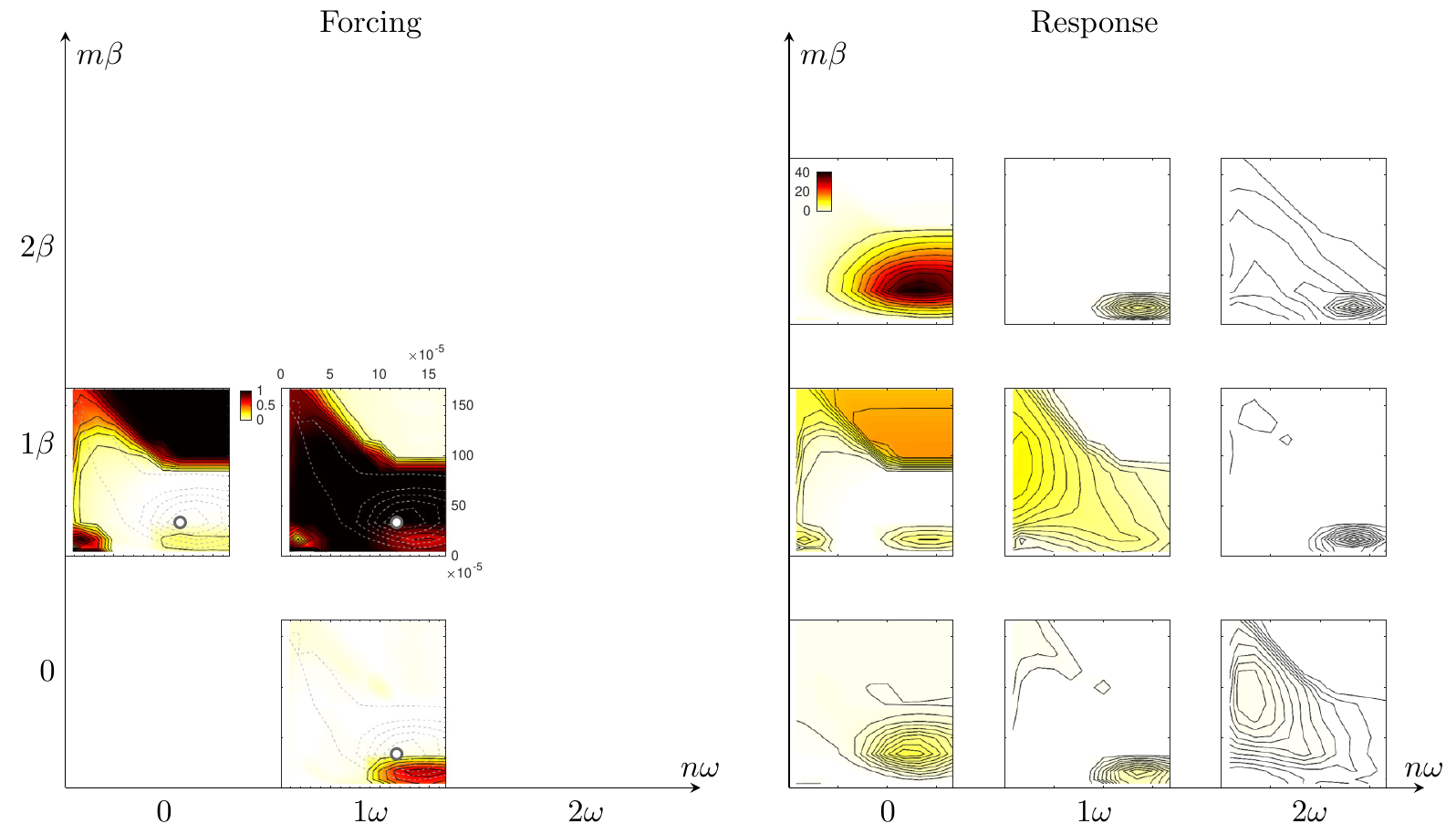}
\caption{Nonlinear optimization for fundamental symmetric forcing with $M=N=2$. Amplitudes of optimal forcing (left) and response (right) for each individual harmonic component $(m,n)$, as depicted in figure~\ref{fig:cases}a. Values have been normalized with the the total forcing amplitude $A=\num{7.07e-5}$. The circle marks the frequency/wavenumber that maximum drag increase is observed. Also, isolines of the cost function (dashed lines) have been added on the forcing components.}
\label{fig:A_fundamental}
\end{figure}

\begin{figure}
\includegraphics[width=0.5\textwidth,trim={1cm 0 1cm 0},clip]{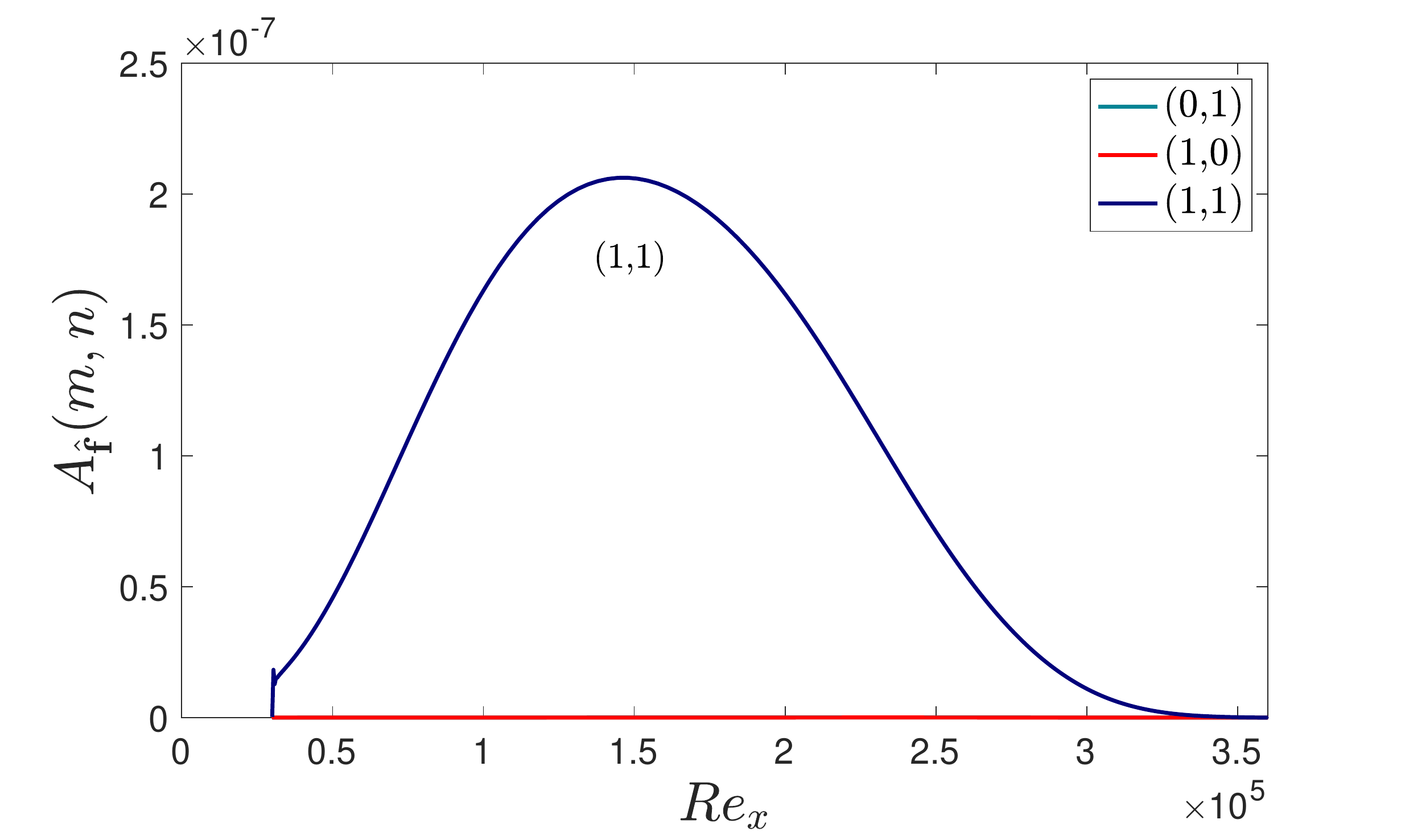}
\includegraphics[width=0.5\textwidth,trim={1cm 0 1cm 0},clip]{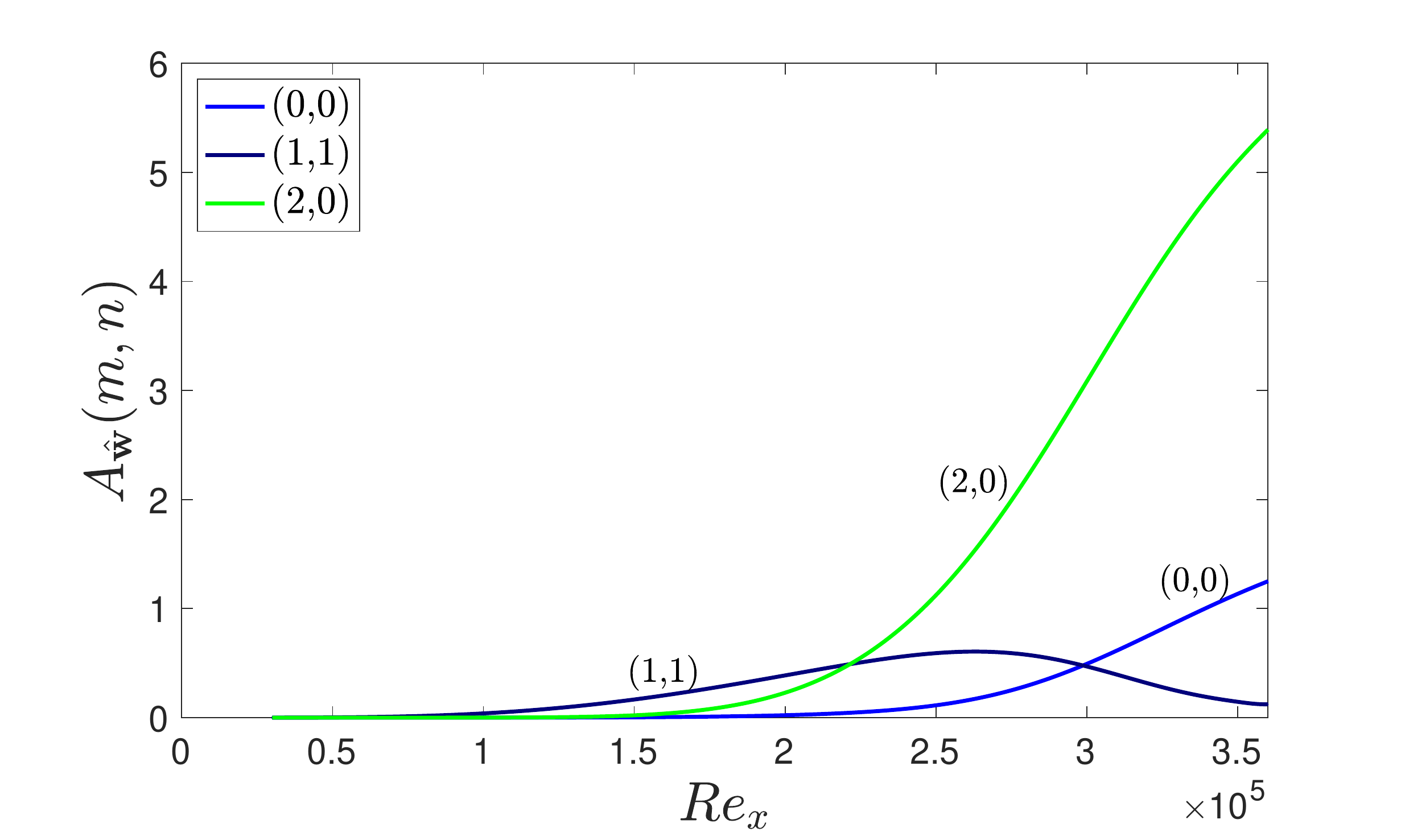}
\includegraphics[width=1\textwidth,trim={30cm 2cm 30cm 0},clip]{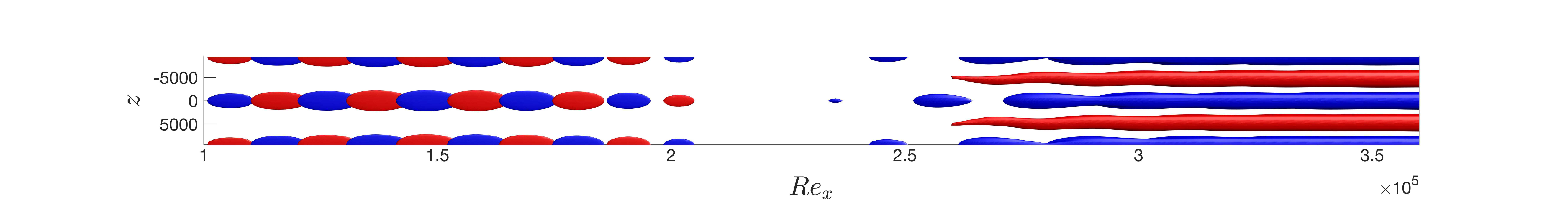}
\caption{Optimal oblique fundamental case with $M=N=2$ at $(\beta,\omega)$=($33.3,11.7)\times10^{-5}$ for $A=\num{7.07e-5}$. This forcing results to the maximum amplification of shear stress for fundamental forcing over all forcing frequencies and wavenumbers (point C in figure \ref{fig:J}a). Isosurfaces of streamwise perturbations $f'_u= \pm \num{8.3e-9}$ (bottom left) and $u'= \pm 0.07$ (bottom right), blue negative iso-value and red positive one. One fundamental wavelenghth is shown in $z$.} 
\label{fig:Oblique}
\end{figure}

Focusing on the fundamental case first (figure~\ref{fig:cases}a) with $z$-reflectional symmetry, we now delve into the optimal forcing and response in greater detail.  To simplify the discussion, we define in appendix \ref{app:HarmonicAmplitudes}, a scalar amplitude $A(m,n)$ of each forcing/response mode, which represents an integral over the spatial domain.
These amplitudes are shown in figure~\ref{fig:A_fundamental}.  Note that upon summation of the forcing modes, this yield the overall forcing amplitude (here $A=\num{7.07e-5}$), and all amplitudes in the plot are normalized by this value. Based on the dominant regions of  the amplitude response on the $\beta-\omega$ planes, three distinct mechanisms can be identified:

\begin{enumerate}
\item
{\bf Oblique waves}.
The maximal drag increase occurs for  $(\beta,\omega)$=($33.3,11.7)\times10^{-5}$, and only involves significant forcing of the oblique wave component $(\beta,\pm\omega)$. In the response, there is some amplification to the response component $(\beta,\pm\omega)$, as expected in a linear framework, but the $(2\beta,0)$ component, which arises from nonlinear interactions between $(\beta,\omega)$ and $(\beta,-\omega)$ components, is highly amplified. The mean flow modification is clearly associated with the amplification of $(2\beta,0)$ streaks via oblique forcing.

\item
{\bf Streamwise vortices}. For high spanwise wavenumbers, $\beta>\num{100e-5}$, the optimal forcing is a streamwise vortex $(m,n)=(\beta,0)$. For these frequencies, the linear amplification of obliques waves is weak and thus the generated streaks through nonlinearity would also be weak. Consequently, for high enough frequencies and wavenumbers, i.e. those that are far from the linear optimal of the oblique waves, the optimal forcing mechanism is the direct amplification of streaks through the lift-up mechanism. 
\item
{\bf K-type mechanism}.
Finally, at $(\beta,\omega)=(16,15)\times10^{-5}$, the optimal forcing is a combination of all three components. Main forcing component is the TS wave followed by the oblique waves. This mechanism is similar to the Klebanov one, describing fundamental K-type transition. 

\end{enumerate}

\begin{figure}
\centering
\includegraphics[width=1\textwidth,trim={30cm 41cm 30cm 20cm},clip]{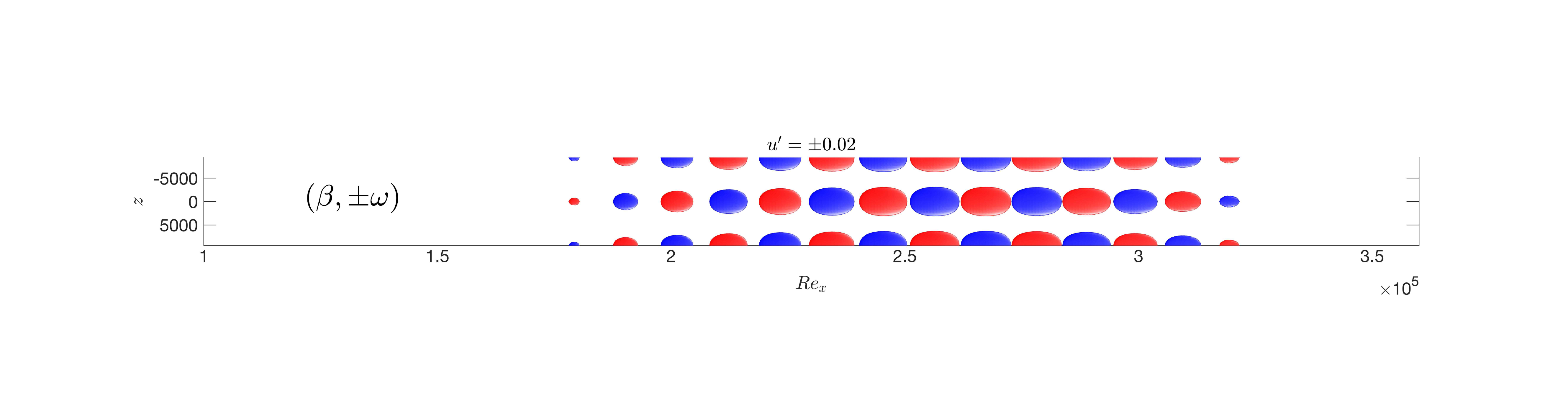}
\includegraphics[width=1\textwidth,trim={30cm 28cm 30cm 20cm},clip]{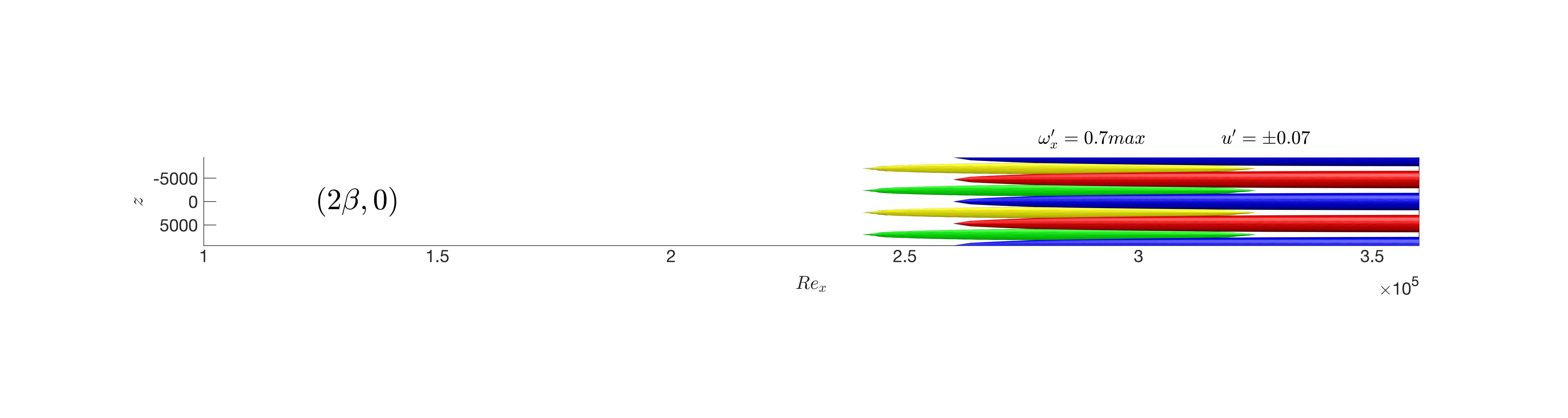}
\caption{Harmonic response components for the optimal oblique fundamental case shown in figure \ref{fig:Oblique}. The response is dominated by the growth of $(\beta,\pm \omega)$ oblique waves ($u'$ shown), followed by the nonlinear generation of $(2\beta,0)$ streamwise vortices ($\omega'_x$ shown) and the linear growth of streaks ($u'$ shown).}
\label{fig:vortex}
\end{figure}

Since the oblique waves are the most dangerous mechanism in terms of drag increase, we examine the structure of the forcing and response fields in greater detail in figure~\ref{fig:Oblique}. Initially, $(\beta,\pm\omega)$ oblique waves\footnote{Only oblique waves with positive wavenumber and frequency are shown due to $z$-symmetry.} are amplified due to linear instability.  The quadratic nonlinearity redistributes the energy of the oblique waves into the $(2\beta,0)$ mode in the form of streamwise vortices  with streamwise vorticity $\omega_x$ ($v,w$ components with $(2\beta,0)$). In turn, the streamwise vortices lead to the growth of the streaks ($u$ component with $(2\beta,0)$) through the linear lift-up mechanism. The spatial shapes of the the harmonic components mentioned above are shown in figure \ref{fig:vortex}, showing the transition sequence from oblique waves to streamwise vortices and streaks.  
These observations are in agreement with previous studies on oblique transition  \citep{schmid1992new,berlin1994spatial}. The link between the nonlinear gain map and that obtained from linear analysis is evident now: the most dangerous nonlinear forcing exploits both linear amplification mechanisms, specifically 3D unsteady oblique waves and 3D steady rolls-streaks, through the redistribution of energy from the first linear mechanism to the second linear mechanism via nonlinearity. The fundamental frequency and spanwise wavenumber, $(\beta,\omega)$=($33.3,11.7)\times10^{-5}$, is very close to the linearly optimal oblique ones, $(\beta,\omega)$=($30,10)\times10^{-5}$, and then nonlinearly generated steady vortices are formed at twice the spanwise wavenumber $(\beta,\omega)$=($66.6,0)\times10^{-5}$. The latter part does not coincide with the maximum linearly amplified lift-up wavenumber, $(\beta,\omega)$=($100,0)\times10^{-5}$, but it is close enough to take advantage of this mechanism in an optimal synergistic way.

\subsubsection{Symmetric superharmonic forcing}

Here, forcing is expanded to consider both the fundamental its first harmonic in both frequency and spanwise wavenumber, see figure \ref{fig:cases}b. Thus, forcing is allowed in 8 forcing components arising from all the combinations of fundamental and first harmonic components.
We retain for now $M=N=2$ and $A=\num{7.07e-5}$, and recall (figure~\ref{fig:J}b) that maximum amplification of shear stress is observed for forcing at $(\beta,\omega)=(50,11.7)\times10^{-5}$.

\begin{figure}
\hspace{-0.8cm}
\includegraphics[width=1.1\textwidth,trim={0cm 0cm 0cm 0},clip]{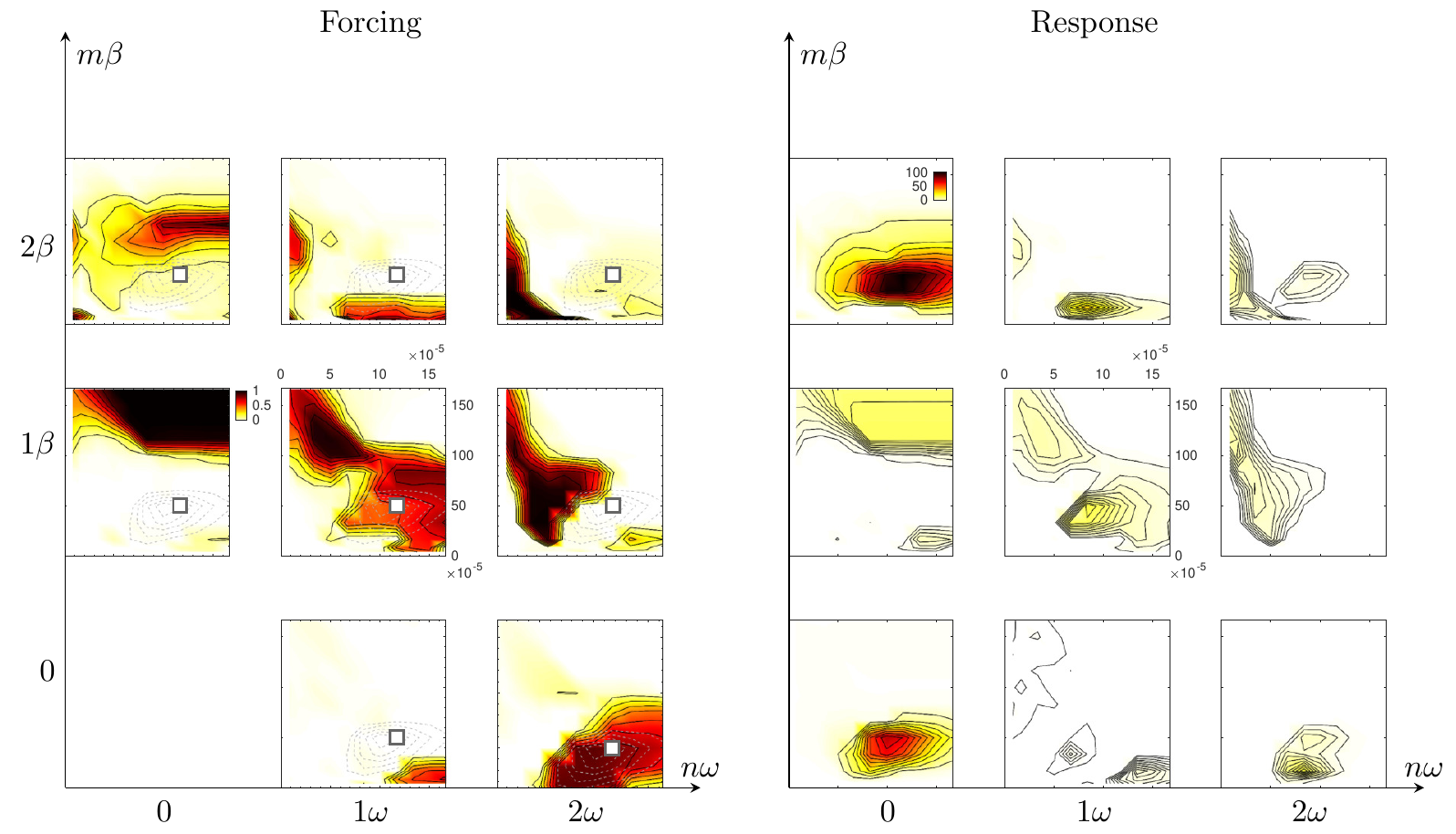}
\caption{Nonlinear optimization for superharmonic symmetric forcing $M=N=2$. Amplitudes of optimal forcing (left) and response (right) for each individual harmonic component $(m,n)$, as depicted in figure~\ref{fig:cases}b. Values are normalized with the the total forcing amplitude $A=\num{7.07e-5}$. The square marks the frequency/wavenumber that maximum drag increase is observed.}
\label{fig:A_subharmonic}
\end{figure}

The optimum in the $(\beta,\omega)$ plane for the superharmonic case is close to the one found for the fundamental case where the oblique waves were the optimal forcing mechanism. However, now the TS waves at twice the frequency of the oblique waves share similar amplitude to the oblique waves.  As can be observed in figure~\ref{fig:A_subharmonic}, only two major forcing components exist at the optimal $(\beta,\omega)$ pair. The optimal forcing corresponds to a three-dimensional oblique wave $(\beta,\omega)$ and a superharmonic two-dimensional TS wave at twice the frequency, $(2\beta,0)$. The optimal superharmonic forcing is in agreement with the typical scenario for H-type transition. In the literature describing H-type transition, typically the TS is called the fundamental wave and the oblique the subharmonic, but our description is equivalent.

\begin{figure}
\includegraphics[width=0.5\textwidth,trim={1cm 0 1cm 0},clip]{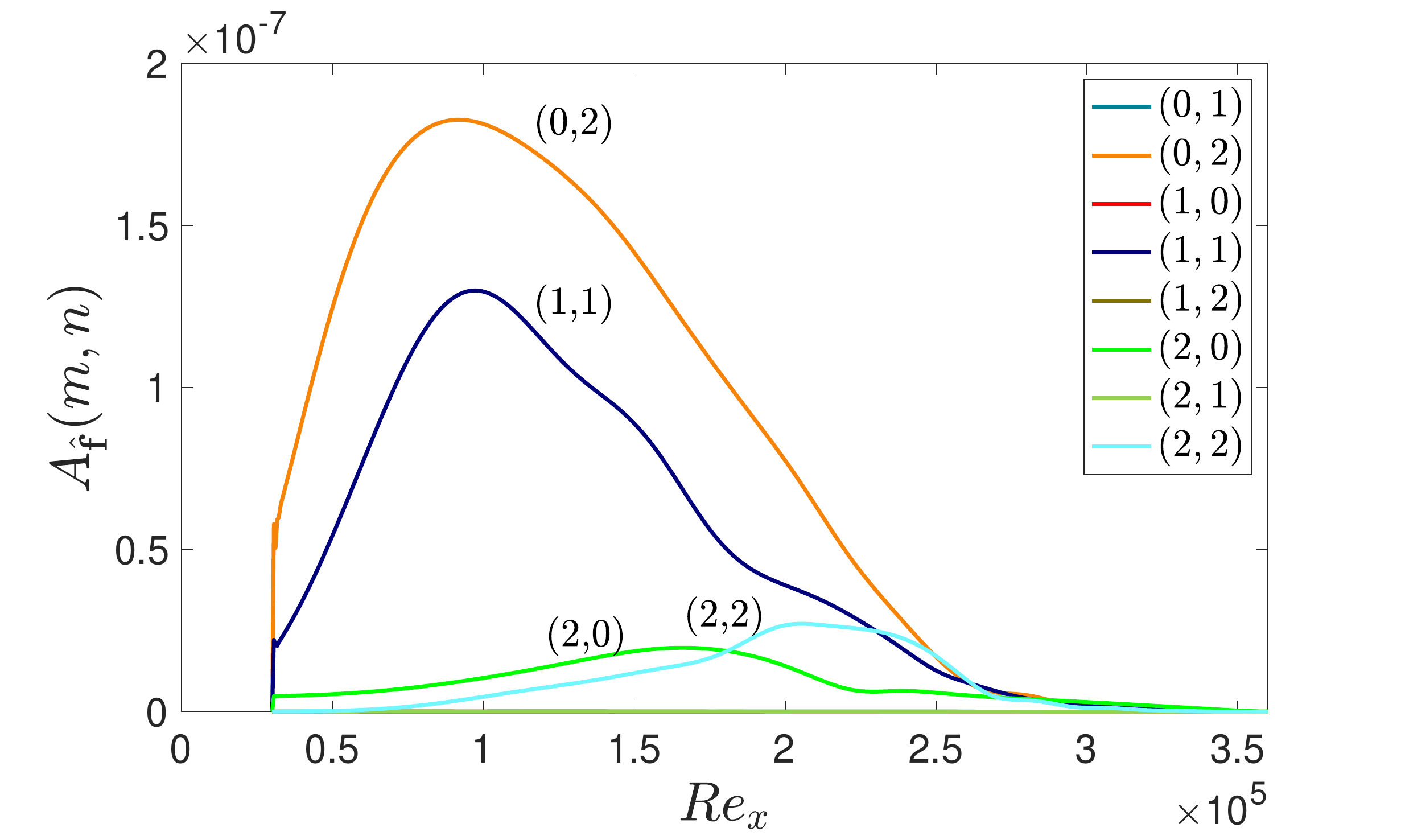}
\includegraphics[width=0.5\textwidth,trim={1cm 0 1cm 0},clip]{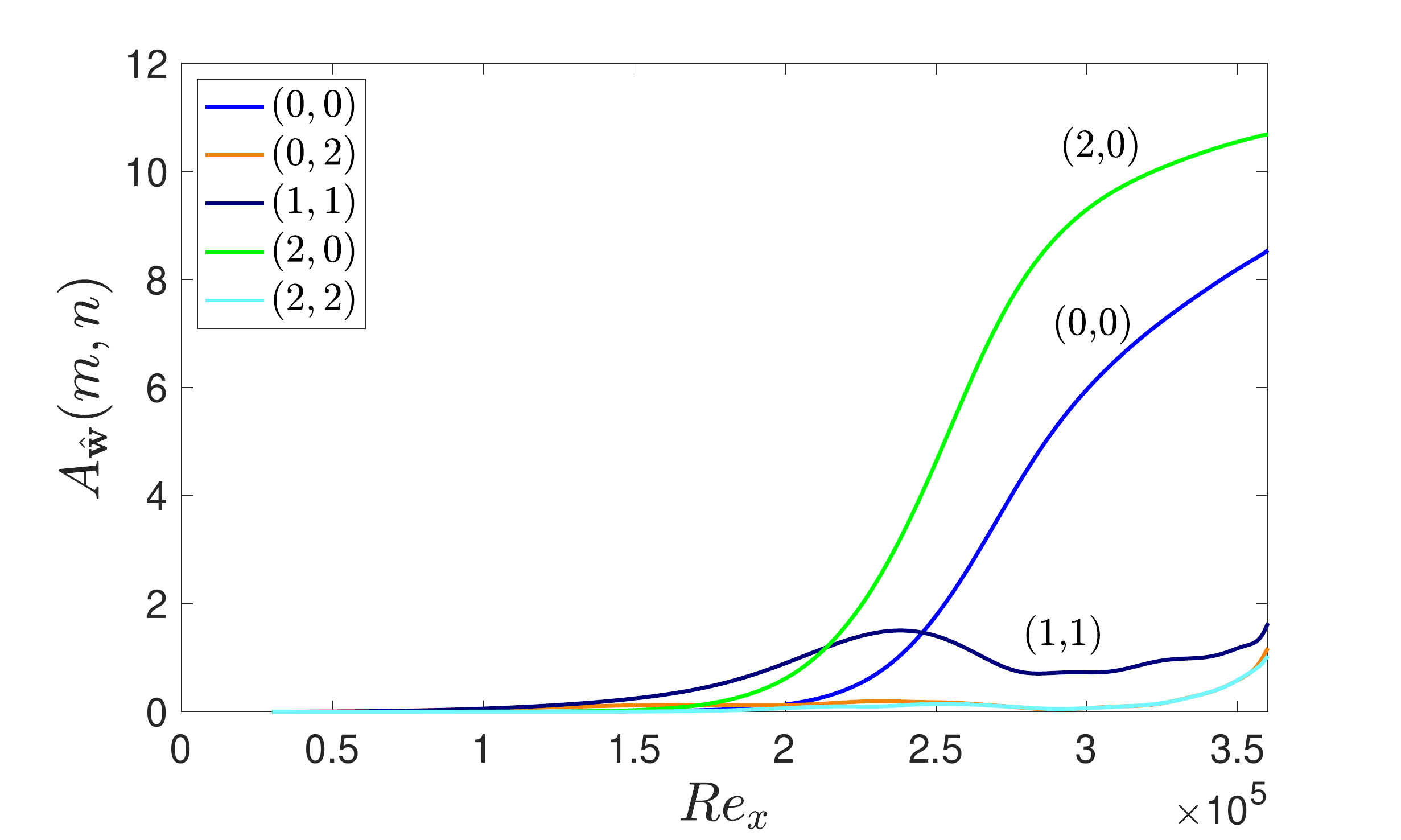}
\includegraphics[width=1\textwidth,trim={30cm 2cm 30cm 0cm},clip]{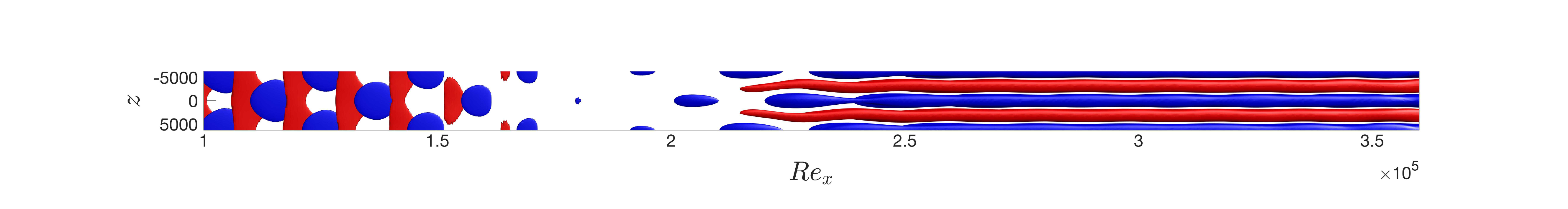}
\caption{Optimal H-type superharmonic case $M=N=2$ at $(\beta,\omega)=(50,11.7)\times 10^{-5}$ for $A=\num{7.07e-5}$. This forcing results to the maximum amplification of shear stress for superharmonic forcing over all forcing frequencies and wavenumbers (point D in figure \ref{fig:J}b). Isosurface of $f'_u= \pm \num{8.3e-9}$ (bottom left) and $u'= \pm 0.07$ (bottom right), blue negative iso-value and red positive one. One fundamental wavelenghth is shown in $z$.}
\label{fig:Htype}
\end{figure}

\begin{figure}
\centering
\includegraphics[width=1\textwidth,trim={30cm 41cm 30cm 20cm},clip]{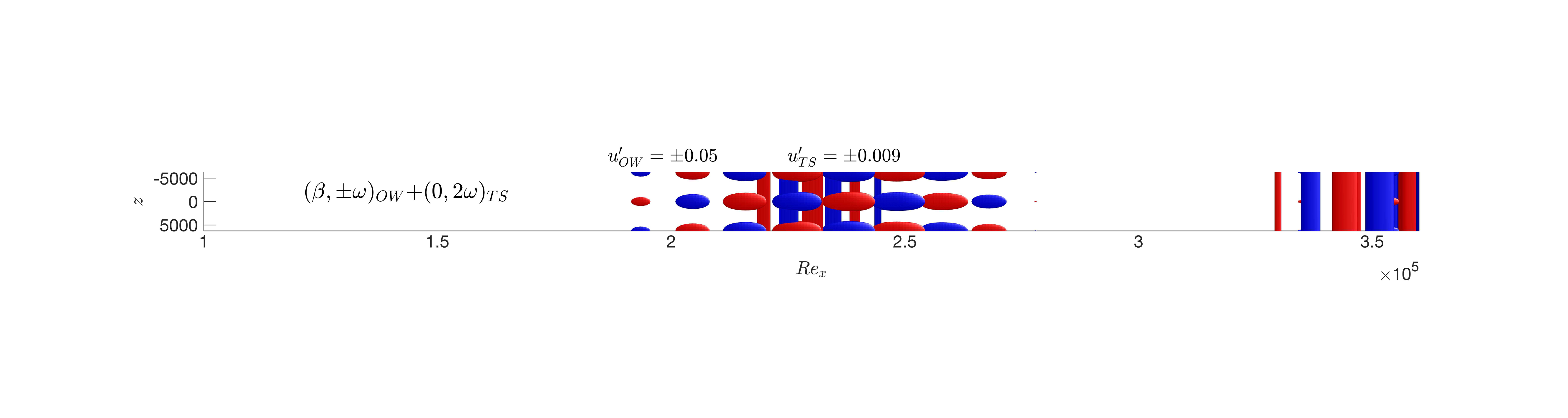}
\includegraphics[width=1\textwidth,trim={30cm 41cm 30cm 20cm},clip]{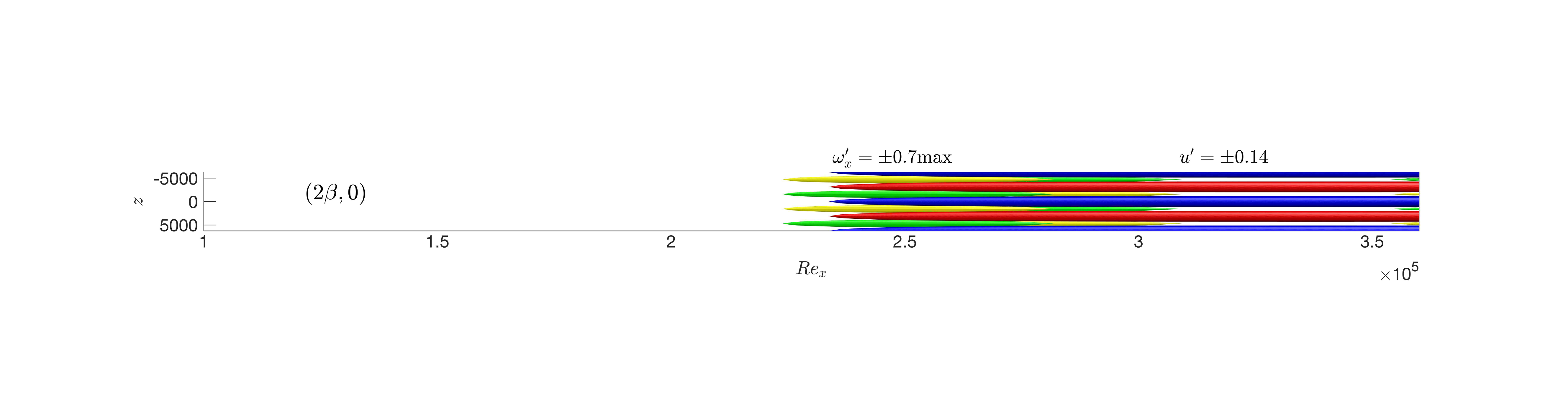}
\caption{Harmonic response components for the optimal superharmonic case shown in figure \ref{fig:Htype}. The response is dominated by the growth of $(\beta,\pm \omega)_{OW}$ oblique waves and planar TS waves $(0,2\pm \omega)_{TS}$ at twice the fundamental frequency ($u'$ overlaid for OW and TS) characteristic of H-type resonance. The oblique waves generate nonlinearly  $(2\beta,0)$ streamwise vortices ($\omega'_x$ shown) which promote the linear growth of streaks ($u'$ shown). Also, initial stages of streak instability are  observed near the domain outlet.}
\label{fig:vortex2}
\end{figure}

The streamwise evolution of each forcing and response harmonic for the optimal superharmonic case is shown in figure \ref{fig:Htype}. The forcing is dominated by the superharmonic two-dimensional TS wave at twice the fundamental frequency, $(0,2\omega)$, and the three-dimensional oblique waves, $(\beta,\pm\omega)$. Since spanwise reflectional symmetry has been enforced, the amplitudes of the $(\beta,+\omega)$ and $(\beta,-\omega)$ oblique waves are equal and only one of those is shown. Despite the differences in the forcing components when compared to the fundamental case where only the oblique waves are present, the amplitude response is qualitatively similar and dominated by streaks.  However, the nonlinearly generated streaks have almost twice as high amplitude when compared to the fundamental case, due to the efficient amplification of the parent oblique waves through the resonant interaction with the planar TS waves (see figure \ref{fig:vortex2}). The subsequent stages are similar to the ones of the fundamental case where streamwise vortices are generated from the nonlinear interactions between $(\beta,\omega)$ and $(\beta,-\omega)$ components, which in turn produce streaks. Finally, towards the end of the domain, low- and high-speed streaks start to undergo streamwise oscillations. These oscillations are stronger than in the fundamental oblique case (compare with figure \ref{fig:Oblique}), since for a given amplitude, the resonant H-type forcing leads to higher streak amplification through the stages described above.

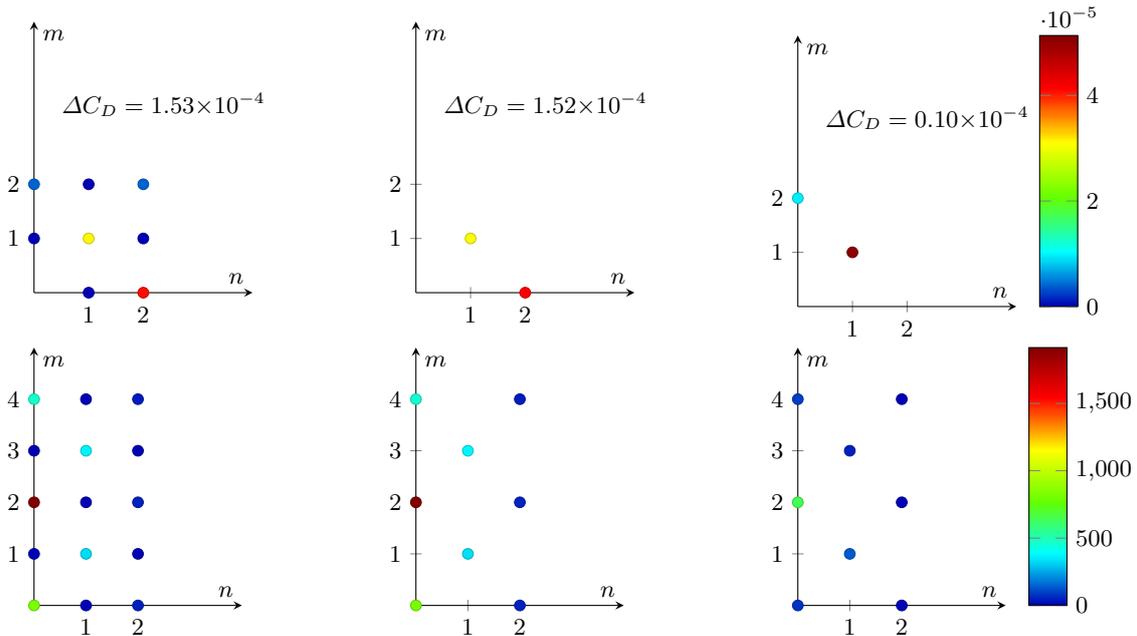
\begin{figure}
\begin{minipage}{0.33\textwidth}
    \begin{tikzpicture}
    \begin{axis}[colormap/bluered,
        axis lines=middle,
        width=6cm,
        unit vector ratio*=1 1 1,
        xmin=0, xmax=4,
        ymin=0, ymax=5,
        xtick={0,1,2}, ytick={0,1,2},
        legend style={draw=none,fill=none,name=legend,at={(0.3,1)},anchor=south west},
        xlabel=$n$,ylabel=$m$,
        point meta=explicit
    ]
       \addplot [scatter, scatter src=explicit, only marks, point meta min=0, point meta max=5.1347e-5] coordinates {
(1,0) [7.389548778278979e-13]
(2,0) [4.057185820012631e-05]
(0,1) [3.465133995037021e-13]
(1,1) [3.096833165628059e-05]
(2,1) [1.432209462557389e-12]
(0,2) [3.997402671136731e-06]
(1,2) [6.842481287399264e-13]
(2,2) [3.926958824971424e-06]
    };
    \end{axis}
    \draw (0.25,2.5) node[right] {$\Delta C_D=\num{1.53e-4}$};
    \end{tikzpicture}
\end{minipage}
\begin{minipage}{0.33\textwidth}
    \begin{tikzpicture}
    \begin{axis}[colormap/bluered,
        axis lines=middle,
        width=6cm,
        unit vector ratio*=1 1 1,
        xmin=0, xmax=4,
        ymin=0, ymax=5,
        xtick={0,1,2}, ytick={0,1,2},
        legend style={draw=none,fill=none,name=legend,at={(0.3,1)},anchor=south west},
        xlabel=$n$,ylabel=$m$,
        point meta=explicit
    ]
       \addplot [scatter, scatter src=explicit, only marks, point meta min=0, point meta max=5.1347e-5] coordinates {
(1,1) [3.0559049e-5]
(2,0) [4.12632667e-5]
    };
    \end{axis}
    \draw (0.25,2.5) node[right] {$\Delta C_D=\num{1.52e-4}$};
    \end{tikzpicture}
\end{minipage}
\begin{minipage}{0.33\textwidth}
    \begin{tikzpicture}
    \begin{axis}[colormap/bluered, colorbar,
        axis lines=middle,
        width=6cm,
        unit vector ratio*=1 1 1,
        xmin=0, xmax=4,
        ymin=0, ymax=5,
        xtick={0,1,2}, ytick={0,1,2},
        legend style={draw=none,fill=none,name=legend,at={(0.3,1)},anchor=south west},
        xlabel=$n$,ylabel=$m$,
        point meta=explicit
    ]
       \addplot [scatter, scatter src=explicit, only marks, point meta min=0, point meta max=5.1347e-5] coordinates {
    (1,1) [5.0439e-5]
    (0,2) [9.6144e-6]
    };
    \end{axis}
    \draw (0.25,2.5) node[right] {$\Delta C_D=\num{0.10e-4}$};
    \end{tikzpicture}
\end{minipage}
\begin{minipage}{0.33\textwidth}
    \begin{tikzpicture}
    \begin{axis}[
        colormap/bluered, axis lines=middle,
        height=5cm,
        unit vector ratio*=1 1 1,
        xmin=0, xmax=4,
        ymin=0, ymax=5,
        xtick={0,1,2}, ytick={0,1,2,3,4},
        legend style={draw=none,fill=none,name=legend,at={(0.3,1)},anchor=south west},
        xlabel=$n$,ylabel=$m$,
    ]
    \addplot [scatter, scatter src=explicit, only marks, point meta min=0, point meta max=1914] coordinates {
(0, 0) [824.8844859382724]
(1, 0) [2.261131743185933e-05]
(2, 0) [47.96084492758992]
(0, 1)  [4.141956006731665e-05]
(1, 1) [329.4789431303861]
(2, 1) [1.41651215946363e-05]
(0, 2) [1914.591614729496]
(1, 2) [2.22950752461426e-05]
(2, 2) [49.31145078993559]
(0, 3) [9.630169953342312e-05]
(1, 3) [362.3873829656392]
(2, 3) [1.574618085313418e-05]
(0, 4)  [461.1134809850595]
(1, 4) [3.687986710372032e-05]
(2, 4) [38.74925998287748]
    };
    \end{axis}
    \end{tikzpicture}
\end{minipage}
\begin{minipage}{0.33\textwidth}
    \begin{tikzpicture}
    \begin{axis}[
        colormap/bluered, axis lines=middle,
        height=5cm,
        unit vector ratio*=1 1 1,
        xmin=0, xmax=4,
        ymin=0, ymax=5,
        xtick={0,1,2}, ytick={0,1,2,3,4},
        legend style={draw=none,fill=none,name=legend,at={(0.3,1)},anchor=south west},
        xlabel=$n$,ylabel=$m$,
    ]
    \addplot [scatter, scatter src=explicit, only marks, point meta min=0, point meta max=1914] coordinates {
(0,0) [815.9]
(2,0) [48.77]
(1,1) [335.3]
(0,2) [1921]
(2,2) [51.74]
(1,3) [368.7]
(0,4) [455.56]
(2,4) [39.84]
    };
    \end{axis}
    \end{tikzpicture}
\end{minipage}
\begin{minipage}{0.33\textwidth}
    \begin{tikzpicture}
    \begin{axis}[
        colormap/bluered, colorbar, axis lines=middle,
        height=5cm,
        unit vector ratio*=1 1 1,
        xmin=0, xmax=4,
        ymin=0, ymax=5,
        xtick={0,1,2}, ytick={0,1,2,3,4},
        legend style={draw=none,fill=none,name=legend,at={(0.3,1)},anchor=south west},
        xlabel=$n$,ylabel=$m$,
    ]
    \addplot [scatter, scatter src=explicit, only marks, point meta min=0, point meta max=1914] coordinates {
(0,0) [74.3156849938601]
(2,0) [5.514317259373511]
(1,1) [134.7652735199577]
(0,2) [640.2949345599598]
(2,2) [3.566519627498638]
(1,3) [49.53349702803237]
(0,4) [92.36027629944658]
(2,4) [0.9874907872324614]
    };
    \end{axis}
    \end{tikzpicture}
\end{minipage}
\caption{Optimal forcing (top) and response (bottom) amplitudes for superharmonic cases $M=4$, $N=2$ at $(\beta,\omega)=(50,11.7)\times 10^{-5}$ for $A=\num{5.13e-5}$. Forcing has been optimised in all fundamental and superharmonic components (left); only at $(\beta,\omega)$ and  $(\beta,2\omega)$ (middle);  only at $(\beta,\omega)$  and $(\beta,0)$.}
\label{fig:analysis_H}
\end{figure}

For the optimal superharmonic forcing (oblique and TS waves), low levels of energy are observed in two other forcing components, the $(2\beta,0)$ and $(2\beta,2\omega)$ components. In figure \ref{fig:analysis_H}, we analyse their importance by plotting the total amplitude of each forcing and response harmonic. Also, to ensure converged results, we have increased the number of response harmonics and lowered the forcing amplitude. The left column shows the superharmonic optimized forcing and response amplitude for each harmonic, when forcing is allowed in all 8 forcing components as above. The $(0,2\omega)$ TS and $(\beta,\omega)$ oblique waves are the dominant forcing components, and the $(2\beta,0)$ and $(4\beta,0)$ streaky structures, the $(\beta,\omega)$ and $(3\beta,\omega)$ oblique waves and the (0,0) MFD in the response. The second column corresponds to an optimization restricted solely to the $(0,2\omega)$ and $(\beta,\omega)$ harmonics: it is seen that it reproduces closely the more complex optimization of the left-column (the reached $\Delta C_D$ is nearly the same). On the contrary (right column), it is seen that if the $(0,2\omega)$ TS wave is replaced by the $(2\beta,0)$ streaky component in the forcing (this also corresponds to a superharmonic forcing, but in spanwise wavenumber), then the optimal response only achieves weak drag increase, in agreement with those shown for the fundamental optimization at similar amplitudes. This validates that it is the interplay between the $(0,2\omega)$ TS and the $(\beta,\omega)$ oblique harmonic that accounts for the strong amplification observed in H-type transition.

The catalytic role of the TS waves in the superharmonic H-type case be also evidenced from a weakly-nonlinear analysis based on scaling arguments. The analysis (see appendix \ref{sec:WNL}) shows that the drag increase, for the two optimal fundamental and superharmonic cases, scales as
\begin{align}
    \Delta C_D &=& \Delta C_{D,2}A^2&+&\Delta C_{D,3}A^3&+&\Delta C_{D,4}A^4&+&\cdots && \mbox{ for superharmonic forcing} \\
    \Delta C_D &=& \Delta C_{D,2}A^2&& &+&\Delta C_{D,4}A^4&+&\cdots && \mbox{ for fundamental forcing}.
\end{align}
Hence, superharmonic resonant forcing allows the presence of additional odd terms in the expansion.
For example, the $A$-order $(0,2\omega)$ TS and $(\beta,-\omega)$ oblique waves generate the $ A^2$-order $(\beta,\omega)$ oblique wave, which may interact with the $ A$-order $(-\beta,-\omega)$ oblique wave to promote an $ A^3$-order (0,0) MFD. Hence, in the case of superharmonic forcing, it is possible to take advantage of the odd-orders to optimize the drag increase, while for fundamental oblique forcing, only even orders exist in the expansion. 

\subsection{Fundamental forcing for higher $ A $ and the effects of truncation} \label{sec:fundconv}

 The results shown above were obtained with a truncated expansion with $M=N=2$ response modes. The impact of the truncation can be preliminary assessed by examining the amplitude of higher or truncated  wavenumber/frequency components. In figure~\ref{fig:A_fundamental}, we observe that the second frequency harmonics, $(m \beta,2\omega)$, have a much smaller amplitude than the fundamental ones,  $(m \beta,\omega)$. However, this is not the case for the truncation in $\beta$ harmonics. As we saw above, a strong response was obtained at $(2\beta,0)$ component through nonlinear interactions.  

\begin{figure}
\includegraphics[width=1\textwidth]{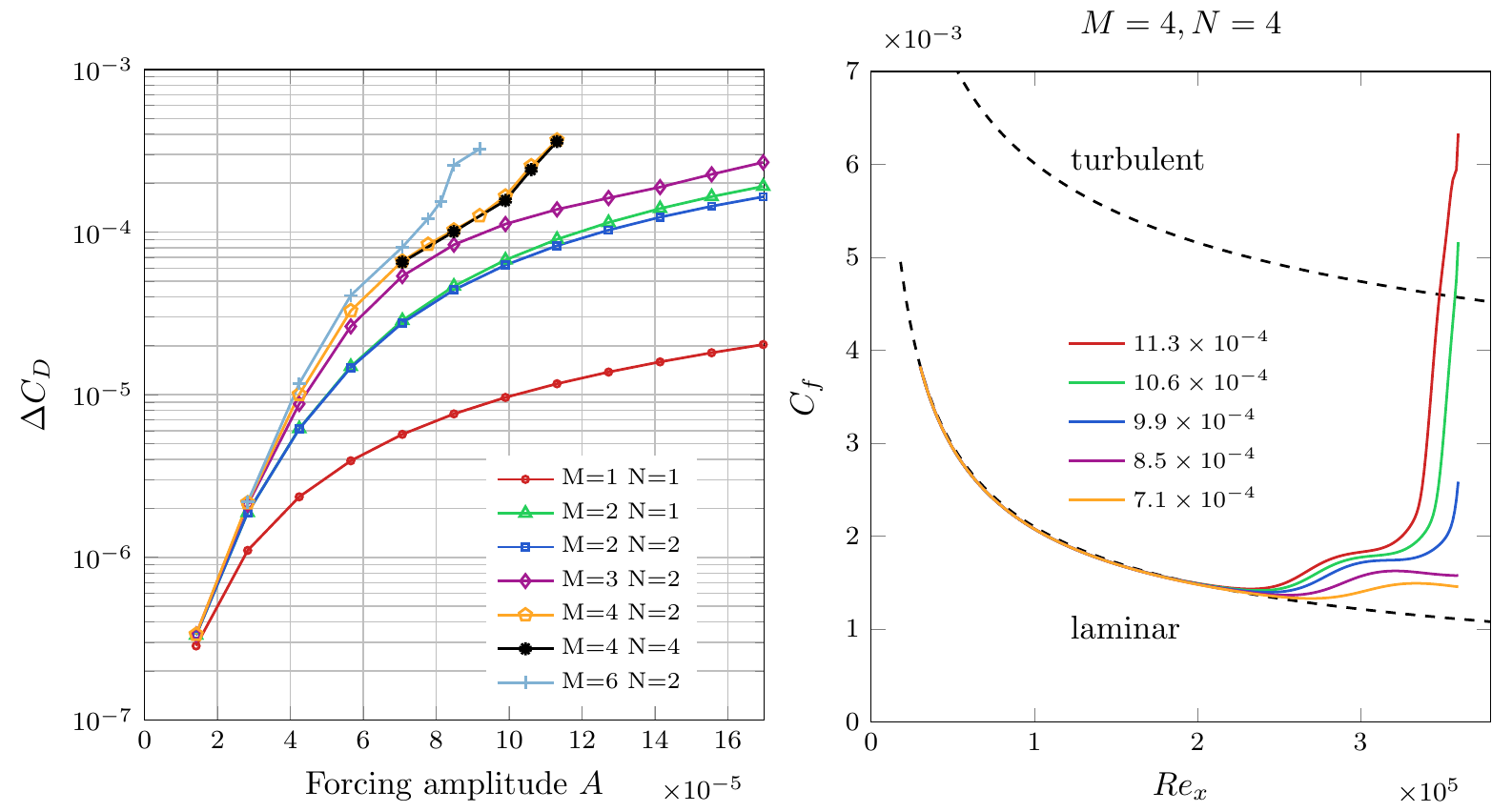}
\caption{Maximum drag increase as a function of forcing amplitude for different truncations in spanwise (M) and frequency (N) components (left). Results shown for the optimal oblique fundamental case $(\beta,\omega)$=($33.3,11.7)\times10^{-5}$, with reflectional symmetry in the spanwise direction. Skin friction coefficient for $M=N=4$ as a function of streamwise distance for different forcing amplitudes (right). For $A>\num{8.5e-5}$, varicose transition of the low-speed streaks is observed.}
\label{fig:convergence_optim_fundamental}
\end{figure}

To directly assess the truncation error, calculations were performed with larger $M$ and $N$.  The resulting maximum in the cost function is shown, as a function of forcing amplitude, in figure~\ref{fig:convergence_optim_fundamental}a for various orders of truncation. Apart from the most highly truncated case, we see a tendency towards convergence for forcing amplitudes $A<\num{7e-05}$. The $M=N=1$ case is clearly too highly truncated--this can be understood physically since the nonlinear amplification mechanisms described above require the generation of streaks at $(2\beta,0)$.

As discussed above, during the initial stages of transition and for a small forcing amplitude, the second and higher $\omega$-harmonics are not as strongly amplified as the $\beta$-harmonics, meaning that the energy spreading occurs faster in $\beta$ than $\omega$. For example, the $M=2,N=1$ case is almost identical to the $M=N=2$. Similarly,  $M=4,N=2$ is close to $M=N=4$. The dominance of the $\beta$-cascade has been observed in various DNS and experimental transition studies (\cite{breuer1997late} and \cite{yeo2010dns} triggered transition with an impulse wavepacket, or K-type controlled transition \citep{rist1995direct}) and it is consistent with the results presented here.

\subsubsection{Streak breakdown}

Increasing further the number $M$ of $\beta$-harmonics, a sudden change is observed in the drag values for $A\approx8\times 10^{-5}$ and for $M\ge 4$, see figure~\ref{fig:convergence_optim_fundamental}a. The skin friction coefficient for various amplitudes is shown in figure~\ref{fig:convergence_optim_fundamental}b for $M=N=4$. The spanwise averaged skin-friction coefficient is calculated from the $(0,0)$ streamwise velocity component:
\begin{equation*}
    C_f = \frac{\tau_{\textrm{wall}}}{\frac{1}{2} U_\infty^2},
    \quad \textrm{with} \quad 
    \tau_{\textrm{wall}} = \nu \left(  \frac{\partial \hat{u}_{00}}{\partial y} \right)_{y=0}.
\end{equation*}
For comparison, the values of the laminar skin friction coefficient ($C_f^{\mathrm{lam}} = 0.664/ \sqrt{Re_x}$) and the empirical one corresponding to fully developed turbulence ($C_f^{\mathrm{turb}} = 0.455 / \ln^2 ( 0.06 Re_x )$) are shown with dashed lines \citep{white2006viscous,yeo2010dns}. The transition is accompanied by an overshoot of the skin friction coefficient up to the empirical turbulent values, for sufficiently high forcing amplitudes. Increasing $M$, the transition threshold moves to lower forcing amplitudes, suggesting that the flow has transitioned to a more complex regime, for which a large number of harmonics would be required to capture quantitatively accurately the solution, as will be discussed in greater detail below.

\begin{figure}
\includegraphics[width=0.5\textwidth,trim={1cm 0 1cm 0},clip]{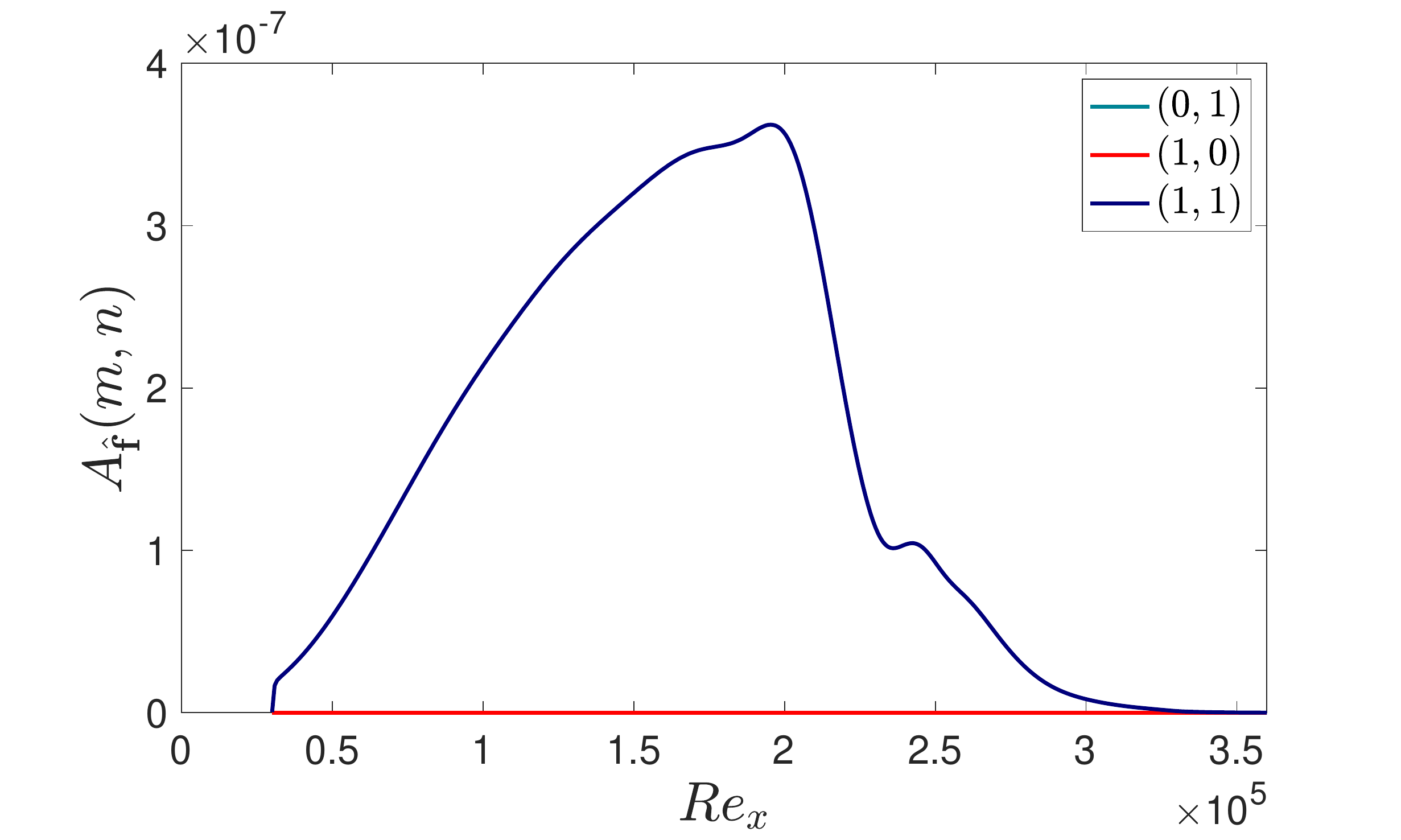}\vspace{1cm}
\includegraphics[width=0.5\textwidth,trim={1cm 0 1cm 0},clip]{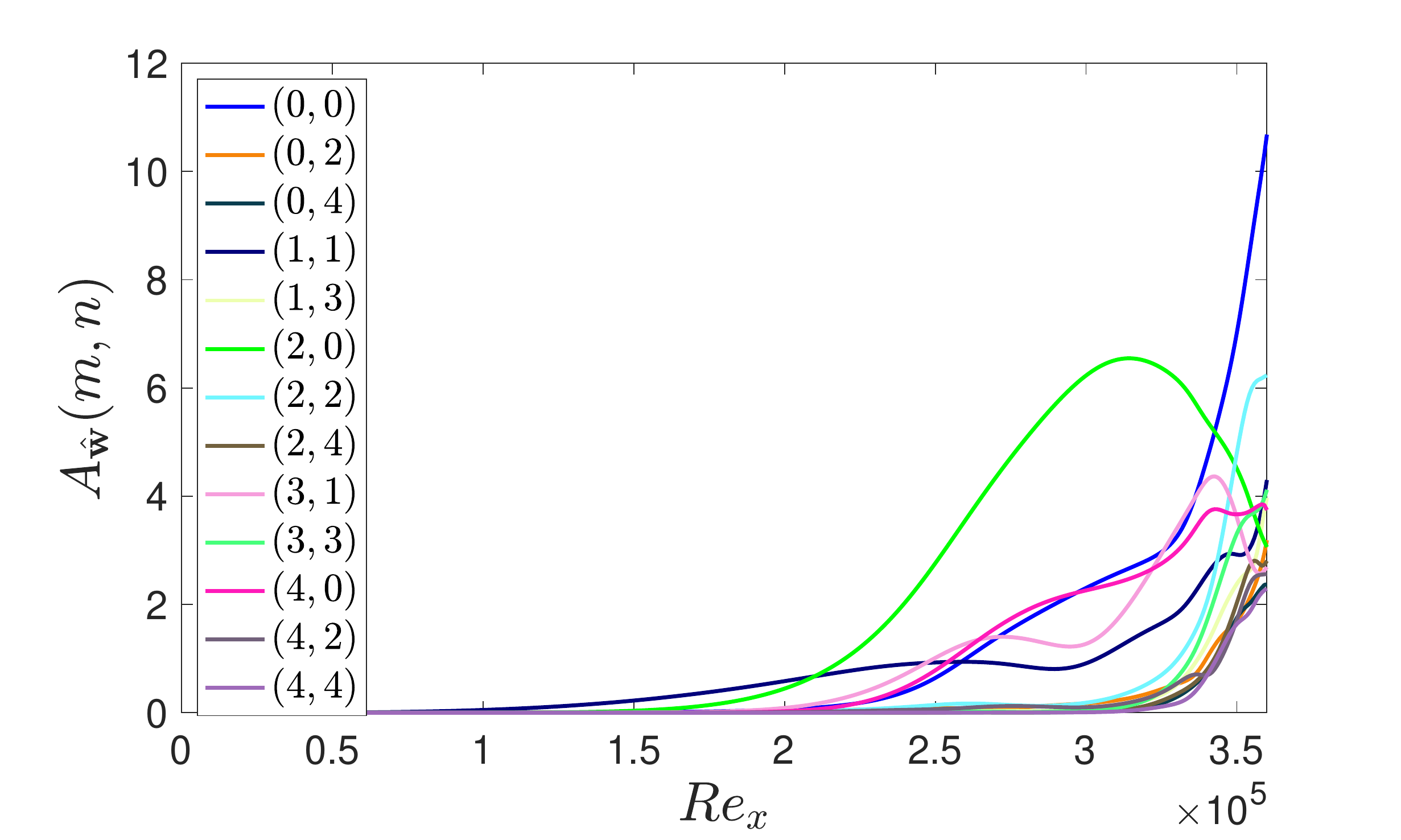}
\includegraphics[width=1\textwidth,trim={35cm 12cm 35cm 25cm},clip]{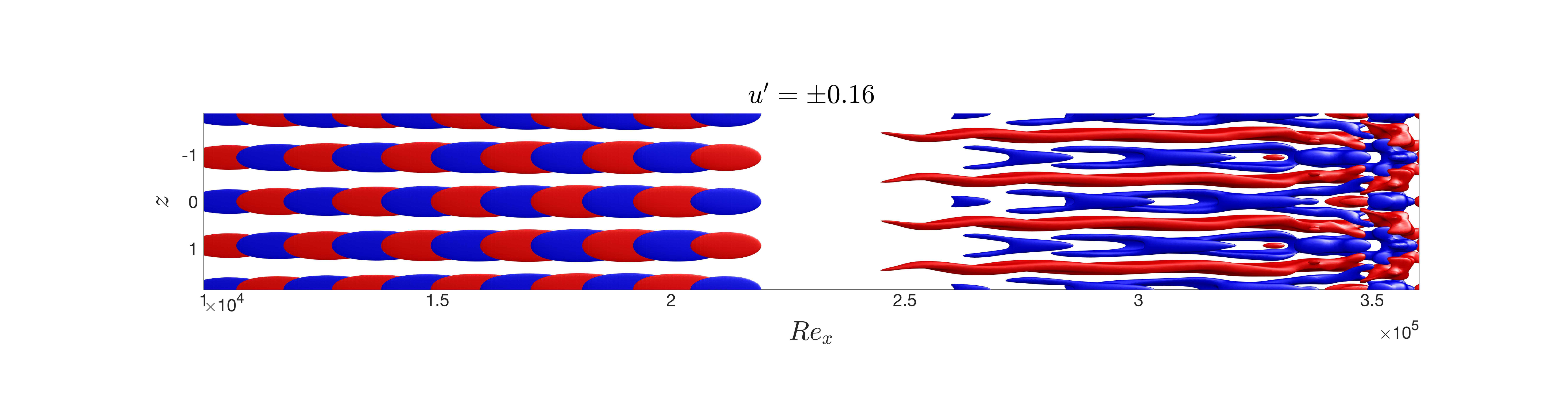}
\includegraphics[width=1\textwidth,trim={30cm 0cm 30cm 13cm},clip]{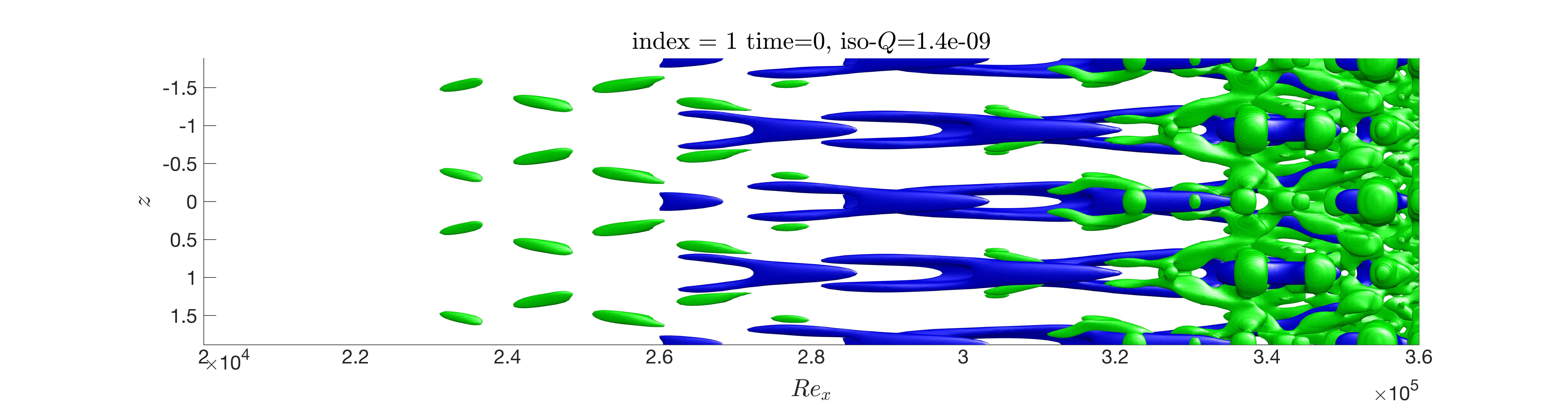}
\caption{Laminar-turbulent transition for optimal oblique fundamental case (symmetry in $z$) with $M=N=4$ at $(\beta,\omega)$=($33.3,11.7)\times10^{-5}$ for $A=\num{11.3e-5}$. Amplitude  for forcing (top left) and response (top right) for each individual harmonic component $(m,n)$. Isosurfaces of streamwise perturbations for $f'_u= \pm \num{8.3e-9}$ (middle left) and $u'= \pm 0.16$ (middle right). Vortical structures visualized with the $Q$-criterion (iso-$Q=\num{1.4e-9}$; green) and low speed streaks ($u'=-0.16$; blue). Two fundamental wavelenghts are shown in $z$ to facilitate the presence of staggered $\Lambda$-structures and hairpins.}
\label{fig:Oblique62_transition}
\end{figure}

The amplitudes of the forcing and response components are shown in figure~\ref{fig:Oblique62_transition} for $A=\num{11.3e-5}$ and the $M=N=4$ case, again for the optimal fundamental forcing. The forcing reaches maximum amplitude further downstream at $Re_x=200,000$, when compared to the lower amplitude case, and also a second distinct region of forcing appears for $Re_x>250,000$. For all the cases examined, we noticed that the second region of forcing triggers streak oscillations in the streamwise direction and they subsequently break down. Regarding the response, once the $(2\beta,0)$ streaks reach sufficiently high amplitude, the harmonic component $(4\beta,0)$ increases up to $Re_x \approx 320,000$ along with the $(3\beta,\omega)$ harmonic. The latter is responsible for the generation and progressive elevation of hairpins from the wall.  The MFD increases monotonically in agreement with the increase in skin friction coefficient. A cascade of nonlinear interactions makes the amplitude of all the harmonic components to increase significantly toward the end of the domain, where the skin friction has exceeded the empirical turbulent value.

For all the cases presented above we have imposed symmetry in $z$. Under this restriction, the low-speed streaks undergo varicose oscillations in $x$ whereas the high-speed streaks undergo sinuous oscillations (subharmonic varicose case in \cite{andersson2001breakdown}) creating a staggered pattern of $\Lambda$-structures and the emergence of hairpin vortices further downstream \citep{asai2002instability}. Similar behavior has been observed in DNS simulations  \citep{berlin1999numerical} where a pair of oblique waves was introduced in the domain inlet and reflectional symmetry in spanwise was imposed. Initial stages of this process are visualized using the Q-criterion. The emergence of the hairpin vortices coincides with the final regime during the transition process and the overshoot of the skin friction coefficient to the turbulent values.

\subsubsection{Breaking the $z$-reflectional symmetry}

\begin{figure}
\centering
\includegraphics[width=1\textwidth]{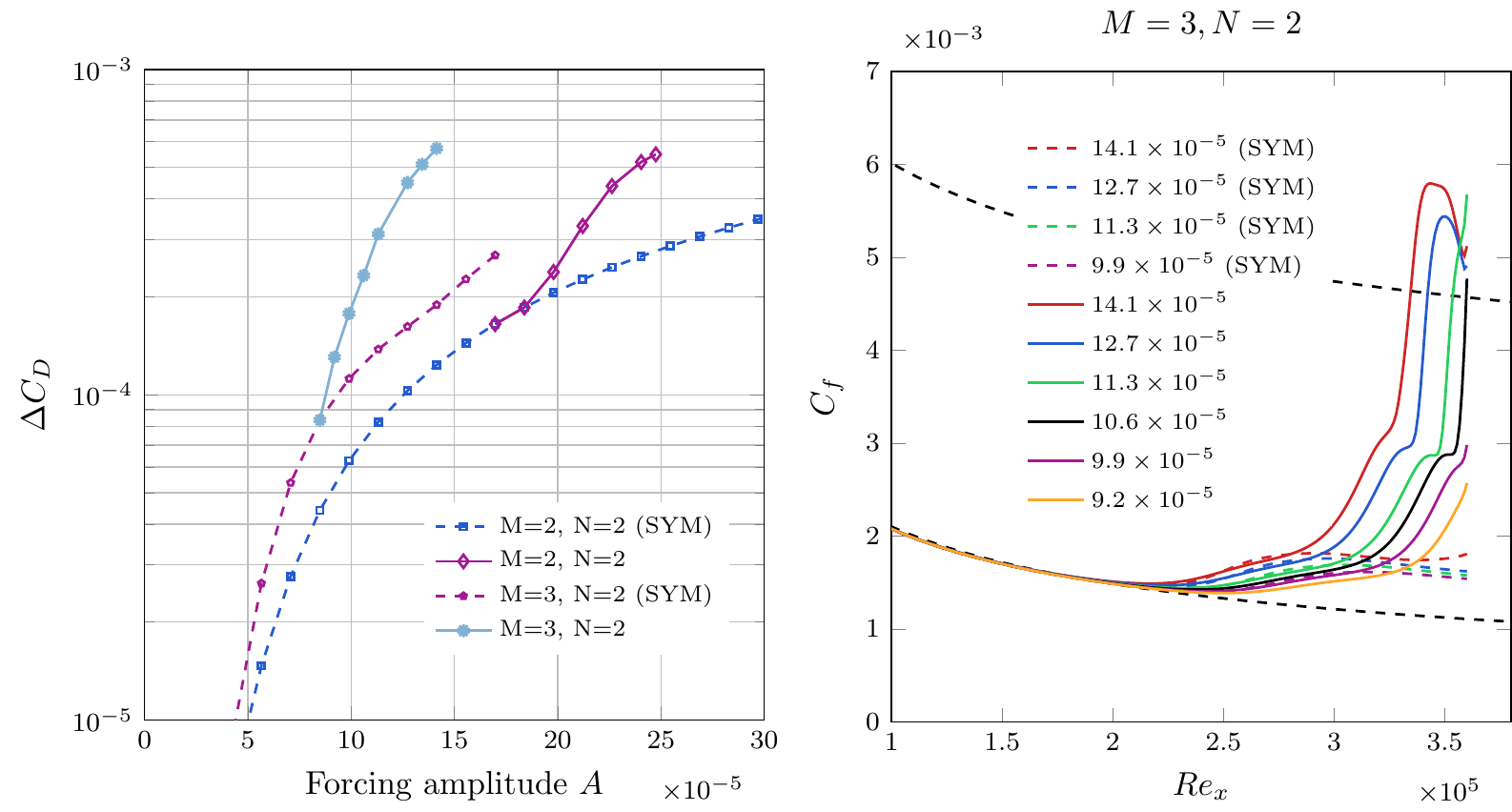}
\caption{Maximum drag increase for optimal oblique fundamental forcing with no imposed symmetry in $z$ (solid lines) and $z$-reflectional symmetry (SYM,
 dashed lines) for various orders of truncation $M,N$ (left). Skin friction coefficient as a function of $Re_x$ for various forcing amplitudes for $M=3,N=2$ (right). For the non-symmetric cases, sinuous-like transition of the low-speed streaks is observed.}
\label{fig:Cf_amplitude}
\end{figure}

\begin{figure}
\includegraphics[width=0.5\textwidth,trim={1cm 0 1cm 0},clip]{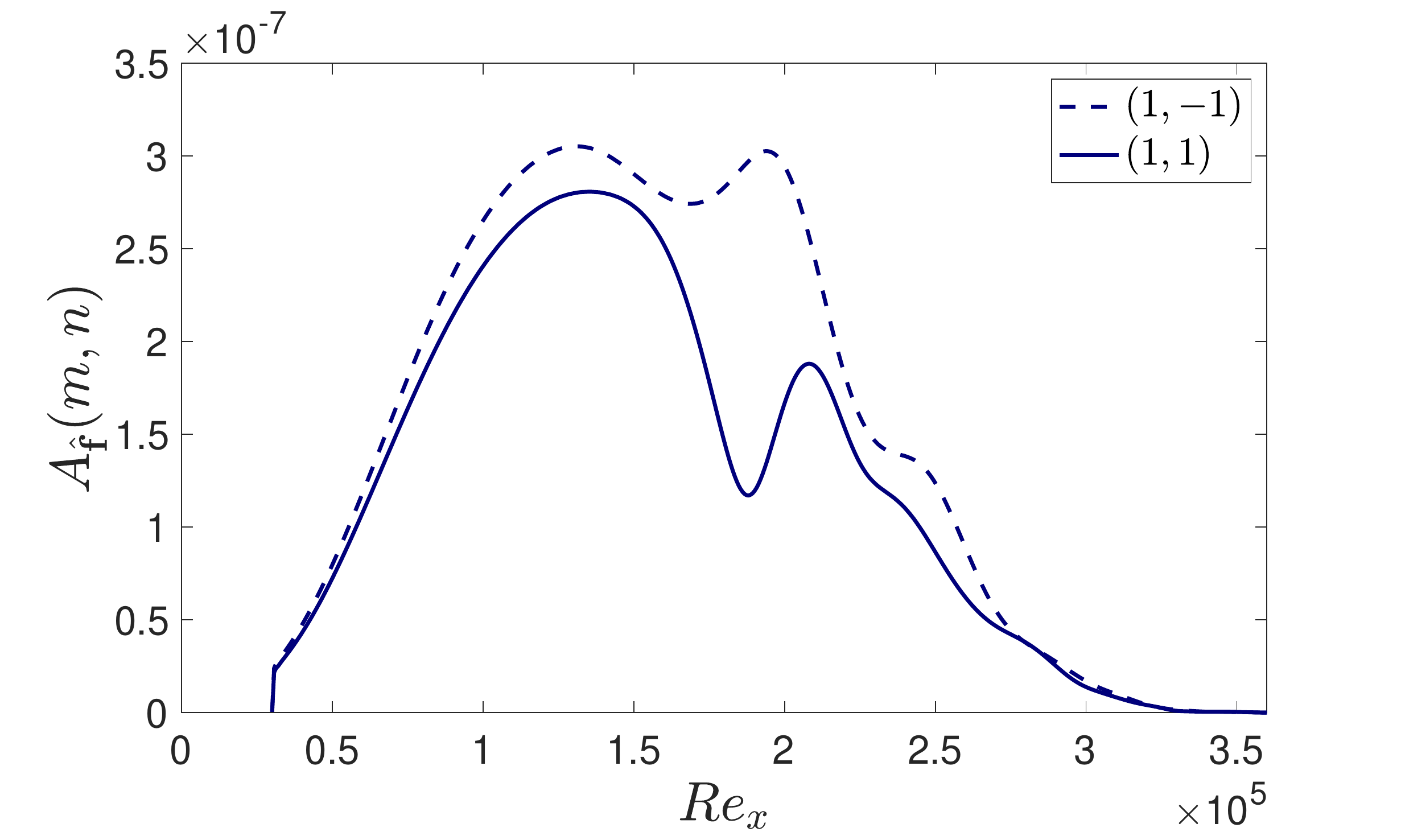}
\includegraphics[width=0.5\textwidth,trim={1cm 0 1cm 0},clip]{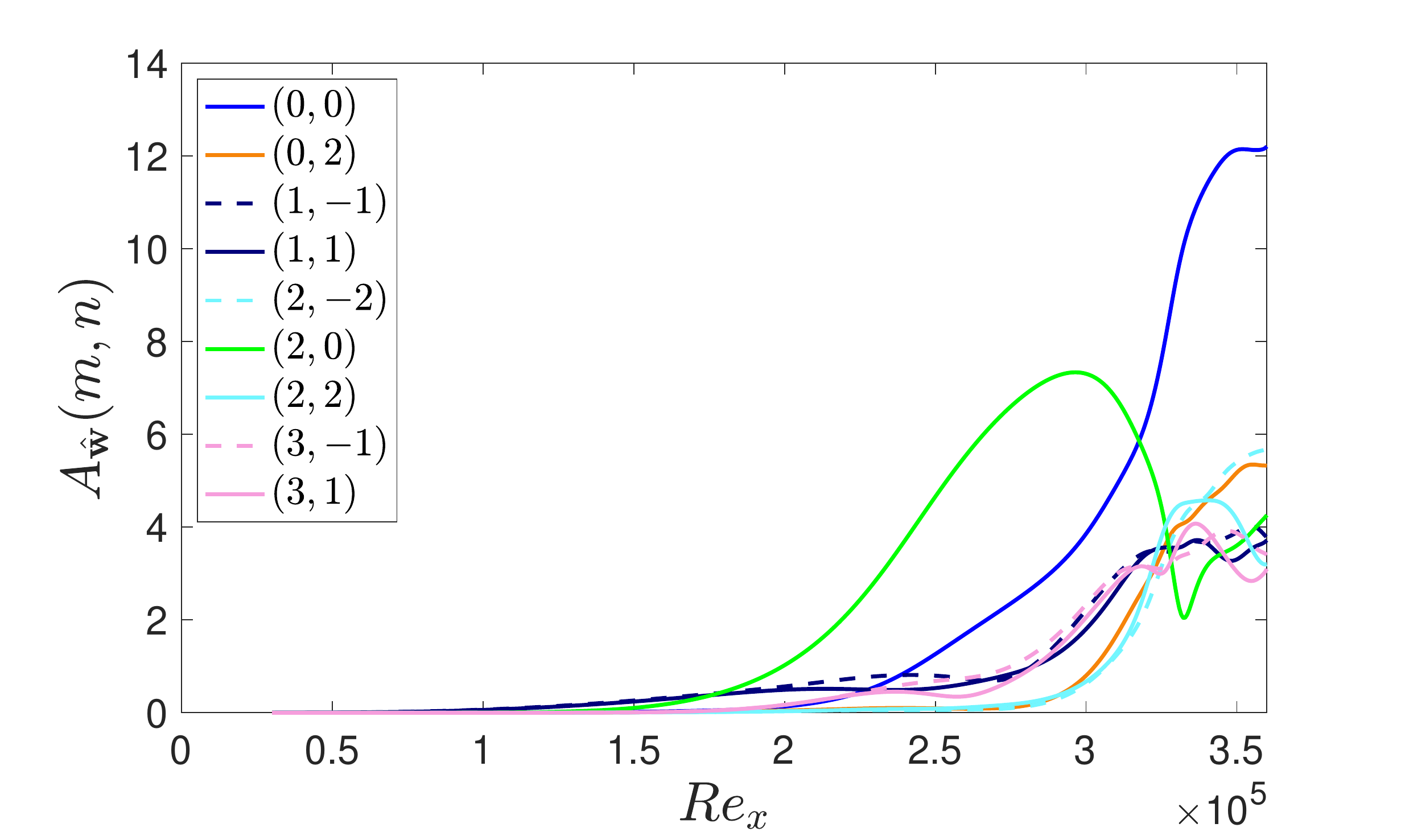}\vspace{1cm}
\includegraphics[width=1\textwidth,trim={36cm 12cm 36cm 25cm},clip]{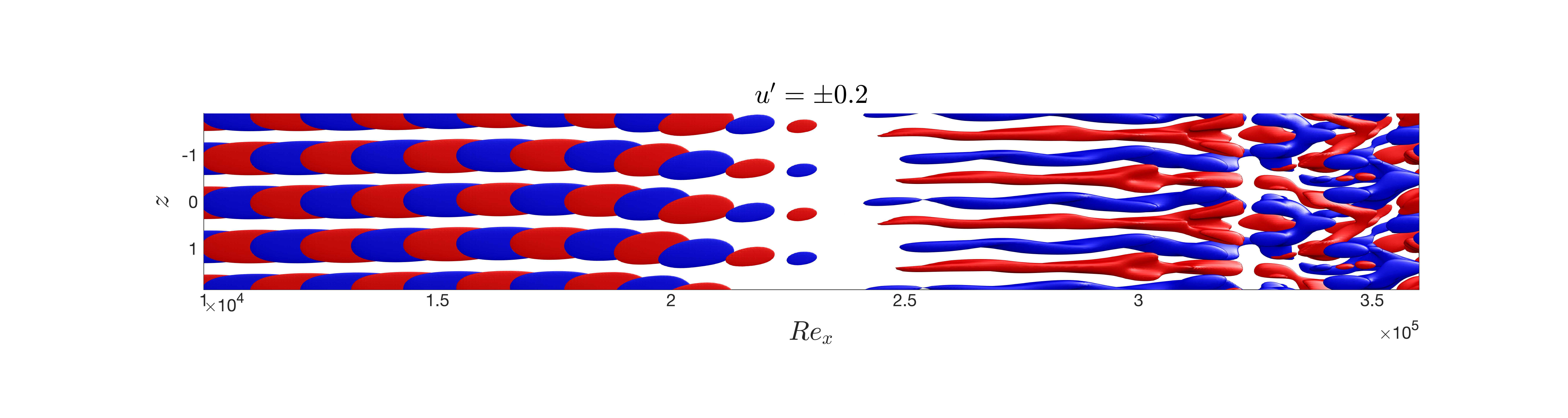}
\includegraphics[width=1\textwidth,trim={30cm 0cm 30cm 13cm},clip]{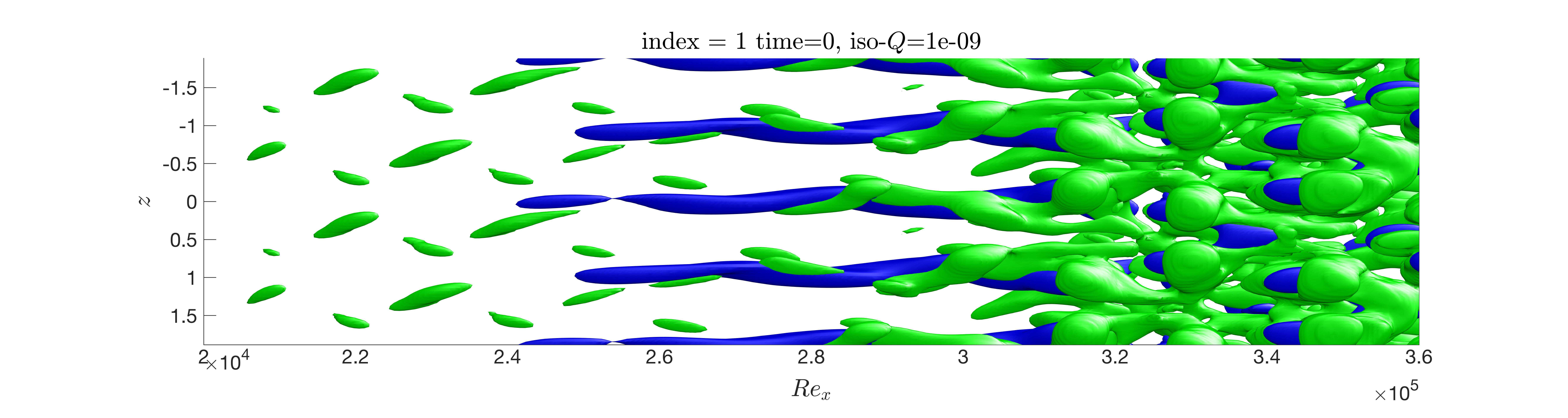}
\caption{Laminar-turbulent for optimal oblique fundamental forcing (no symmetry in $z$) with $M=3,N=2$, $(\beta,\omega)$=($33.3,11.7)\times10^{-5}$, $A=\num{14.1e-5}$. Maximum amplitudes of optimal forcing (top left) and response (top right) for each individual harmonic component $(m,n)$.  Isosurfaces of streamwise perturbations for $f'_u$ and $u'$ (middle). Vortical structures visualized with the $Q$-criterion along with low-speed-streaks (bottom). Two fundamental wavelenghts are shown in $z$ ($f'_u= \pm \num{8.3e-9}, u'= \pm 0.2, Q=10^{-9}$).}
\label{fig:ObliqueGEN}
\end{figure}

In this section we relax the reflectional symmetry assumption in $z$ that was imposed above. The computational cost increases since we have to account for  almost twice the number of harmonics. We focus again here on the optimal fundamental forcing at $(\beta,\omega)$=($33.3,11.7)\times10^{-5}$ that is initiated through a pair of equal amplitude oblique waves. 



The dependence of the maximum drag increase on the forcing amplitude with and without $z$-reflectional symmetry is shown in figure~\ref{fig:Cf_amplitude}a for $M=N=2$ and $M=3,N=2$. The dashed lines correspond to the values obtained in the previous section imposing reflectional symmetry (SYM cases). We repeated the optimization for each forcing amplitude and restricted the forcing to act only on the oblique $(\beta,\omega)$ component without imposing symmetry in $z$. The initial guess was the symmetric solution with random noise of 10\% of the maximum value of each forcing component added to break the symmetry.  Up to a critical forcing amplitude, $A_{c}=\num{18e-5}$ for $M=N=2$ and $A_{c}=\num{9.2 e-5}$ for $M=3,N=2$, the solution converges to the one satisfying the reflectional symmetry. At the critical amplitude the solution bifurcates to a different equilibrium with approximately two times higher drag increase than the one for the symmetric case. 

In figure~\ref{fig:Cf_amplitude}b, the skin friction coefficients of the two cases with and without $z$-reflectional symmetry are shown for $M=3,N=2$ for various forcing amplitudes. For the symmetric cases, the skin friction values saturate to values close and above the laminar curve for low forcing amplitudes. Only the highest amplitude shows a tendency for departuture from the trend of the lower amplitude curves, indicating that the streaks are on the verge of symmetric breakdown. Relaxing the symmetry assumption, and for the same amplitudes as the symmetric case, the skin friction reaches values significantly higher than the turbulent ones. For the two highest amplitudes, after the overshoot to the turbulent values, the skin friction drops. In contrast, for the symmetric case a monotonic increase for similar values beyond the threshold of the turbulent skin friction values was observed (see figure \ref{fig:convergence_optim_fundamental}b).

The amplitude of the forcing and response harmonic components is shown in figure~\ref{fig:ObliqueGEN} for the  $M=3,N=2$ case. The oblique forcing components, $(\beta,+\omega)$ and $(\beta,-\omega)$, break their symmetry and are characterized by different amplitudes now. Also, two new local maxima appear for $Re_x>\num{2e5}$ in the amplitude forcing curves. This is similar to the one that appeared in the symmetric case that promoted the varicose streak breakdown, but here it is more pronounced. The amplitude response of the different harmonic components shows that the initial stages are similar to the ones observed in the case with  imposed spanwise symmetry. The oblique waves $(\beta,\pm\omega)$ interact nonlinearly to promote the growth of rolls-streaks at twice the spanwise wavenumber, $(2\beta,0)$.  The $(3\beta,\pm\omega)$ components are amplified as well, similar to the symmetric case. Immediately after that, all the harmonics appear to attain high energy values, due to the more effective energy spread through the symmetry break of the forcing.

Despite the similaties in the amplitude response, the flow is qualitatively different to the symmetric case. The reflectional symmetry break of the forcing can be observed in the isorfurfaces of the streamwise velocity perturbation. Towards the decaying phase of the forcing, the dominance of the left-travelling $(1,-1)$ oblique wave is evident. This mechanism promotes in an optimal way the sinuous-like breakdown of the low-speed streaks. The sinuous low-speed streak breakdown occurs for lower forcing thresholds compared to the varicose breakdown. This is in accordance with previous results in the literature examining streak breakdown \citep{andersson2001breakdown}. This regime is not associated with hairpin vortices, but with quasi-streamwise vortices of alternate sign, in accordance with the findings of \cite{asai2002instability}. Visualization of the vortices using the $Q$-criterion shows longitudinal vortices staggered on each side of the low speed streaks up to $Re_x=300,000$. At this location the the low-speed streaks come close together in an alternate staggered manner and merge. In the same time they break and then create individual $\Lambda$-like staggered structures. Exactly at this stage, the skin friction coefficient has reached the turbulent value.

\subsection{Superharmonic forcing for high $\textit{MN}$ and high $ A $} \label{sec:superconv}

\begin{figure}
\includegraphics[width=1\textwidth]{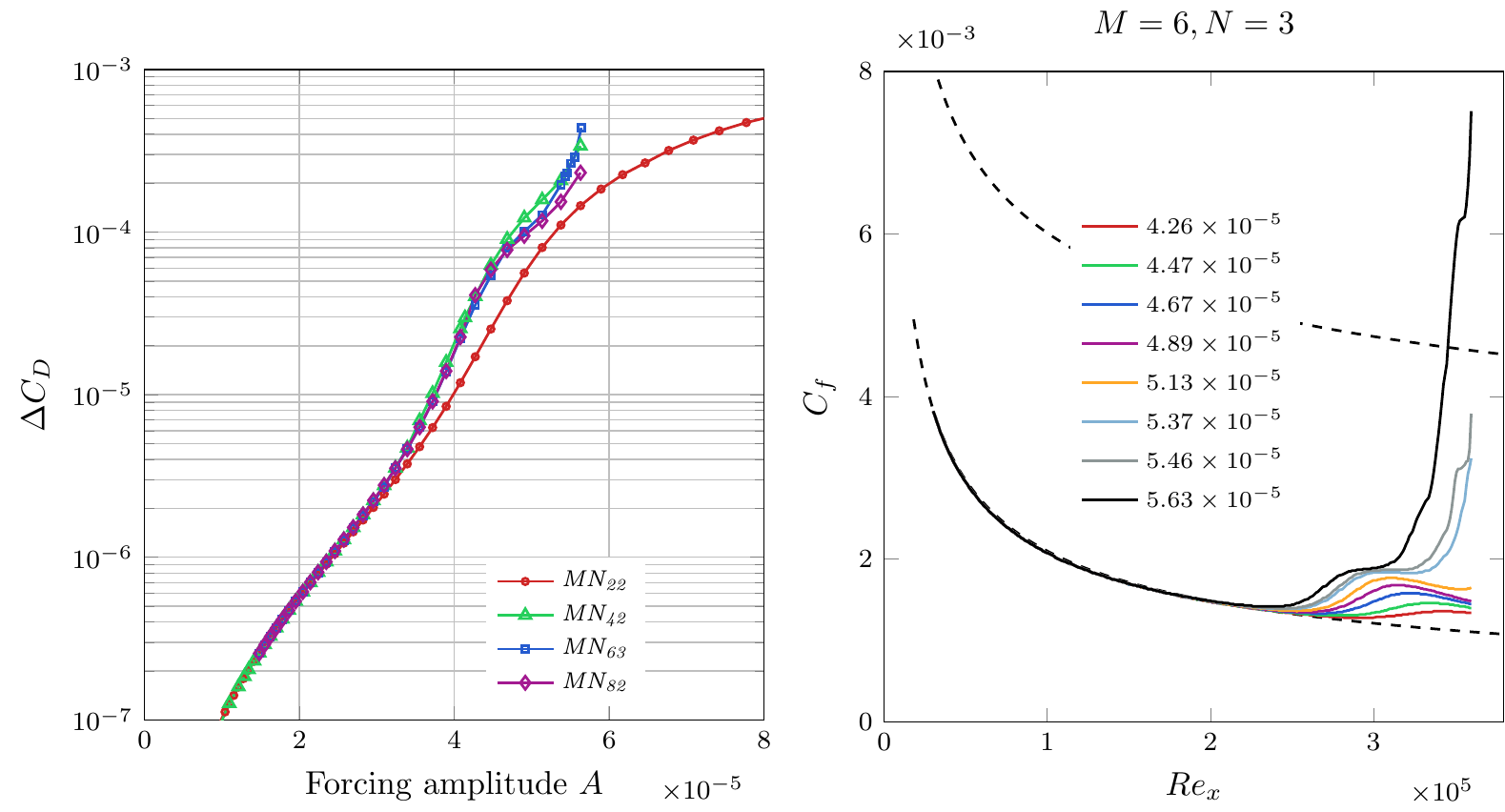}
\caption{Maximum drag increase for optimal H-type superharmonic forcing at $(\beta,\omega)=(50,11.7)\times10^{-5}$ with $z$-symmetry as a function of forcing amplitude (left). Various orders of truncation $\mathit{MN}$ are shown. Skin friction coefficient (right) as a function of $Re_x$ for $M=6,N=3$.}
\label{fig:J_sub}
\end{figure}

A convergence study of the truncated HBM expansion was performed for the superharmonic case with imposed $z$-reflectional symmetry. Up to a forcing amplitude $A=\num{3e-05}$, the solution appears converged, for the $M=N=2$ case.  Increasing the forcing amplitude, the flow transitions. For $A> \num{5.13e-05}$ and $M=6,N=3$, the skin friction coefficient overshoots towards the turbulent values, see figure~\ref{fig:J_sub}b. Similarly to the symmetric fundamental case, a monotonic increase of the skin friction coefficient is observed by increasing the forcing amplitude.

\begin{figure}
\includegraphics[width=0.5\textwidth,trim={1cm 0 1cm 0},clip]{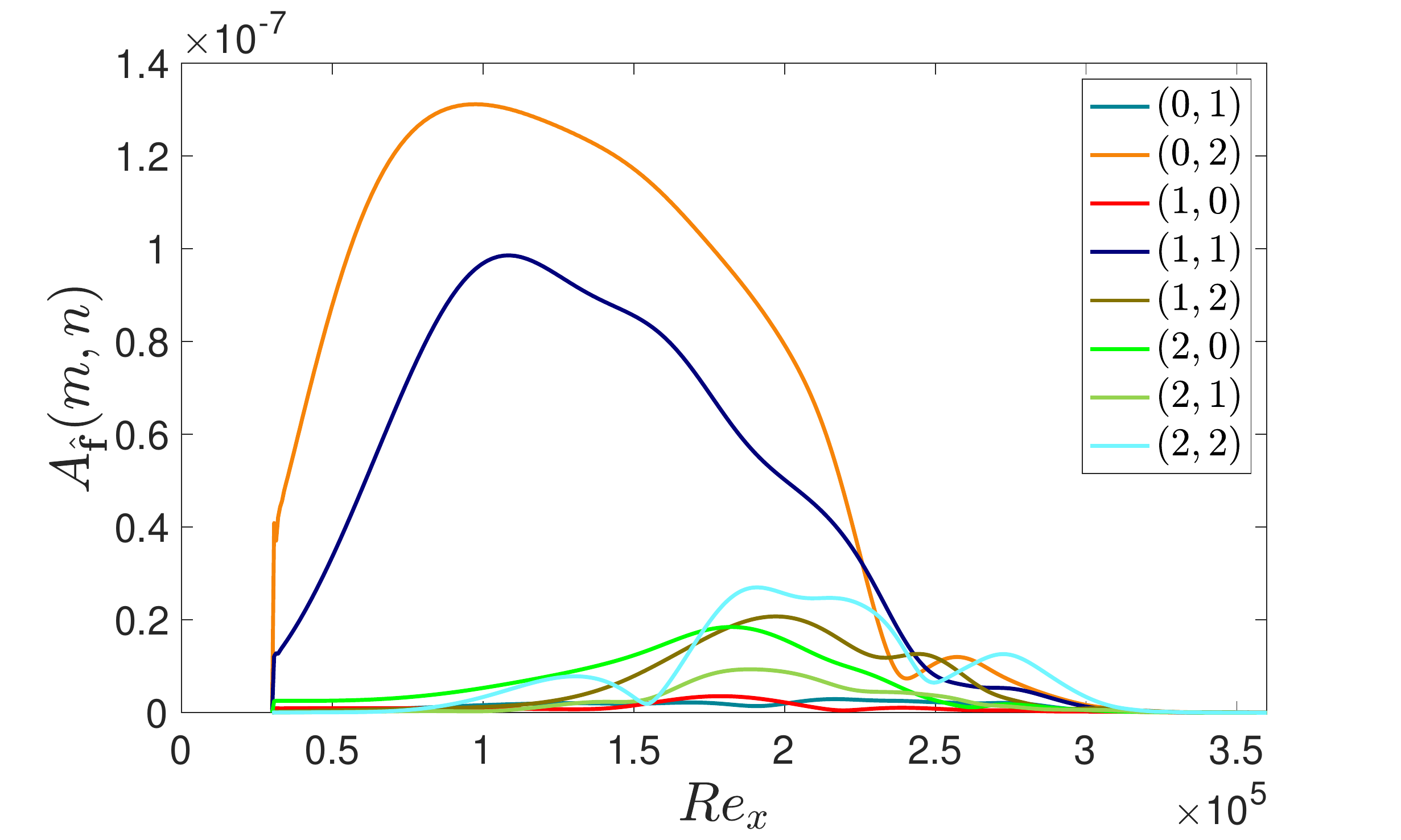}
\includegraphics[width=0.5\textwidth,trim={1cm 0 1cm 0},clip]{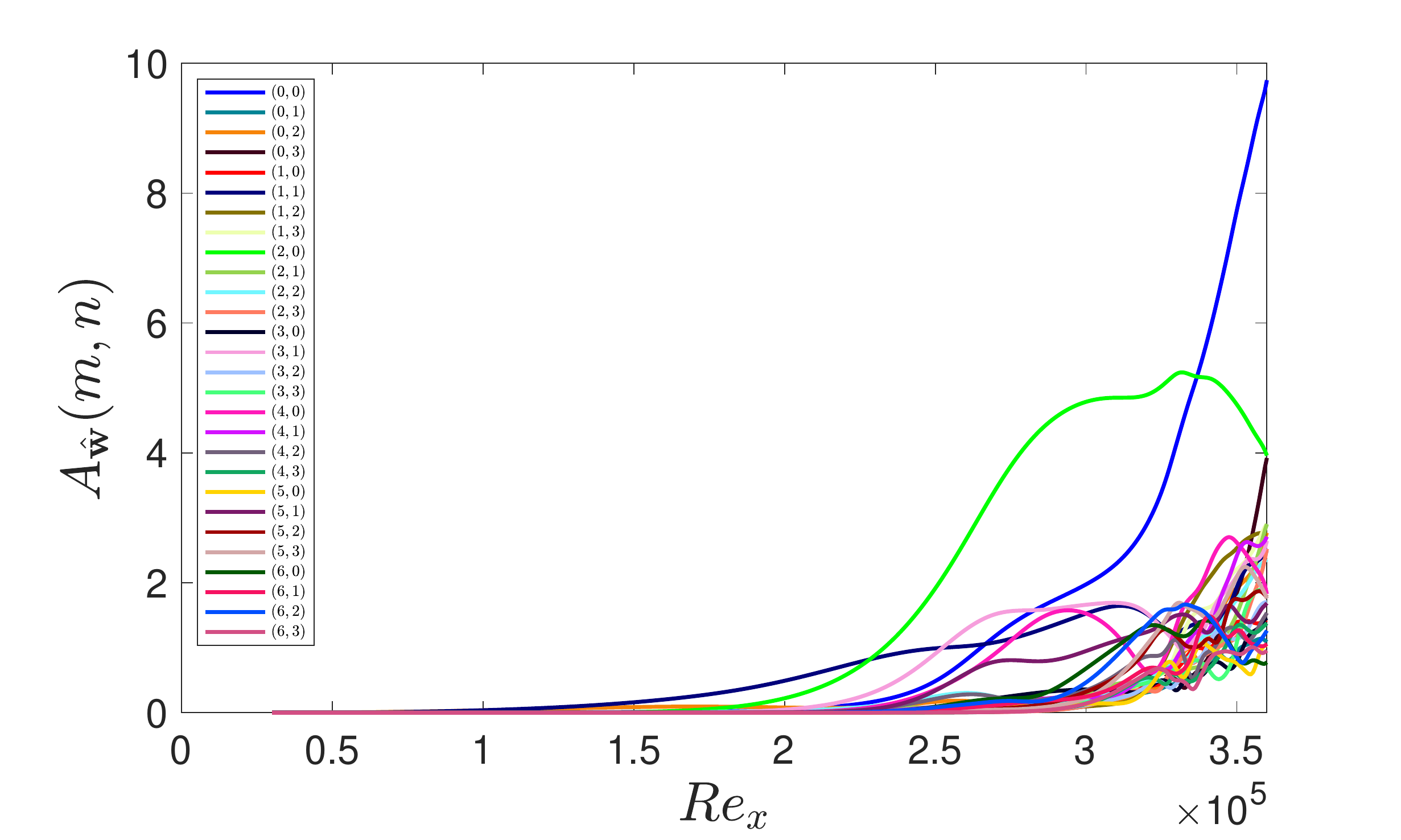}\vspace{1cm}
\includegraphics[width=1\textwidth,trim={36cm 12cm 36cm 32cm},clip]{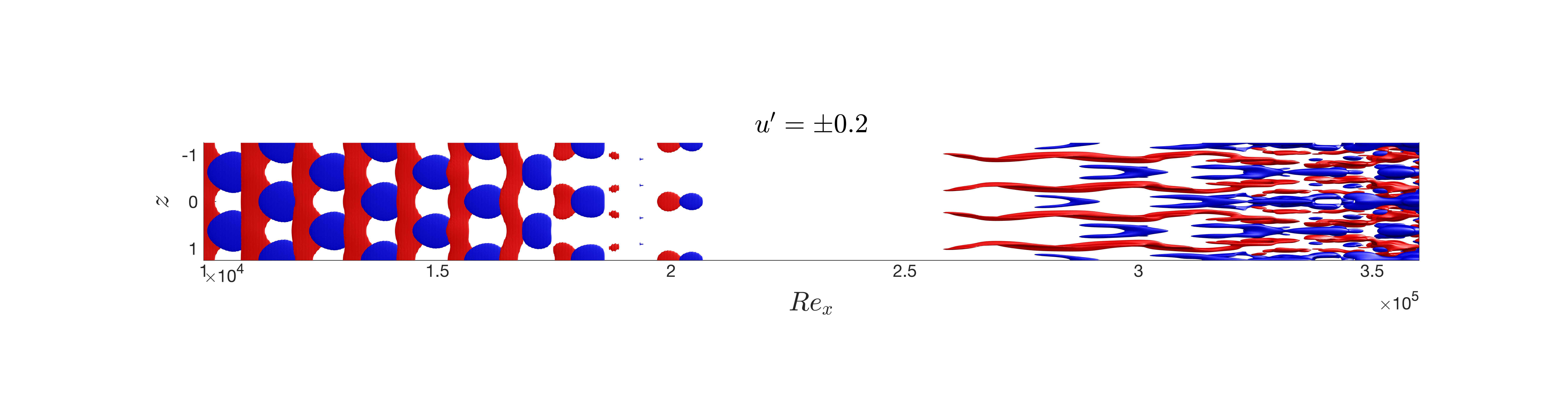}
\includegraphics[width=1\textwidth,trim={30cm 10cm 30cm 24cm},clip]{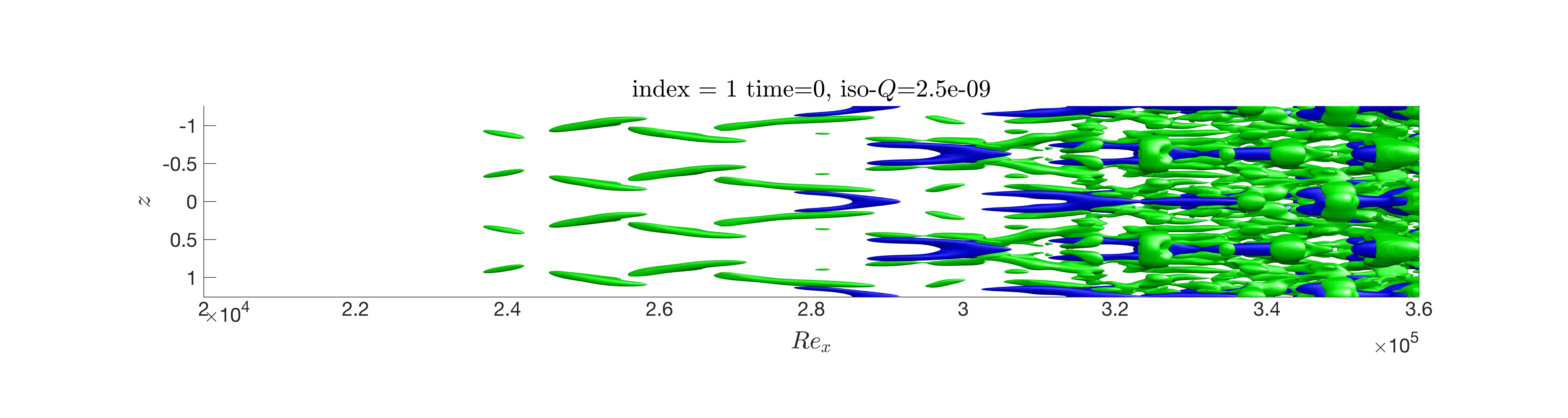}
\caption{Laminar-turbulent transition for optimal H-type superharmonic forcing with $M=6,N=3$,  $(\beta,\omega)=(50,11.7)\times10^{-5}$, $A=\num{5.65e-5}$. Total energy  for forcing (top left) and response (top right) for each individual harmonic component $(m,n)$. Isosurfaces of streamwise perturbations $f'_u$ (middle left) and $u'$ (middle right). Vortical structures visualized with the $Q$-criterion along with low-speed-streaks (bottom). Two fundamental wavelenghts are shown in $z$ ($f'_u= \pm \num{6.2e-9}, u'= \pm 0.2$, Q=\num{5.5e-9}).}
\label{fig:Htype_highampl63}
\end{figure}

The amplitude of the forcing and response components are shown in figure~\ref{fig:Htype_highampl63}, for the high amplitude forcing case with $M=6,N=3$. The dominant forcing component is the $(0,2\omega)$ mode followed by the $(\beta,\pm\omega)$ components. The nonlinear interaction of $(\beta,\pm\omega)$ response components create a strong response in the $(2\beta,0)$ component. This process continues resulting in the progression of energy along the $\beta$-axis and the emergence of $(4\beta,0)$ and $(6\beta,0)$ components. Although higher harmonics are also created by nonlinear interactions, they are less energetic since they are not amplified by the transient growth to the same degree as the low-wavenumber modes \citep{breuer1997late}.
 The low-speed streaks undergo symmetric varicose type of oscillations, whereas the high-speed streaks oscillate in a sinuous mode in the streamwise direction. The response appears similar to the one for the fundamental oblique forcing, where spanwise reflectional symmetry is imposed. The low-speed streaks attain a $\Lambda$-shape, which creates a staggered pattern of $\Lambda$ vortices. These vortices are identified using the $Q$-criterion.

\subsection{Summary and implications for turbulent dynamics}

\begin{figure}
\vspace{0.5cm}
{\scriptsize \hspace{0.7cm} (a) Fundamental ($z$-symmetry) \hspace{0.7cm} (b) Fundamental (no symmetry)  \hspace{0.7cm} (c) Superharmonic ($z$-symmetry) }\\
\includegraphics[width=0.35\textwidth,trim={0.cm 0cm 1cm 0cm},clip]{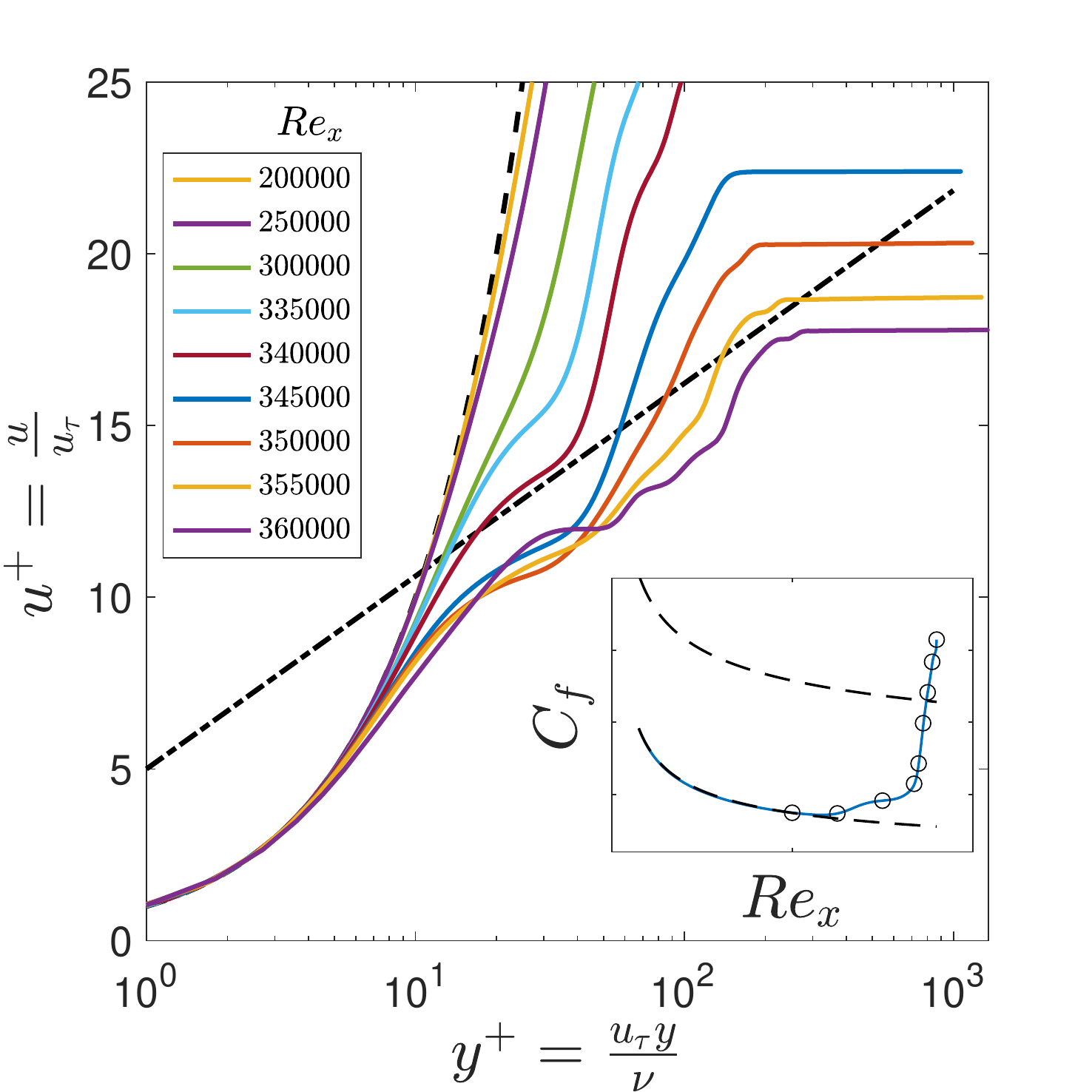}
\includegraphics[width=0.31\textwidth,trim={1.5cm 0cm 1cm 0cm},clip]{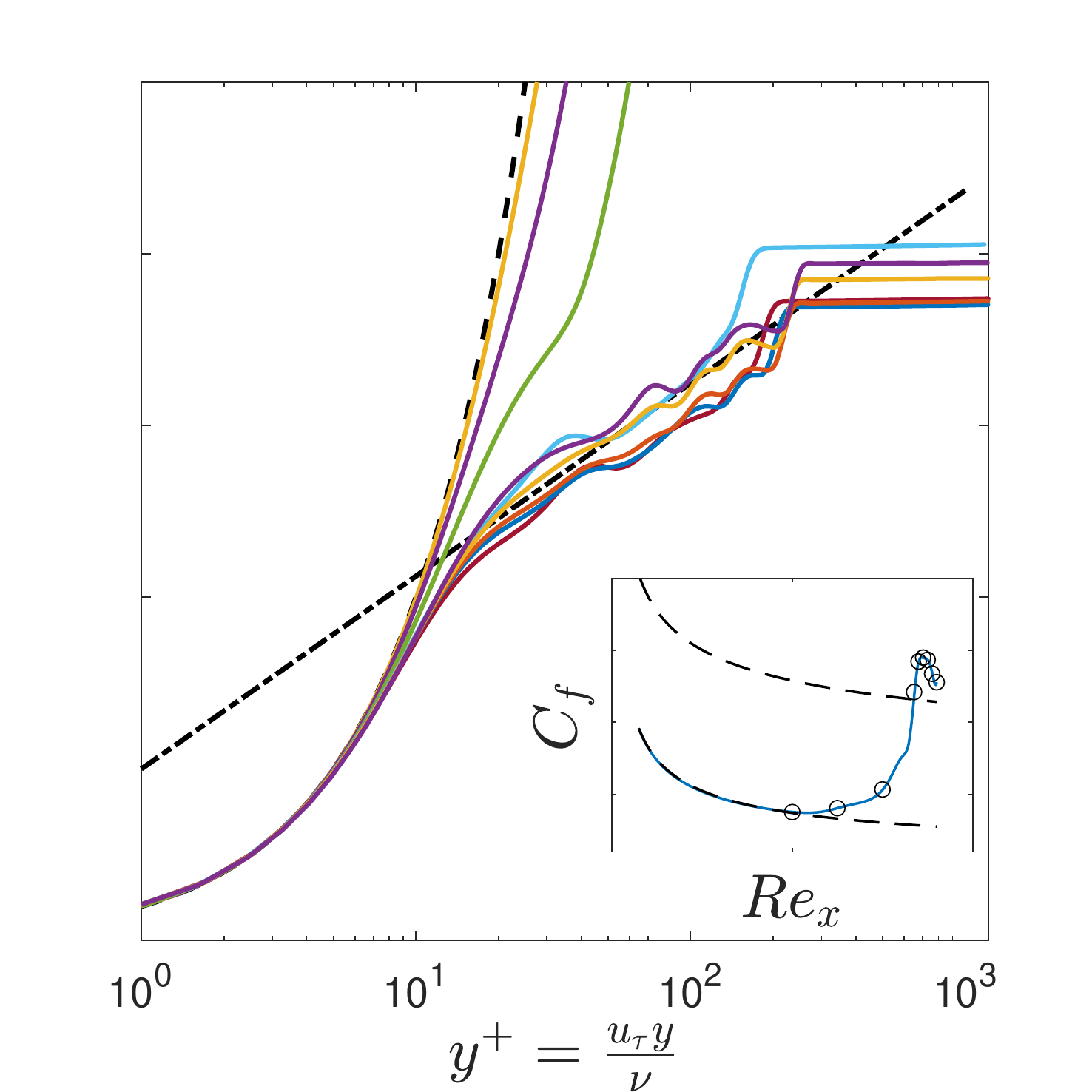}
\includegraphics[width=0.31\textwidth,trim={1.5cm 0cm 1cm 0cm},clip]{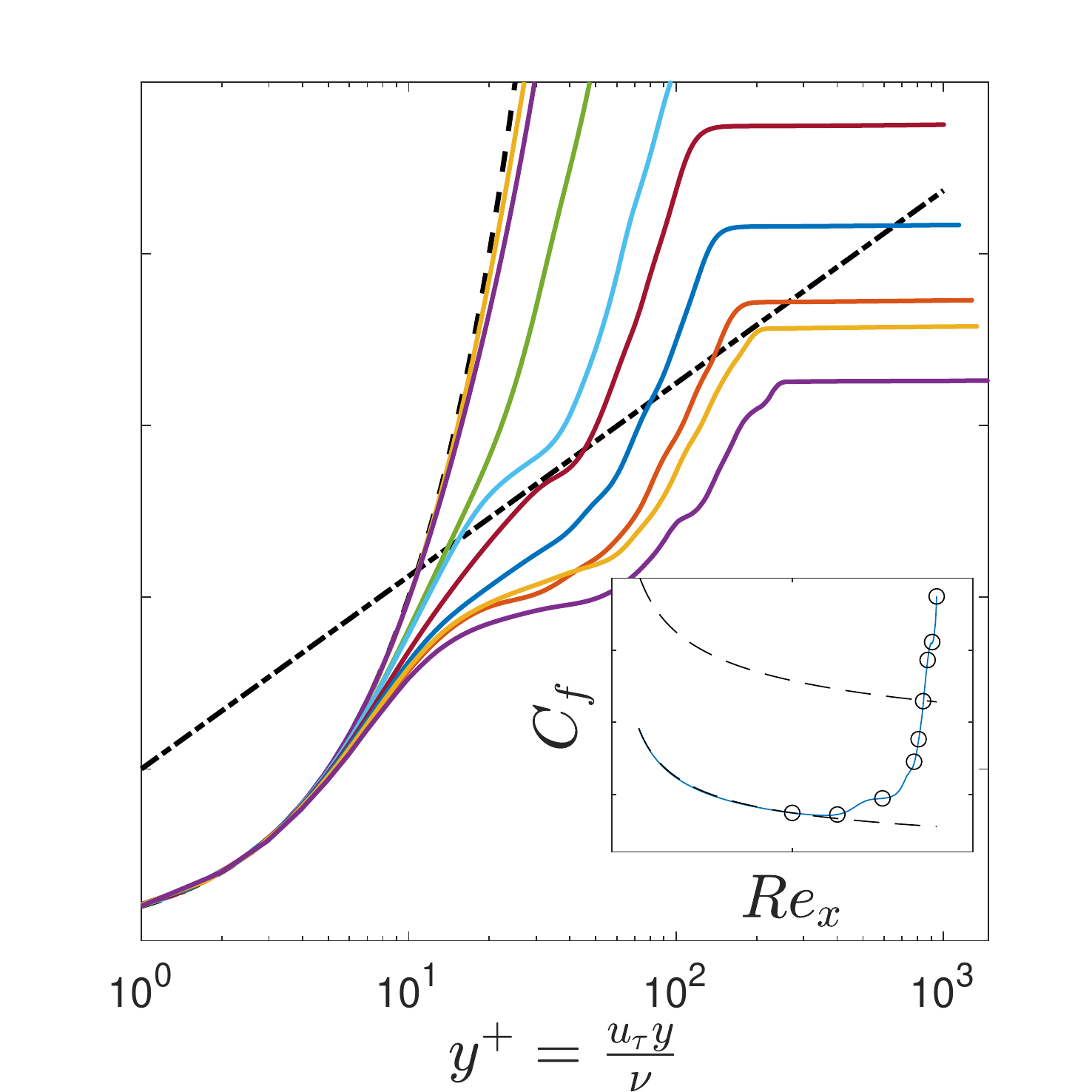}
\caption{Mean velocity profiles during transition in innner units based on the local friction velocity $u_\tau$. Linear ($u^+=y^+$; dashed) and log laws ($u^+=\frac{1}{0.41}\log y^+ + 5$; dashed-dotted) are also shown. Then insets show the skin friction coefficient as a function of $Re_x$ and the location where the velocity profiles are plot are marked with circles.}
\label{fig:velocity_profiles}
\end{figure}

Three high-amplitude forcing cases have been identified above as the worst case nonlinear disturbances that reach values of the skin friction coefficient that are close to and above the empirical turbulent values. These cases  were obtained by restricting the forcing to specific harmonic components, with or without spanwise symmetry. For the three cases considered, 
we plot the mean velocity profile at various streamwise locations for the highest forcing amplitude in figure \ref{fig:velocity_profiles}. Distinct regimes can be identified in accordance with the transition sequences observed in the previous sections.

\begin{itemize}

\item
At the very early stages of transition up to  $Re_x=200,000$ the velocity profiles obey the linear wall law $u^+=y^+$ for all three cases. This stage is characterised by linear growth of perturbations. Transition has been triggered optimally with a pair of oblique waves (fundamental cases). In the case of subharmonic instability (superharmonic case), the TS waves are also excited. 

\item
The second stage of transition is associated with the generation of streaks through nonlinear interactions of the oblique waves. At this regime the skin friction coefficient departs from the laminar Blasius values. This new regime is reflected as well through the modification of the local velocity profile outside of the viscous sublayer for $y^+>5$, in accordance with the increase of the skin friction coefficient (recall that $u_\infty^+=\sqrt{2/C_f}$). Depending on the symmetry of the forcing, varicose $\Lambda$-shaped (symmetry in $z$) or sinuous (no symmetry in $z$) low-speed streaks have been clearly identified for $Re_x>260,000$. 

\item
A third regime is observed where a distinct plateau is formed in the buffer region, $15<y^+<30$ for  all three cases. In the symmetric cases, hairpin-like vortical structures grow around the $\Lambda$-shaped low-speed streaks at $Re_x\approx 330000$. In the case without symmetry, alternate quasi-streamwise vortices grow around the sinuous low-speed streaks, i.e. $Re_x\approx 300000$. Immediately after the vortical structures are formed, the skin friction coefficient overshoots to the turbulent values.

\item
The final transition regime is associated with the breakdown. At this regime, the skin friction coefficient reaches the empirical turbulent values and energy is transferred to all the higher harmonics. 
\end{itemize}

When $z$-reflectional symmetry is imposed, the velocity profiles show qualitatively  similar characteristics as $Re_x$ increases, both for the fundamental and superharmonic cases. A monotonic decrease of the local streamwise velocity is observed in accordance with the monotonic increase of the skin friction coefficient. For the fundamental case with symmetry, a small logarithmic region is observed for $Re_x=360000$; however, the velocities are lower than those associated with the turbulent profile, which is in accordance with the increased skin friction coefficient beyond the turbulent values. For the superharmonic case and the forcing amplitudes examined here, the behavior is similar without the observation of logarithmic region. The symmetric high amplitude solutions show striking similarities with the optimal nonlinear solutions calculated by \cite{cherubini2011minimal,duguet2012self} in the time domain. However, this is not surprising since their calculations were obtained by using a symmetric initial condition as a guess for the optimization \citep{cherubini2011minimal} or spanwise symmetry was imposed \citep{duguet2012self}.

Interestingly, for the fundamental case with no symmetry in $z$ (figure \ref{fig:velocity_profiles}b), the velocity profiles at the final stages of transition show characteristics similar to the ones observed in turbulent boundary layers. Specifically, the velocity profile appears to develops a nascent logarithmic region, $u^+=\frac{1}{0.41}\log y^+ + 5$, that extends in $y^+$ as $Re_x$ increases. The skin friction coefficient, after an initial overshoot above the turbulent empirical values, drops to values close to the turbulent ones.  For this specific case, we observed sinuous low-speed streaks and quasi-streamwise staggered vortices, which are fundamental building blocks in the self-sustaining process (SSP) in a variety of streamwise homogeneous flows \citep{waleffe1997self}, in contrast to the varicose streak instability and the hairpins that were observed for the two symmetric cases.

\section{Conclusion}\label{sec:conclusions}

The nonlinear optimal mechanisms for wall-bounded laminar-turbulent transition have been investigated through solution of the frequency-domain Harmonic-Balanced Navier-Stokes equations by projecting the governing Navier-Stokes equations on to a limited number of harmonics whose triadic interactions are considered. The new framework complements previous methods that seek nonlinear optimal initial conditions in the time domain within a finite time horizon. The proposed \emph{nonlinear input/output analysis} identifies the most dangerous nonlinear forcing mechanisms that trigger transition and can be viewed as \emph{the minimal forcing seed in the frequency domain}.


Optimal nonlinear forcing mechanisms that lead to transition and maximize the skin-friction coefficient have been identified based on a variational method using direct-adjoint looping. By increasing the finite forcing amplitude, we identified the key-mechanisms that distort the laminar flow and lead to transition.  We showed that for fundamental forcing, the most amplified disturbances correspond to a pair of oblique waves with frequency and spanwise wavenumber close to the linear optimal one. Nonlinearity is responsible for redistributing the energy to the streamwise uniform vortex component which leads to the amplification of streaks through the lift-up mechanism. Depending on the imposed spanwise symmetry, the low-speed streaks break down to turbulence through varicose oscillations (imposing reflectional symmetry in spanwise) or sinuous-like ones (no symmetry in spanwise), with the latter being more efficient in promoting transition. In each case, hairpin vortices and quasi-streamwise vortices are observed prior to breakdown. When multi-harmonic forcing is allowed, the resonant interaction between oblique waves and TS waves at twice the frequency allows for even more rapid transition.  At the final stages of transition, the skin-friction coefficient reaches the empirical turbulent values and the velocity profiles depart from the law of the wall, for all cases examined here. However, only for the non-symmetric sinuous-like streak breakdown the velocity profiles develop a clear logarithmic region similar to the one observed for turbulent boundary layers. 


\vspace{0.5cm}

 We would like to thank U. Rist for providing  the details for the boundary conditions used in the DNS \citep{rist1995direct}. This work was initiated while D. Sipp was Visiting Associate at Caltech. G.R. and T.C. also acknowledge the support of the Boeing Company through a Strategic Research and Development Relationship Agreement CT-BA-GTA-1.

\appendix
\section{Mesh and domain sensitivity for K-type transition}\label{app:sensitivityKtype}

A sensitivity analysis of the domain length, the finite element discretization, and the number of retained harmonics has been performed for the K-type controlled transition. The amplitudes of the first few harmonics obtained by the HBM method are shown in figure~\ref{fig:comP_{2}}. Mesh 1 extends to $ x_o=2.52 \times 10^5 $  whereas mesh 2 is for a longer domain up to $ x_o=3.00 \times 10^5 $. For both meshes, the elements close to the plate are based on split rectangular elements, which exhibit a uniform streamwise length of 400 and a height at the plate of 40. The height of the split rectangles is stretched in the vertical direction by a factor 1.04 up to the point where the rectangles become squares. From this height, the mesh is gradually stretched isotropically up to the upper boundary. The number of triangles and degrees of fredom of the discretized problem for the two meshes and for different choices of finite elements ($[P_{1b},P_{1b},P_{1b},P_{1}] $ and $ [P_{2},P_{2},P_{2},P_{1}] $) are given in table \ref{tab:grid_parameters}.

\begin{table}
\begin{center}
\begin{tabular}{c c c c c c l}
        & MN & $x_i$  & $x_o$   & Triangles  & DOF      & Elements\\ 
 Mesh 1 & 22 & 30,000 & 252,000 & 71,207     & 357,469 & P$_{1b}$-P$_1$ \\
 Mesh 1 & 44 & 30,000 & 252,000 & 71,207     & 357,469 & P$_{1b}$-P$_1$ \\
 Mesh 1 & 22 & 30,000 & 252,000 & 71,207     & 465,352 & P$_2$-P$_1$ \\
 Mesh 2 & 22 & 30,000 & 300,000 & 86,595     & 434,653 & P$_{1b}$-P$_1$ \\
\end{tabular}
\caption{Parameters for sensitivity analysis.}
\label{tab:grid_parameters}
\end{center}
\end{table}

\begin{figure}
\centering
\includegraphics[width=0.6\textwidth]{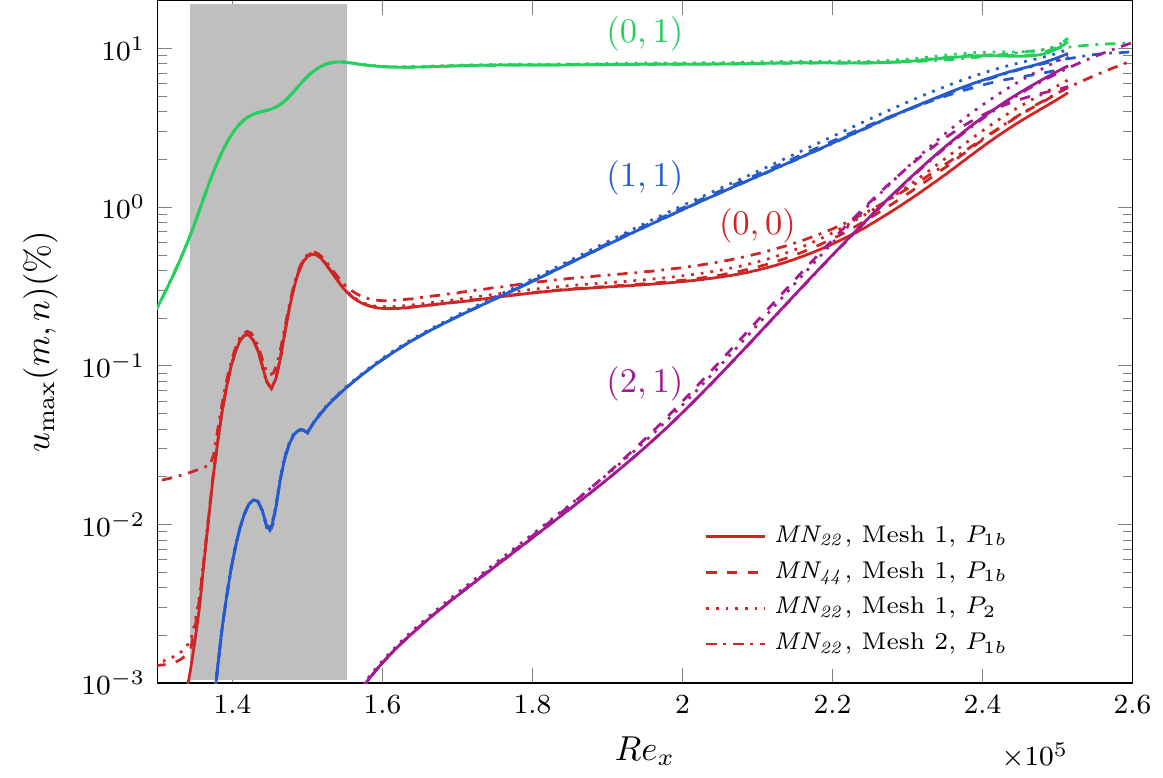}
\caption{Sensitivity of various harmonics for various choices of the numerical parameters. Note that to ease representation, we have plotted one fifth of the amplitude of harmonics $(0,\omega)$ and $(2\beta,\omega)$.}
\label{fig:comP_{2}}
\end{figure}



\section{Amplitudes of harmonics for HBM}\label{app:HarmonicAmplitudes}
In the $z-$symmetric case, the full solution may be rewritten under the form:
\begin{eqnarray}
\hat{\mathbf{w}}&=&{\mathbf{w}}_b+(\hat{\mathbf{w}}_{00}-{\mathbf{w}}_b)
+\sum_{m=1}^M (\hat{\mathbf{w}}_{m0} e^{im\beta z}+\mbox{c.c.})
+
\sum_{n=1}^N (\hat{\mathbf{w}}_{0n} e^{in\omega t}+\mbox{c.c.}) \nonumber \\
&+&
\sum_{m=1}^M \sum_{n=1}^N \left(\hat{\mathbf{w}}_{mn} e^{im\beta z+in\omega t}+\hat{\mathbf{w}}_{-mn} e^{-im\beta z+in\omega t}+\mbox{c.c.}\right)
\end{eqnarray}
The domain-integrated amplitudes of the response harmonics may be defined according to:
\begin{eqnarray}\label{eq:amplitude}
  A_{\hat{\mathbf{w}}}(m,n) =\left\{ \begin{array}{ll}
        \sqrt{(\hat{\mathbf{w}}_{00}-\mathbf{w}_b)^*\mathbf{Q}'(\hat{\mathbf{w}}_{00}-\mathbf{w}_b)} &\mbox{ if } (m,n)=(0,0) \\
         \sqrt{2\hat{\mathbf{w}}_{mn}^*\mathbf{Q}'\hat{\mathbf{w}}_{mn}}                             &\mbox{ if } (m,n) \in (0,1\cdots N) \cup (1\cdots M,0) \\
         \sqrt{4\hat{\mathbf{w}}_{mn}^*\mathbf{Q}'\hat{\mathbf{w}}_{mn}}                             &\mbox{ if } (m,n) \in (1 \cdots M,1\cdots N)
\end{array}\right.
\end{eqnarray}
where $\mathbf{Q}'$ has been defined in eq. (\ref{eq:Qp}).
The overall amplitude of all the harmonics is:
\begin{equation} \label{eq:defampl}
    A_{\hat{\mathbf{w}}}=\sqrt{\sum_{m\geq 0, n\geq 0} A_{\hat{\mathbf{w}}}(m,n)^2}
\end{equation}

The overall amplitude of the forcing $ \hat{\mathbf{f}} $ was defined in eq. \eqref{eq:surface} by the $ \mathbf{Q} $ matrix:
$$
A = \sqrt{ \hat{\mathbf{f}}^*\mathbf{Q}\hat{\mathbf{f}}}
= \sqrt{ \sum_{\substack{|m|\leq M\\ |n|\leq N\\(m,n) \neq 0}} {\hat{\mathbf{f}}_{mn}^*\mathbf{Q}\hat{\mathbf{f}}_{mn}}}.
$$
In the symmetric case, noting that $\hat{\mathbf{f}}_{00}={\mathbf{f}}_b=\mathbf{0} $, following Eq. \eqref{eq:defampl}, we have $ A_{\hat{\mathbf{f}}}=\sqrt{\hat{\mathbf{f}}^*\mathbf{Q}\hat{\mathbf{f}}}$.
The amplitudes of the individual harmonics may be represented as well with the quantity $A_{\hat{\mathbf{f}}}(m,n)$. 

The maximum amplitude of any velocity or pressure component of the state vector can be calculated in accordance with \eqref{eq:amplitude}. For example, for the $u$ component:
\begin{eqnarray}
  u_{max}(m,n) =\left\{ \begin{array}{ll}
        \max \sqrt{(\hat{u}_{00}-u_b)^*(\hat{u}_{00}-u_b)}      &\mbox{ if } (m,n)=(0,0) \\
        \max  \sqrt{2\hat{u}_{mn}^*\hat{u}_{mn}}     &\mbox{ if } (m,n) \in (0,1\cdots N) \cup (1\cdots M,0) \\
        \max  \sqrt{4\hat{u}_{mn}^*\hat{u}_{mn}}     &\mbox{ if } (m,n) \in (1 \cdots M,1\cdots N)
\end{array}\right.
\end{eqnarray}

\section{Link with weakly nonlinear analysis} \label{sec:WNL}

In this appendix, we will analyse weakly nonlinear expansions of the HBNS solutions at low amplitudes $ A \ll 1 $. We will consider two cases: section \ref{sec:one} will consider the case of a fixed forcing structure composed of a single harmonic (as obtained in the case of fundamental forcing) and section \ref{sec:two} the case with two harmonics (as obtained in the case of superharmonic forcing at point D for high forcing amplitude $A$).

\subsection{Single harmonic forcing} \label{sec:one}

Suppose that the forcing only comprises a $(\beta,\omega)$ oblique harmonic of amplitude $A$ (plus the 3 others resulting from the $z$-reflectional symmetry and the real-value contraints). This forcing will be noted $A \mathbf{\hat{f}}_{11}$ in the following.
For $A\ll1$, considering the solution under the form \eqref{eq:expansion}, the various harmonics may be expanded as:
\begin{align}
  \mathbf{\hat{w}}_{11}&=\underbrace{A\mathbf{\hat{w}}_{11}^A}_{A\mathbf{\hat{f}}_{11}} & & & &+\underbrace{A^3\mathbf{\hat{w}}_{11}^{AAA}}
  _{\substack{
  A\mathbf{\hat{w}}_{11}^{A} \times A^2\mathbf{\hat{w}}_{00}^{AA}\\
 +A\mathbf{\hat{w}}_{-1,-1}^{A} \times A^2\mathbf{\hat{w}}_{22}^{AA}\\
 +\cdots}}  & &  & & + O(A^5) 
  \\
\mathbf{\hat{w}}_{02}&=& &  \underbrace{A^2\mathbf{\hat{w}}_{02}^{AA}}_{A\mathbf{\hat{w}}_{11}^{A} \times A\mathbf{\hat{w}}_{-1,1}^{A}}& &   & &+O(A^4) & & \\
    \mathbf{\hat{w}}_{20}&=&&\underbrace{A^2\mathbf{\hat{w}}_{20}^{AA}}_{A\mathbf{\hat{w}}_{11}^{A} \times A\mathbf{\hat{w}}_{1,-1}^{A}}&&&&+O(A^4) && \\
  \mathbf{\hat{w}}_{22}&= && \underbrace{A^2\mathbf{\hat{w}}_{22}^{AA}}_{A\mathbf{\hat{w}}_{11}^{A} \times A\mathbf{\hat{w}}_{11}^{A}} && &&+ O(A^4) && \\
  \mathbf{\hat{w}}_{13}&= &&&& \underbrace{A^3\mathbf{\hat{w}}_{13}^{AAA}}_{A\mathbf{\hat{w}}_{11}^{A} \times A^2\mathbf{\hat{w}}_{02}^{AA}} && &&+ O(A^5) && \\
    \mathbf{\hat{w}}_{31}&= &&&& \underbrace{A^3\mathbf{\hat{w}}_{31}^{AAA}}_{A\mathbf{\hat{w}}_{11}^{A} \times A^2\mathbf{\hat{w}}_{20}^{AA}} && &&+ O(A^5) && \\
  \mathbf{\hat{w}}_{33}&= &&&& \underbrace{A^3\mathbf{\hat{w}}_{33}^{AAA}}_{A\mathbf{\hat{w}}_{11}^{A} \times A^2\mathbf{\hat{w}}_{22}^{AA}} && &&+ O(A^5) && \\
\mathbf{\hat{w}}_{00}&=&&\underbrace{A^2\mathbf{\hat{w}}_{00}^{AA}}_{\substack{A\mathbf{\hat{w}}_{11}^{A} \times A\mathbf{\hat{w}}_{-1,-1}^{A}\\+\cdots}}&&&&
  +\underbrace{A^4\mathbf{\hat{w}}_{00}^{AAAA}}_{\substack{A\mathbf{\hat{w}}_{11}^{A} \times A^3\mathbf{\hat{w}}_{-1,-1}^{AAA}\\+A^2\mathbf{\hat{w}}_{20}^{AA} \times A^2\mathbf{\hat{w}}_{-2,0}^{AA}\\+
  \cdots}}&&+O(A^6).
\end{align}
All non-zero terms up to order $A^3$ have been indicated for $(m,n)\neq(0,0)$, while the development is complete up to order $A^5$ for the mean-flow harmonic $(0,0)$. We have shown in the underbraces a sample of forcings  that trigger the considered term. 
Hence, the mean-friction (being a linear operator acting on  $\mathbf{\hat{w}}_{00}$) scales as:
\begin{equation} \label{eq:one}
    \Delta C_D = \Delta C_{D,2}A^2+\Delta C_{D,4}A^4+\cdots.
\end{equation}

\subsection{Two-harmonic forcing} \label{sec:two}

Suppose that the forcing lies in the $(\beta,\omega)$ oblique and $(0,2\omega)$ TS harmonics (plus the ones due to symmetry).
Similarly to the previous section, these forcings will be noted $A \mathbf{\hat{f}}_{11}$
and $A \mathbf{\hat{f}}_{02}$.
We obtain the following expansions:
\begin{align}
  \mathbf{\hat{w}}_{11}&=\underbrace{A\mathbf{\hat{w}}_{11}^A}_{A\mathbf{\hat{f}}_{11}}&&+\textcolor{red}{\underbrace{A^2\mathbf{\hat{w}}_{11}^{AA}}
  _{\substack{
  A\mathbf{\hat{w}}_{02}^{A} \times A\mathbf{\hat{w}}_{1,-1}^{A}\\
 +\cdots}}}&&+O(A^3)&& \\
\mathbf{\hat{w}}_{02}&=\color{red}{\underbrace{A\mathbf{\hat{w}}_{02}^{A}}_{A\mathbf{\hat{f}}_{02}}}&&+\underbrace{A^2\mathbf{\hat{w}}_{02}^{AA}}_{A\mathbf{\hat{w}}_{11}^{A} \times A\mathbf{\hat{w}}_{-1,1}^{A}}&&+O(A^3)&&\\
\mathbf{\hat{w}}_{20}&=&&\underbrace{A^2\mathbf{\hat{w}}_{20}^{AA}}_{A\mathbf{\hat{w}}_{11}^{A} \times A\mathbf{\hat{w}}_{1,-1}^{A}}&&+O(A^3)&&\\
\mathbf{\hat{w}}_{22}&=&&\underbrace{A^2\mathbf{\hat{w}}_{22}^{AA}}_{A\mathbf{\hat{w}}_{11}^{A} \times A\mathbf{\hat{w}}_{11}^{A}}&&+O(A^3)&&\\
\mathbf{\hat{w}}_{13}&=&&\textcolor{red}{\underbrace{A^2\mathbf{\hat{w}}_{22}^{AA}}_{A\mathbf{\hat{w}}_{11}^{A} \times A\mathbf{\hat{w}}_{02}^{A}}}&&+O(A^3)&&\\
\mathbf{\hat{w}}_{04}&=&&\textcolor{red}{\underbrace{A^2\mathbf{\hat{w}}_{04}^{AA}}_{A\mathbf{\hat{w}}_{02}^{A} \times A\mathbf{\hat{w}}_{02}^{A}}}&&+O(A^3)&&\\
\mathbf{\hat{w}}_{00}&=&&\underbrace{A^2\mathbf{\hat{w}}_{00}^{AA}}_{\substack{A\mathbf{\hat{w}}_{11}^{A} \times A\mathbf{\hat{w}}_{-1,-1}^{A}\\+A\mathbf{\hat{w}}_{02}^{A} \times A\mathbf{\hat{w}}_{0,-2}^{A}\\+\cdots}}
  &&+\textcolor{red}{\underbrace{A^3\mathbf{\hat{w}}_{00}^{AAA}}_{\substack{A\mathbf{\hat{w}}_{11}^{A} \times A^2\mathbf{\hat{w}}_{-1,-1}^{AA}\\
  A\mathbf{\hat{w}}_{02}^{A} \times A^2\mathbf{\hat{w}}_{0,-2}^{AA}\\+\cdots}}}&&+O(A^4).
\end{align}
These expansions are the same as for the single forcing harmonic case described in the previous section with additional terms marked in red. All terms that scale as $A^2$ are given explicitely for $(m,n)\neq(0,0)$, while the development is valid up to order $A^3$ for the mean-flow harmonic $(0,0)$.
The drag increase now follows
\begin{equation} \label{eq:two}
    \Delta C_D = \Delta C_{D,2}A^2+\Delta C_{D,3}A^3+\cdots.
\end{equation}

\subsection{Scalings}

\begin{figure}
\centering
\includegraphics[width=0.6\textwidth]{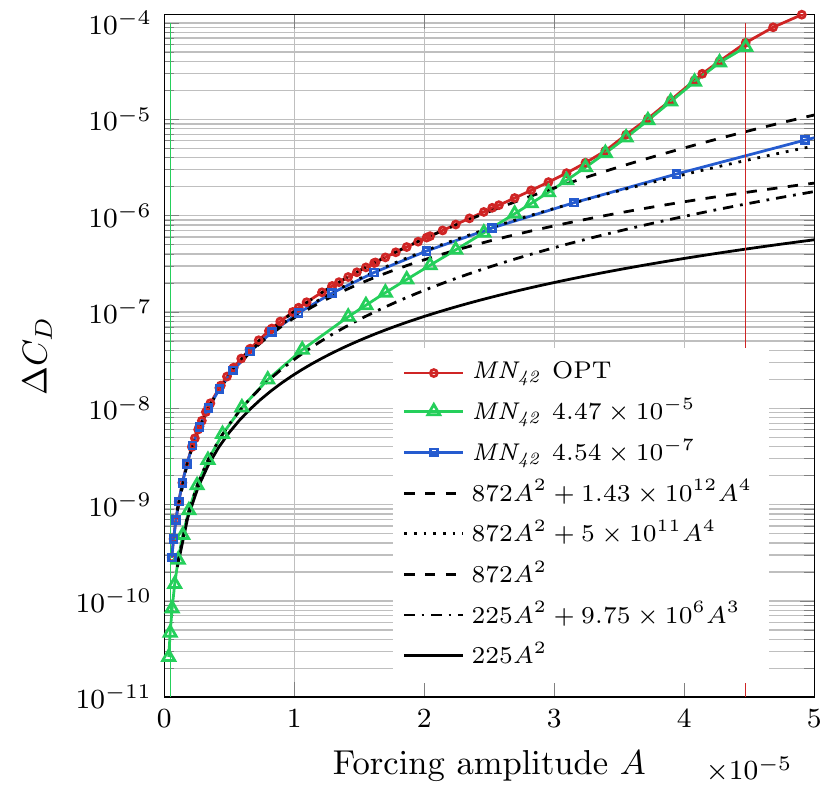}
\caption{Low amplitude $A$ scalings of drag increase in the case superharmonic forcing at point D, $(\beta,\omega)=(50,11.7)\times10^{-5}$ with $MN_{42}$. 
The red curve corresponds to the optimized solution at all amplitudes. The green curve corresponds to HBNS solutions with a fixed forcing structure corresponding to the optimal one obtained at $ A=\num{4.47e-5}$ (a $(0,2\omega)$ TS wave plus a $(\beta,\omega)$ oblique wave essentially).
 The blue curve is similar than the green curve, except that the chosen forcing structure corresponds to the one obtained at $ A=\num{5.54e-07}$ (a pure $(\beta,\omega)$ oblique forcing). Dashed lines correspond to fitting polynomial expansions for $A\ll1$.}
\label{fig:WNAH}
\end{figure}

Such scalings have been verified in figure \ref{fig:WNAH} for superharmonic forcing at poind D. The red curve corresponds to the optimized solution at all amplitudes, as presented in section \ref{sec:superconv}.

For very low amplitudes, e.g. for the left-most point at $A=\num{4.54e-7}$ on that curve in the graph, the
optimal forcing is a pure $(\beta,\omega)$ oblique forcing (for very low amplitudes, the cooperation between forcing harmonics vanishes and the optimal forcing converges to a single harmonic).
The blue curve 
then corresponds to HBNS solutions with a forcing structure frozen and equal to the one obtained for $ A=\num{4.54e-7}$, ie the above mentioned pure oblique $(\beta,\omega$ wave. Only the amplitude $ A$ was adjusted in the various computations. For low amplitudes $A$, a curve fitting technique yields the following scaling:$$
 \Delta C_D=872A^2+\num{5e11}A^4+\cdots,
$$
which is consistent with the weakly nonlinear expansion
given in eq. \eqref{eq:one}. 

For higher amplitudes, say $A=\num{4.47e-5}$ (see red vertical line on the graph), the
optimal forcing 
is essentially a combination of a $(0,2\omega)$ TS wave plus a $(\beta,\omega)$ oblique wave.
The green curve is similar than the blue curve, except that the chosen forcing structure corresponds to the one obtained at $ A=\num{4.47e-05}$, ie the just mentioned combination between a TS and an oblique wave.
Fitting this curve yields for small amplitudes:
$$
\Delta C_D=225A^2+\num{9.75e6}A^3+\cdots,
$$
which exhibits a cubic term, consistent with the development presented in eq. \eqref{eq:two}.

 It is interesting to note that the structure of the optimal forcing at $ A=\num{4.47e-5}$ is strongly sub-optimal at very low amplitudes: it becomes optimal only at high-amplitudes due to the additional odd terms in the $ \Delta C_D$ development. On the contrary (blue curve), the structure of the optimal forcing at very low amplitude only benefits from the even terms in the $\Delta C_D$ development and becomes strongly suboptimal at high amplitudes.

Finally, note that the optimal drag increase (red curve) scales as:
$$
\Delta C_D=872A^2+\num{1.43e12}A^4+\cdots,
$$
which does not exhibit a cubic term. It is more difficult to justify this expansion theoretically (as done in \S \ref{sec:one} and \S \ref{sec:two}) since the forcing structure is adjusted at all amplitudes $A$.

\bibliographystyle{jfm}
\bibliography{jfmbib}

\end{document}